%% file: main.tex
\input{preamble}

\begin{document}

\def\spacingset#1{\renewcommand{\baselinestretch}%
{#1}\small\normalsize} \spacingset{1}

\maketitle

\bigskip
\begin{abstract}
Clustering observations across partially exchangeable groups of data is a routine task in Bayesian nonparametrics. Previously proposed models allow for clustering across groups by sharing atoms in the group-specific mixing measures. However, exact atom sharing can be overly rigid when groups differ subtly, introducing a trade-off between clustering and density estimates and fragmenting across-group clusters, particularly at larger sample sizes.
We introduce the hierarchical shot-noise Cox process (HSNCP) mixture model, where group-specific atoms concentrate around shared centers through a kernel. 
This enables accurate density estimation within groups and flexible borrowing across groups, overcoming the density–clustering trade-off of previous approaches.
Our construction, built on the shot-noise Cox process, remains analytically tractable: we derive closed-form prior moments and an inter-group correlation, obtain the marginal law and predictive distribution for latent parameters, as well as the posterior of the mixing measures given the latent parameters. We develop an efficient conditional MCMC algorithm for posterior inference. We assess the performance of the HSNCP model through simulations and an application to a large galaxy dataset, demonstrating balanced across-group clusters and improved density estimates compared with the hierarchical Dirichlet process, including under model misspecification.
\end{abstract}

\vfill

\newpage
\spacingset{1.8} 

\section{Introduction}
Data arising from multiple related groups or populations, often referred to as \emph{partially exchangeable data}, are common in numerous scientific and applied fields, including genetics and bioinformatics \citep{Teh_JASA_2006, Rodriguez2008}, neuroscience \citep{Durante2017}, linguistics and text modeling \citep{Goldwater2011, Doyle2009}, econometrics \citep{Griffin2006, Bassetti2014}, and machine learning \citep{Jordan2010, Orbanz2011}. These data typically exhibit heterogeneity across groups, reflecting inherent differences in underlying experimental or observational conditions, measurement processes, or environmental contexts, yet also often share some commonalities in their distributional structure. Proper statistical modeling of such partially exchangeable data is crucial to accurately borrow strength across groups, leveraging shared information while flexibly accounting for differences. Bayesian nonparametric (BNP) approaches have emerged as powerful tools to tackle these modeling challenges, due to their intrinsic flexibility and their capacity to adapt complexity to data without imposing restrictive parametric assumptions \citep{Muller2013, Hjort2010}.

BNP models for partially exchangeable data date back to the dependent Dirichlet process introduced by \citet{MacEachern1999}; see \citet{Quintana2022} for a recent review. 
By partially exchangeable data from different groups, we mean that their law is invariant with respect to permutations that maintain the grouping structure. For infinite sequences of observations, de Finetti's representation theorem \citep[see][]{regazzini1991coherence} implies that partially exchangeable data are independent across groups and independent and identically distributed (i.i.d.) within groups, conditioning on a set of group-specific random measures, often modeled with mixtures.
That is, if $\Obs_{\GroupIdx} = (\Obs_{\GroupIdx 1}, \ldots, \Obs_{\GroupIdx \ObsNum_{\GroupIdx}})$ are data in group $\GroupIdx$, as $\GroupIdx=1, \ldots, \GroupNum$, we assume 
$\Obs_{\GroupIdx\ObsIdx} \mid \ChildProcNorm_{\GroupIdx} \iid \int_{\ChildSpace} \ObsMixt(\cdot \mid x) \ChildProcNorm_{\GroupIdx}(\dd x)$, where $(\ChildProcNorm_1, \ldots, \ChildProcNorm_\GroupNum)$ is a vector of almost surely discrete random distributions. 
The hierarchical Dirichlet process (HDP, \citealp{Teh_JASA_2006}) is a particularly notable and widely adopted example of prior for $(\ChildProcNorm_1, \ldots, \ChildProcNorm_\GroupNum)$, extensively employed in various applied contexts including topic modeling \citep{Blei2010}, genetic clustering \citep{wang2013hierarchical}, and image analysis \citep{Sudderth2008}.
Other nonparametric priors include hierarchical NRMIs \citep{Camerlenghi_AoS_2019,Argiento20}, hierarchical species sampling processes \citep{Bassetti2018}, the common atoms model of \citet{Denti_JASA_2023}, hierarchical nested models \citep{Lijoi_ScanJouStat_2023}, hierarchical finite mixture models \citep{Colombi25}, and the latent factor models of \citet{beraha2023normalized}.
Despite their methodological diversity, these models assume that the $\ChildProcNorm_{\GroupIdx}$'s are supported on the same atoms.
Alternative constructions, such as GM-dependent models \citep{lijoi2014bayesian}, latent nested models  \citep{CamerlenghiLN2019}, and the semi-HDP \citep{Beraha2021} allow for the possibility of group-specific atoms, but achieve borrowing of information across groups only through common atoms. 

\begin{figure}[t]
    \centering
    \includegraphics[width=0.8\linewidth]{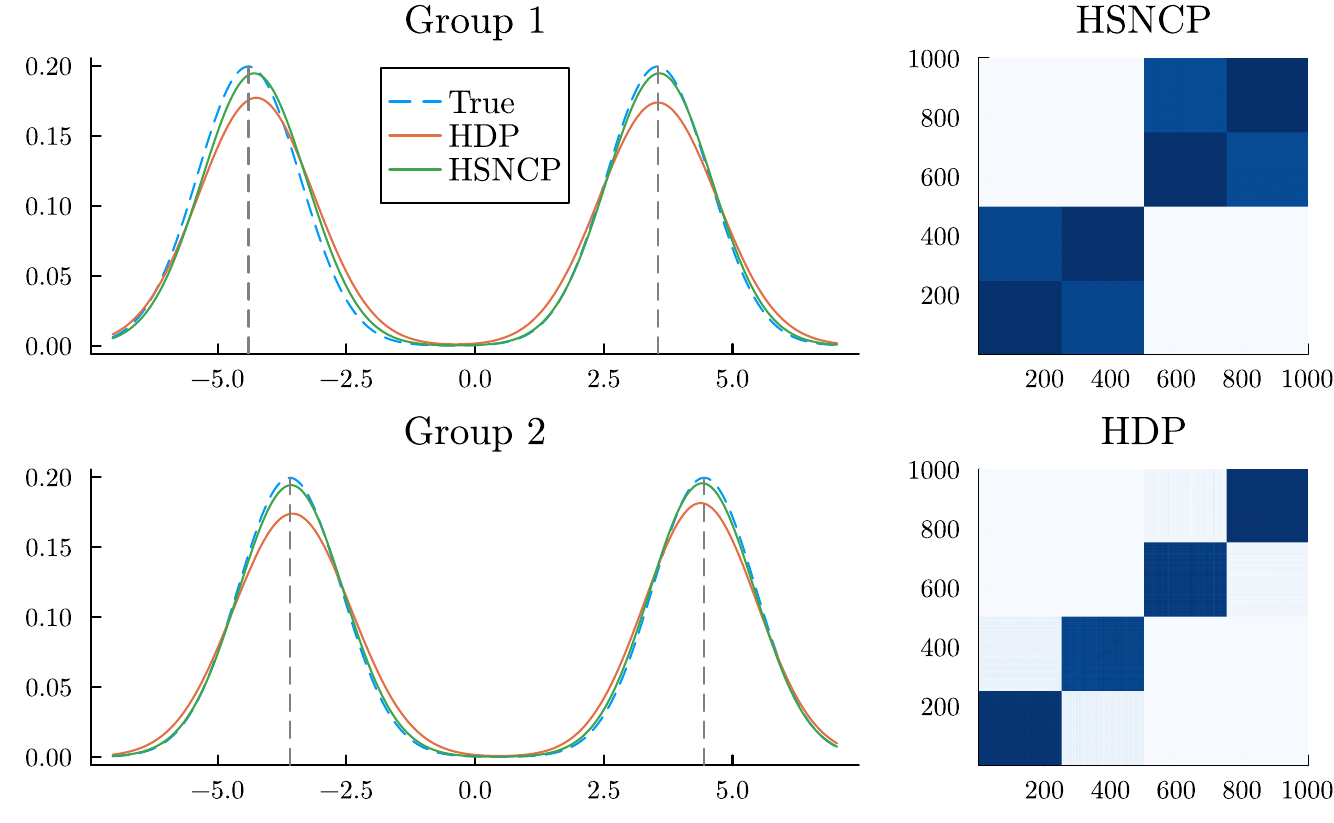}
    \caption{Illustrative simulated example with 500 observations for each group. 
    The plots on the left display the true and the estimated densities from HSNCP and HDP mixtures. 
    The right plots show the posterior similarity matrix of all data in the two models.
    }
    \label{fig:illustrative}
\end{figure}


However, exact atom-sharing has proven successful in many scenarios 
but imposes rigidity, potentially limiting flexibility when subtle differences across groups matter. 
Data in different groups could be collected with different instruments or subject to slightly diverse experimental setups.
Such subtle differences can make clustering across groups less reliable if clusters are forced to depend on identical atoms. To illustrate this point, consider the following experiment: generate $500$ observations in each of two groups from a mixture of two Gaussians, whose means are the dashed grey vertical lines in \Cref{fig:illustrative}. In the two groups of data, the cluster means differ only slightly, so we expect to identify only two clusters in the whole dataset.
When using the HDP mixture model to fit the data, we  obtain an excellent density estimate, but we recognize four clusters (see the posterior similarity matrix in the bottom row of \Cref{fig:illustrative}), due to the exact atom sharing of the HDP, i.e., there is no borrowing of information across groups.

Motivated by these practical considerations, we propose a novel framework based on the shot-noise Cox process \citep{Møller_AdvApplProb_2003} to introduce a more flexible yet structured approach to modeling and clustering partially exchangeable data. 
Our model, termed hierarchical shot-noise Cox process (HSNCP) mixture model, departs from traditional BNP models by allowing group-specific mixture distributions to have components that are not necessarily identical, but rather ``close by'' in a sense that will be made precise later. 
This new model overcomes the limitations inherent to exact atom sharing: looking at \Cref{fig:illustrative}, it is clear that the HSNCP mixture recovers exactly two clusters (see the posterior similarity matrix in the top row) and estimates perfectly the true data generating density.

The HSNCP prior defines group-specific mixing distributions $(\ChildProcNorm_1,\dots,\ChildProcNorm_{\GroupNum})$ by normalizing (dependent)
completely random measures, in the same spirit of \cite{Jam(02),Pru(02),Reg(03)}.
The dependence among the group-specific completely random measures is achieved by introducing a shared ``mother'' Poisson process which drives the borrowing of information across groups. The mother process' points are shared centers around which the group-specific atoms lay down. This improves the HSNCP prior's flexibility w.r.t.~the HDP or hierarchical NRMIs, which, in contrast,  constrain group-specific atoms to be identical.
In addition to improved flexibility, the HSNCP maintains attractive analytical and computational properties. Its structure ensures theoretical tractability, allowing for the explicit calculation of key statistical properties, such as expected values, correlation structures, and a priori clustering behavior. 
Furthermore, posterior inference remains computationally manageable, facilitated by an efficient conditional Markov Chain Monte Carlo (MCMC) algorithm.

The paper is organized as follows. \Cref{sec:model} defines our prior, and elucidates the induced clustering. In \Cref{sec:theory} we derive several analytical properties of our model: a priori moments and correlation structure, the marginal distribution of a sample from the HSNCP, the posterior distribution of the HSNCP given such a sample, and the predictive distribution which allows for a simple interpretation in terms of a restaurant franchise process with thematic rooms.
Beyond shedding light on our prior, these theoretical results are the backbone of the sampling algorithm presented in \Cref{sec:mcmc}.
Applications of our model to simulated data, including those of the illustrative example introduced above, are described in Section~\ref{sec:simulations}, and to galaxies measurements are reported in Section~\ref{sec:sloan}. We conclude with a discussion section. The supplementary materials collect all the proofs of our theoretical results, further simulation studies, and additional plots not reported in the main paper.

\section{The normalized HSNCP and associated mixtures}\label{sec:model}

An anticipated in the Introduction, we assume a mixture model for data in each group. Specifically, let $\Obs_{\GroupIdx \ObsIdx}$ be the $i$-th observation, $\ObsIdx=1, \ldots, \ObsNum_\GroupIdx$, in group $\GroupIdx=1, \ldots, \GroupNum$,  with each $\Obs_{\GroupIdx \ObsIdx} \in \R^d$. Let $(\ChildSpace, \mathcal B(\ChildSpace))$ be a Polish space equipped with the corresponding Borel $\sigma$-field, and $\ObsMixt(\cdot \mid \cdot): \R^d \times \ChildSpace \rightarrow \Rp$ be a probability density kernel, where $f(\cdot\mid x)$  is a density on $\R^d$, for every $x\in \ChildSpace$, w.r.t. the Lebesgue measure. 
Then, we assume
\begin{equation}\label{eq:likelihood}
    \Obs_{\GroupIdx \ObsIdx} \mid \ChildProcNorm_{\GroupIdx} \iid \int_\ChildSpace \ObsMixt(\cdot \mid x) \ChildProcNorm_{\GroupIdx}(\dd x),
\end{equation}
where $\ChildProcNorm_{\GroupIdx} = \sum_{\ChildIdx \geq 1} \ChildJump_{\GroupIdx \ChildIdx} \delta_{\ChildLoc_{\GroupIdx \ChildIdx}}$ is the almost sure discrete mixing probability measure in the $\GroupIdx$-th group. Conditioning to $(\ChildProcNorm_1, \ldots, \ChildProcNorm_\GroupNum)$, data in different groups are assumed to be independent.
To discuss our construction for the prior of $(\ChildProcNorm_1, \ldots, \ChildProcNorm_\GroupNum)$ we need some preliminaries.

\subsection{Background on normalized random measures}
\label{sec:background}

A fruitful approach to constructing random probability measures is by normalization of completely random measures, i.e., setting $\GenNRMI(\cdot) = \GenCRM(\cdot) / \GenCRM(\GenSpace)$, which was originally introduced in \cite{Reg(03)}.
A random measure $\GenCRM$ is a random element taking values in the space of locally finite measures over $\GenSpace$. The measure $\GenCRM$ is \emph{completely random} if for any $n > 1$ and disjoint sets $B_1, \ldots, B_n \in \mathcal B(\GenSpace)$, the random variables 
$\GenCRM(B_1), \ldots, \GenCRM(B_n)$ are mutually independent \citep{Kingman_PoissProc_1993}.
As customary in BNP, we restrict our attention to random measures of the kind $\GenCRM(B) = \int_{\Rp \times B} s \GenPP(\dd s, \dd x)$ where $\GenPP$ is a Poisson process \citep{Kingman_PoissProc_1993} on $\Rp \times \GenSpace$ with mean measure $\GenIntensity(\dd s, \dd x) = \GenMeanMass \GenMeanLoc(\dd x) \GenMeanJump( s)\dd s$ where $\GenMeanMass > 0$, $\GenMeanLoc$ is a non-atomic probability measure on $\GenSpace$ and $\GenMeanJump:\Rp\to\Rp$ is a measurable function, henceforth the L\'evy density. We write $\GenPP \sim \PP(\GenIntensity)$ and $\GenCRM \sim \CRM(\GenIntensity)$. In the context of 
point process theory, $\GenIntensity$ is the mean measure.
To ensure that the normalization of $\GenCRM$, i.e. $\GenNRMI$, is well defined, one must require $0 < \GenCRM(\GenSpace) < +\infty$ almost surely. 
As shown in \cite{Reg(03)}, sufficient conditions are 
\begin{equation}\label{eq:condNorm}
    \int_{\Rp} \GenMeanJump(s)\dd s = +\infty \quad \text{and} \quad \int_{\Rp} \min\{1, s\} \GenMeanJump(s)\dd s < +\infty.
\end{equation}
If \eqref{eq:condNorm} is enforced, $\ChildProc$ has an infinite number of support points almost surely.

\subsection{The normalized hierarchical shot-noise Cox process}
\label{sec:normHSNCP}

We now define the vector $(\ChildProcNorm_1, \ldots, \ChildProcNorm_\GroupNum)$ by shifting our focus to the construction of a  vector of random measures $(\ChildProc_1, \ldots, \ChildProc_{\GroupNum})$, and then we set $\ChildProcNorm_{\GroupIdx} = \ChildProc_{\GroupIdx} / \ChildProc_{\GroupIdx}(\ChildSpace)$.
By the discussion above, this is equivalent to specifying the law of $(\tilde N_1, \ldots, \tilde N_g)$, where each $\tilde N_\ell = \sum_{j \geq 1} \delta_{(S_{\ell h}, \phi_{\ell h})}$ is a Poisson process on $\R_+ \times \ChildSpace$. In the following, we will refer to the $\ChildJump_{\GroupIdx \ChildIdx}$'s as the \emph{jumps} and $\ChildLoc_{\GroupIdx \ChildIdx}$'s as the \emph{locations} of $\ChildProcPP_{\GroupIdx}$ (and of $\ChildProc_{\GroupIdx}$).

As in the HDP, we introduce borrowing of information across the $\tilde \mu_\ell$'s through a hierarchical formulation. Specifically, we assume that conditionally to a random measure $\eta_\Lambda$ on $\ChildSpace$, the $\tilde N_\ell$'s are i.i.d. Poisson processes with mean measure $\rho(s) \dd s \eta_\Lambda(\dd x)$ where $\rho$ is a L\'evy density. However, contrary to the HDP, we do not assume that $\eta_{\Lambda}$ is  atomic, which would cause the same exact atom sharing property, but rather that it is almost surely diffuse.
Let $\Kernel(\cdot, \cdot): \ChildSpace \times \MotherSpace \rightarrow \Rp$ be a probability density kernel and $\Lambda$ be a Poisson process on $\Rp \times \MotherSpace$ shared across groups, with mean measure $\MotherIntensity(\dd \MotherJump, \dd \MotherLoc) = \MotherMeanJump( \MotherJump)\dd \MotherJump \MotherMeanLoc(\dd \MotherLoc)$ where $\rho_0$ is a L\'evy density and $G_0$ is diffuse. Then we set $\ChildMeanLoc(\dd x) = \int_{\Rp \times \MotherSpace} \MotherJump \Kernel(x, \MotherLoc) \MotherProc(\dd \MotherJump, \dd \MotherLoc)\ \dd x$.
Note that $\ChildMeanLoc$ is not a probability measure, but can be normalized since $\int_\ChildSpace \ChildMeanLoc(\dd x) <+\infty$ by \eqref{eq:condNorm}.

Spelling out $\Lambda = \sum_{j \geq 1} \delta_{(\gamma_j, \tau_j)}$, we observe that $\ChildMeanLoc(\dd x) = \sum_{j \geq 1} \gamma_j \Kernel(\dd x, \tau_j)$.  Hence, one can easily realize that, given $\Lambda$, the locations $\phi_{\ell, h}$'s of each $\ChildProcPP_{\GroupIdx}$
are i.i.d. with density proportional to $\sum_{j \geq 1} \gamma_j \Kernel(\dd x, \tau_j)$ so that they are generated from a mixture density. 
It is worth noticing that this construction induces a random clustering of the locations of the $\tilde p_\ell$'s, while ensuring that the $\phi_{\ell h}$'s are all different almost surely, thus overcoming the drawbacks inherited by the exact-atom sharing of the HDP. Moreover, marginally, the locations of each $\ChildProcPP_{\GroupIdx}$ form a shot-noise Cox process as introduced in \cite{Møller_AdvApplProb_2003}, and the joint distribution of all locations can be seen as an instance of the multivariate shot-noise Cox process in \cite{Moller_StatInfSpat_2003}.

In the sequel, we will write $(\ChildProcNorm_1,\dots,\ChildProcNorm_{\GroupNum})\sim \NormHSNCP( \ChildMeanJump, \MotherMeanJump, \MotherMeanLoc, \Kernel(\cdot,\cdot))$ and $(\ChildProc_1,\ldots,\ChildProc_{\GroupNum})\sim \HSNCP( \ChildMeanJump, \MotherMeanJump, \MotherMeanLoc, \Kernel(\cdot,\cdot)$). For notational purposes, we also introduce a generalization of this prior, to the case where  
$\ChildProcPP_{\GroupIdx} \mid \MotherProc \ind \PP(\ChildMeanJump_\GroupIdx(s)\dd s \ChildMeanLoc(\dd x))$, being $\ChildMeanJump_\GroupIdx$ a L\'evy density on $\Rp$, $\GroupIdx=1,\ldots,\GroupNum$. In this case we write $(\ChildProcNorm_1,\dots,\ChildProcNorm_{\GroupNum})\sim \NormHSNCP((\ChildMeanJump_1,\ldots, \ChildMeanJump_\GroupNum), \MotherMeanJump, \MotherMeanLoc, \Kernel(\cdot,\cdot))$

The $\NormHSNCP$ prior shares similarities with several models previously considered. 
First, letting $\ChildSpace = \MotherSpace$ and $\Kernel(x, \MotherLoc)= \delta_{\MotherLoc}(x)$ we recover the hierarchical random measures recently considered by \cite{catalano2025hierarchical}, which encompass the HDP with a random total mass parameter.  
This highlights the importance of the kernel $\Kernel(\cdot,\cdot)$ in our model.
If instead we let $\MotherMeanJump(s) = \delta_{\MotherJump}(s)$ (i.e., the jumps of the mother process are all equal), we obtain the shot-noise Cox process mixture recently considered in \cite{Beraha_arXiv_2024}, who however consider a single group of data. Note that $\ChildSpace$ and $\MotherSpace$ will always be assumed different in the Gaussian HSNCP mixture model of Section~\ref{sec:prior_elicitation}.

\subsection{Clustering in the HSNCP mixture model}
\label{sec:hsncp_mom}

For the prior above, by exploiting the superposition property of Poisson processes it is possible to write $\ChildProcPP_{\GroupIdx} \mid \MotherProc = \sum_{\MotherIdx \geq 1} \ChildProcPP_{\GroupIdx \MotherIdx}$, where, if $\MotherProc = \sum_{\MotherIdx \geq 1} \delta_{(\MotherJump_\MotherIdx, \MotherLoc_\MotherIdx)}$, we have $\ChildProcPP_{\GroupIdx \MotherIdx} \mid \MotherProc \ind \PP(\MotherJump_\MotherIdx \ChildMeanJump(s)\dd s \Kernel(x, \MotherLoc_\MotherIdx) \dd x)$. That is to say, $\ChildProcPP_\GroupIdx$ is a \emph{cluster point process} where $\ChildProcPP_{\GroupIdx \MotherIdx}$'s locations are concentrated around the points $\MotherLoc_\MotherIdx$'s. 
Equivalently, $\ChildProc_\GroupIdx =  \sum_{\MotherIdx \geq 1} \ChildProc_{\GroupIdx \MotherIdx}$ with $\ChildProc_{\GroupIdx \MotherIdx}=\sum_{\ChildIdx \geq 1} \ChildJump_{\GroupIdx \MotherIdx\ChildIdx} \delta_{\ChildLoc_{\GroupIdx \MotherIdx \ChildIdx}}$ being the CRM associated with the Poisson process $\ChildProcPP_{\GroupIdx \MotherIdx}$.
Then, the mixture model \eqref{eq:likelihood} with an HSNCP prior is equivalent to 
\begin{equation}\label{eq:mix_of_mix}
\Obs_{\GroupIdx \ObsIdx} \mid \ChildProcNorm_\GroupIdx \iid \tilde\ObsMixt_\GroupIdx(\cdot) = \frac{1}{\ChildJump_{\GroupIdx\middot\middot}} \sum_{\MotherIdx \geq 1} S_{\GroupIdx\MotherIdx\middot} \tilde \ObsMixt_{\GroupIdx \MotherIdx}(\cdot), \ \ \text{for } \tilde \ObsMixt_{\GroupIdx\MotherIdx}(\cdot) := \frac{1}{\ChildJump_{\GroupIdx\MotherIdx\middot}} \sum_{\ChildIdx \geq 1} \ChildJump_{\GroupIdx\MotherIdx\ChildIdx} \ObsMixt(\cdot \mid \ChildLoc_{\GroupIdx \MotherIdx \ChildIdx}),
\end{equation}
where $\ChildJump_{\GroupIdx\MotherIdx\middot} = \sum_{\ChildIdx \geq 1} \ChildJump_{\GroupIdx \MotherIdx \ChildIdx}$ and $\ChildJump_{\GroupIdx\middot\middot} := \sum_{\MotherIdx \geq 1} \ChildJump_{\GroupIdx \MotherIdx\middot} = \sum_{\MotherIdx, \ChildIdx \geq 1} \ChildJump_{\GroupIdx \MotherIdx \ChildIdx}$. 
That is, \eqref{eq:likelihood} is a mixture model where the mixture components $\tilde \ObsMixt_{\GroupIdx \MotherIdx}$ are represented as nonparametric mixtures themselves.
If $\GroupNum=1$, this is akin to the mixture of mixture model in \cite{Malsiner_Walli_JCGS_2017}, who consider a fixed finite number of atoms in each mixture component. A similar model was also considered in \cite{aragam2020} in a frequentist setting, who establish consistency and identifiability properties.
Moreover, this two-level modeling bears similarities with the post-processing approach proposed by \cite{do2024dendrogram}, and the fusing of localized densities method developed by \cite{Dombowsky2025}.

\begin{figure}[!ht]
    \centering
    \includegraphics[width=\linewidth]{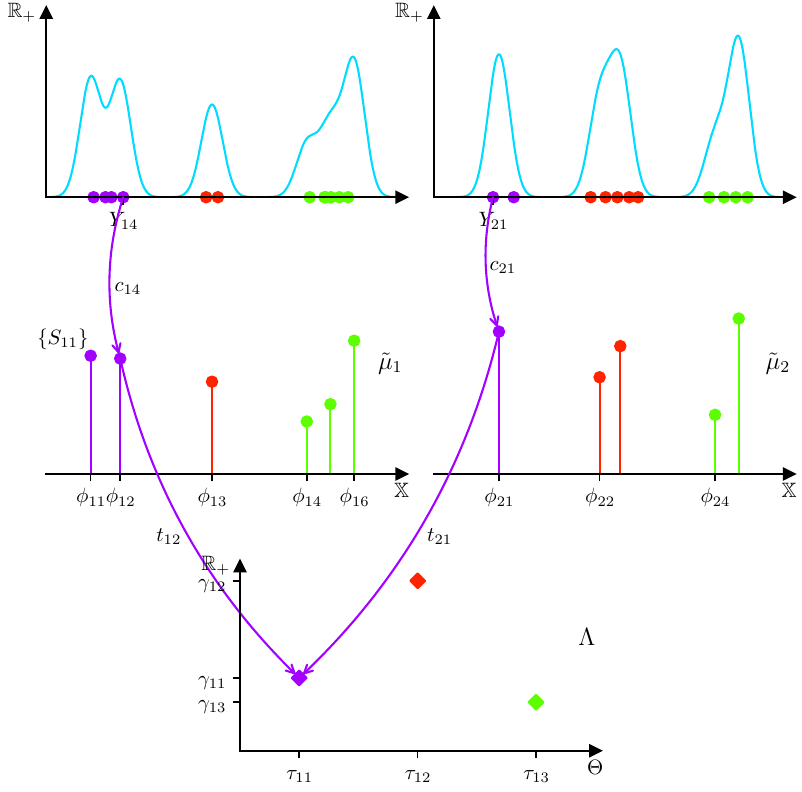}
    \caption{A (truncated) draw from a HSNCP mixture model for $\GroupNum=2$. Each stick denotes an atom of $\ChildProc_{\GroupIdx}$, with jumps given by the length of the sticks and locations denoted with $\ChildLoc_{\GroupIdx\ChildIdx}$'s. The diamonds represent the points of the mother process. The color of the sticks, the diamonds and the observations are coherent with the across-group clusters. The arrows represent the paths that obtain the association between two observations and the same point of the mother process.}
    \label{fig:HSNCP_mixture_model_example}
\end{figure}

The HSNCP mixture model induces an interesting two-level clustering structure.
First, consider \eqref{eq:likelihood} with $(\ChildProcNorm_1,\dots,\ChildProcNorm_{\GroupNum})\sim \NormHSNCP( \ChildMeanJump, \MotherMeanJump, \MotherMeanLoc, \Kernel(\cdot,\cdot))$ and introduce variables $\LatentMixt_{\GroupIdx \ObsIdx} \mid \ChildProcNorm_\GroupIdx$  $\iid \ChildProcNorm_\GroupIdx$, so that $\Obs_{\GroupIdx \ObsIdx} \mid \LatentMixt_{\GroupIdx \ObsIdx} \ind \ObsMixt(\cdot \mid \LatentMixt_{\GroupIdx \ObsIdx})$.
Since $\ChildProcNorm_\GroupIdx$ is almost surely discrete, with positive probability $\LatentMixt_{\GroupIdx \ObsIdx} = \LatentMixt_{\GroupIdx\ObsIdxAlt}$ for $\ObsIdx \neq \ObsIdxAlt$. This induces the standard clustering in mixture models by the equivalence relation $\Obs_{\GroupIdx \ObsIdx} \sim \Obs_{\GroupIdx \ObsIdxAlt}$ if and only if $\LatentMixt_{\GroupIdx \ObsIdx} = \LatentMixt_{\GroupIdx \ObsIdxAlt}$. Introducing latent cluster variables $\LatentMixtClusLabels_{\GroupIdx \ObsIdx}$, this is indicated by $\LatentMixtClusLabels_{\GroupIdx \ObsIdx} = \LatentMixtClusLabels_{\GroupIdx \ObsIdxAlt}$.
Next, let $\ChildProc_\GroupIdx = \sum_{\ChildIdx \geq 1} \ChildJump_{\GroupIdx\ChildIdx} \delta_{\ChildLoc_{\GroupIdx\ChildIdx}}$ and consider the double-summation formulation discussed in \eqref{eq:mix_of_mix}, and define latent indicator variables $\LatentChildMother_{\GroupIdx \ChildIdx}$ such that $\LatentChildMother_{\GroupIdx \ChildIdx} = \MotherIdx$ if and only if $\ChildProc_{\GroupIdx \MotherIdx}(\ChildLoc_{\GroupIdx \ChildIdx}) > 0$, i.e., the $\ChildIdx$-th atom of $\ChildProc_\GroupIdx$ belongs to $\ChildProc_{\GroupIdx \MotherIdx}$. 
Then, the clustering in the HSNCP mixture model is defined by $(\LatentChildMother_{\GroupIdx \LatentMixtClusLabels_{\GroupIdx \ObsIdx}} \mid \GroupIdx=1,\ \ldots, \GroupNum, \ \ObsIdx=1, \ldots, \ObsNum_\GroupIdx)$. That is, we say that two points $\Obs_{\GroupIdx \ObsIdx}$ and $\Obs_{\GroupIdxAlt \ObsIdxAlt}$ belong to the same cluster if $\LatentChildMother_{\GroupIdx  \LatentMixtClusLabels_{\GroupIdx \ObsIdx}} = \LatentChildMother_{\GroupIdxAlt \LatentMixtClusLabels_{\GroupIdxAlt \ObsIdxAlt}}$. 
In particular, since $\LatentChildMother_{\GroupIdx \ChildIdx} = \LatentChildMother_{\GroupIdxAlt\ChildIdxAlt}$ with positive probability even when $\GroupIdx \neq \GroupIdxAlt$, the HSNCP allows for clustering across  groups of data. 
See \Cref{fig:HSNCP_mixture_model_example} for a stylized draw from the HSNCP mixture and the associated two-level clustering.

In addition to allowing for clustering datapoints across groups, the HSNCP model also introduces a very flexible notion of cluster, whereby a cluster is not equivalent to a mixture component.
This is beneficial in applied settings, where the kernel $\ObsMixt(\cdot \mid x)$ in \eqref{eq:likelihood} does not necessarily agree with the true data generating density. In such cases, standard nonparametric mixture models provide asymptotically consistent estimators for the density but not for the clustering. That is, we face a density-clustering trade-off when dealing with misspecified models, with standard mixture models favoring density estimation over clustering. See \cite{Beraha_arXiv_2024} for a detailed discussion and \cite{guha2021posterior} for asymptotic consistency results.
The normHSNCP model sidesteps this trade-off by relying on a different and more flexible notion of clustering. As we show in our simulations, even if we let $\ObsMixt(\cdot \mid \cdot)$ and $\Kernel(\cdot, \cdot)$ be the Gaussian probability density function (defined on appropriate spaces), our model is able to correctly identify clusters also when the true data generating process is a mixture of skewed or heavy-tailed kernels, contrary to what happens with traditional mixture models.

\section{Theoretical properties of the HSNCP prior}
\label{sec:theory}

The direct study of model \eqref{eq:likelihood} under the joint normHSNCP prior is challenging. Therefore, as usually done in BNP  and mentioned at the beginning of Section~\ref{sec:hsncp_mom}, we introduce the latent variables $\LatentMixt_{\GroupIdx \ObsIdx}$ for $\GroupIdx = 1, \ldots, \GroupNum$ and $\ObsIdx = 1, \ldots, \ObsNum_{\GroupIdx}$, with values in $\ChildSpace$, such that $\Obs_{\GroupIdx \ObsIdx} \mid \LatentMixt_{\GroupIdx \ObsIdx} \ind \ObsMixt(\cdot \mid \LatentMixt_{\GroupIdx \ObsIdx})$, and study the model at the level of the latent variables. That is we consider the latent variable model
\begin{equation}\label{eq:latent_model}
\begin{aligned}
\LatentMixt_{\GroupIdx \ObsIdx} \mid \ChildProcNorm_\GroupIdx & \ind \ChildProcNorm_\GroupIdx, \ \ObsIdx=1,\ldots,\ObsNum_\GroupIdx, \textrm{ for any }\GroupIdx=1,\ldots,\GroupNum  \\
    (\ChildProcNorm_1, \ldots, \ChildProcNorm_\GroupNum) & \sim 
    \NormHSNCP( \ChildMeanJump, \MotherMeanJump, \MotherMeanLoc, \Kernel(\cdot,\cdot)).
\end{aligned}
\end{equation}
The study of \eqref{eq:latent_model} is useful for prior elicitation in the mixture model, by studying, e.g., prior moments and correlation structures.
Moreover, the marginal distribution of the $\LatentMixt_{\GroupIdx \ObsIdx}$'s and the posterior of $(\ChildProc_1, \ldots, \ChildProc_\GroupNum)$, given the $\LatentMixt_{\GroupIdx \ObsIdx}$'s, are the fundamental building blocks to design numerical algorithms for posterior inference in the full model.

We introduce some notation for ease of presentation. Let $\ChildLaplaceExp(u)=\int_{\Rp}(1-\e^{-us})\ChildMeanJump(s)\dd s$ be the Laplace exponent of (the jumps of) the $\ChildProc_\GroupIdx$'s and define $\ChildMoment(u,n)=\int_{\Rp}\e^{-us}s^n\ChildMeanJump(s)\dd s$ for $u \in \Rp, n \in \N$. Define analogously $\MotherLaplaceExp$ and $\MotherMoment$ starting from $\MotherMeanJump$. 
For a sample $\LatentMixtVec=(\LatentMixtVec_1, \ldots, \LatentMixtVec_\GroupNum)$, $\LatentMixtVec_\GroupIdx=(\LatentMixt_{\GroupIdx 1},\ldots,\LatentMixt_{\GroupIdx \ObsNum_\GroupIdx})$, of size $\ObsNum= \ObsNum_1 + \cdots + \ObsNum_\GroupNum$ from \eqref{eq:latent_model}, ties occur within each $\LatentMixtVec_\GroupIdx$ with positive probability due to the almost-sure discreteness of the $\ChildProcNorm_\GroupIdx$'s. Let $\LatentMixtUnique_{\GroupIdx 1}, \ldots, \LatentMixtUnique_{\GroupIdx \LatentMixtNumUnique_\GroupIdx}$ be the $\LatentMixtNumUnique_\GroupIdx$ unique values displayed in group $\GroupIdx$, 
with value $\LatentMixtUnique_{\GroupIdx \ChildIdx}$ appearing exactly $\LatentMixtCounter_{\GroupIdx \ChildIdx}$ times.
Finally, for $x_1, \ldots, x_p \in \ChildSpace$ let 
\begin{equation}\label{eq:m_def}
    m(\dd x_1, \ldots, \dd x_p) =  \int_{\MotherSpace} \prod_{i=1}^p \Kernel(x_i, \MotherLoc) \dd x_i \, \MotherMeanLoc(\dd \MotherLoc).
\end{equation}
That is, $m(\cdot)$ is the measure associated with the marginal distribution of a sample from the Bayesian hierarchical model $x_1, \ldots, x_p \mid \MotherLoc \iid \Kernel(\cdot, \MotherLoc)$ with prior $\MotherLoc\sim \MotherMeanLoc$.

\subsection{Prior moments}
\label{sec:a_priori_moments}
Consider the marginal law of $\ChildProcNorm_\GroupIdx$. An application of the tower rule of conditional expectations and  Proposition 1 in \cite{James_ScanJouStat_2006}  yield for any $A \in \mathcal B(\ChildSpace)$:
\begin{equation}\label{eq:prior_expectation}
    \E[\ChildProcNorm_\GroupIdx(A)] =
m(A)=\int_A\int_{\MotherSpace}\Kernel(x,\MotherLoc)\MotherMeanLoc(\dd \MotherLoc)\dd x,
\end{equation}
for any choice of $\ChildMeanJump$ and $\MotherMeanJump$.
An expression for the second moment of $\ChildProcNorm_\GroupIdx(A)$ is provided in \Cref{sec:corr_proof} of the Supplementary material. 

Beyond the first (and second) moment of $\ChildProcNorm_\GroupIdx$, it is interesting to study the correlation between $\ChildProcNorm_\GroupIdx$ and $\ChildProcNorm_{\GroupIdxAlt}$ for $\GroupIdx\neq \GroupIdxAlt$. Indeed, such a correlation sheds light on the amount of borrowing of strength induced a priori, which is often of significant interest in complex models such as ours. See, e.g., \cite{Camerlenghi_AoS_2019, Denti_JASA_2023, beraha2023normalized}.
Other indexes to measure the dependence between $\ChildProcNorm_\GroupIdx$ and $\ChildProcNorm_{\GroupIdxAlt}$ have recently been proposed in \cite{catalano2021measuring, catalano2024wasserstein}, but their construction does not directly apply to our model.
\begin{theorem}\label{teo:corr}
Let $(\ChildProcNorm_1,\dots,\ChildProcNorm_\GroupNum)\sim \NormHSNCP(\ChildMeanJump, \MotherMeanJump, \MotherMeanLoc, \Kernel(\cdot,\cdot))$. Then, for $A\in\mathcal{B}(\ChildSpace)$ and  $\GroupIdx, \GroupIdxAlt=1,\dots,\GroupNum$ with $\GroupIdx\neq\GroupIdxAlt$,
\begin{equation}\label{eq:prior_corr}
\Cor(\ChildProcNorm_\GroupIdx(A),\ChildProcNorm_{\GroupIdxAlt}(A))=\left\{\frac{\pEPPFFirstInt}{\MotherEPPFInt}+\frac{m(A)-m(A)^2}{m(A, A)-m(A)^2}\frac{\pEPPFSecondInt}{\MotherEPPFInt}\right\}^{-1},
\end{equation}
where $m(\cdot)$ is defined in \eqref{eq:m_def}, $\MotherEPPFInt = \int_{\Rp}u\e^{-\MotherLaplaceExp(u)}\MotherMoment(u,2)du$, $\pEPPFFirstInt=\int_{\Rp}u\ChildMoment(u,1)^2\MotherMoment(\ChildLaplaceExp(u),2)\e^{-\MotherLaplaceExp(\ChildLaplaceExp(u))}du$ and $\pEPPFSecondInt=\int_{\Rp}u\ChildMoment(u,2)\MotherMoment(\ChildLaplaceExp(u),1)\e^{-\MotherLaplaceExp(\ChildLaplaceExp(u))}du$.
\end{theorem}   
Although not completely evident from \eqref{eq:prior_corr}, the correlation is always positive. Indeed, it can be shown that $m(A)\ge m(A, A)\ge m(A)^2$, and hence that both addends in \eqref{eq:prior_corr} are positive.

Clearly, the correlation increases as $\MotherEPPFInt$ increases and decreases as $\pEPPFFirstInt$ or $\pEPPFSecondInt$ increase. We give an intuitive interpretation of these three quantities here, while Section~\ref{sec:corr_interpretation} of the Supplementary material reports a rigorous treatment.
Considering two observations from distinct groups $\LatentMixt_{\GroupIdx\ObsIdx}, \LatentMixt_{\GroupIdxAlt\ObsIdxAlt}$, $\MotherEPPFInt$ is the probability that they belong to the same cluster (i.e., refer to the same atom of the mother process, cf. \Cref{fig:HSNCP_mixture_model_example}). On the other hand, considering two observations in the same group, $\LatentMixt_{\GroupIdx\ObsIdx}, \LatentMixt_{\GroupIdx\ObsIdxAlt}$, $\pEPPFFirstInt$ (resp. $\pEPPFSecondInt$) is the probability that they belong to the same cluster and $\LatentMixt_{\GroupIdx\ObsIdx}\neq\LatentMixt_{\GroupIdx\ObsIdxAlt}$ (resp. $\LatentMixt_{\GroupIdx\ObsIdx}= \LatentMixt_{\GroupIdx\ObsIdxAlt}$).

We specialize our treatment to two cases of interest, namely when $\ChildMeanJump(s)\dd s$ and $\MotherMeanJump(\MotherJump)\dd\MotherJump$ are the mean measures of the gamma and generalized gamma processes, respectively. In these cases, some or all of the integrals admit closed-form expressions. 
%

\begin{figure}[h]
\centering
\includegraphics[width=\textwidth]{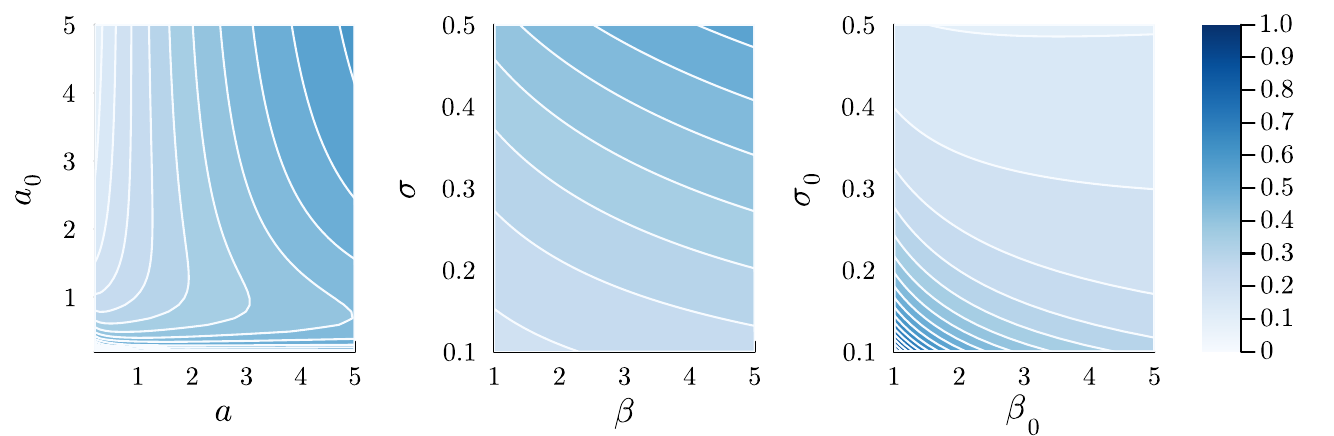}
\caption{Contour plots of the correlation between  $\ChildProcNorm_\GroupIdx(A)$ and $\ChildProcNorm_{\GroupIdxAlt}(A)$, under different prior specifications.
From left to right: Gamma processes as in \Cref{ex:gamma} as $a$ and $a_0$ vary, Generalized gamma processes as in \Cref{ex:gen_gamma} with $a_0 = 1$, $\sigma_0=0$ as $\sigma$ and $\beta$ vary, Generalized gamma processes with $a=1$, as $\sigma_0$ and $\beta_0$ vary. In all cases,
$A=(-1,1)$, $\MotherMeanLoc(\cdot)=\Gaussian(\cdot\mid0,10)$ and $\Kernel(\cdot,\MotherLoc_{\MotherIdx})=\Gaussian(\cdot\mid\MotherLoc_{\MotherIdx},1)
$.}
\label{fig:corr_examples}
\end{figure}

\begin{example}[Gamma processes]\label{ex:gamma}
If $\ChildMeanJump(s)=\ChildProcTotMass s^{-1}\e^{-s}$, where $\ChildProcTotMass>0$
then $\ChildProc_1,\dots,\ChildProc_{\GroupNum}$ 
are, conditionally on $\MotherProc$, gamma processes, and hence  
$\ChildProcNorm_1,\dots,\ChildProcNorm_{\GroupNum}$ are conditionally i.i.d.~Dirichlet processes. Further, if $\MotherMeanJump(\MotherJump)=\MotherProcTotMass \MotherJump^{-1}\e^{-\MotherJump}$,  where $\MotherProcTotMass>0$,  
then $\MotherEPPFInt = \frac{1}{1+\MotherProcTotMass}$, while $\pEPPFFirstInt$ and $\pEPPFSecondInt$ must be computed numerically.  Figure~\ref{fig:corr_examples} (left) displays the contour plot of the correlation between $\ChildProcNorm_\GroupIdx(A)$ and $\ChildProcNorm_{\GroupIdxAlt}(A)$
for different values of $a$ and $a_0$.
\end{example}

\begin{example}[Generalized gamma processes]\label{ex:gen_gamma}
If  $\ChildMeanJump(s) = \tfrac{\ChildProcTotMass}{\Gamma(1-\ChildProcSigma)} s^{-1-\ChildProcSigma}\e^{-\ChildProcTau s}$ where   $\ChildProcTotMass>0$, $\ChildProcTau\ge0$, and $\ChildProcSigma\in[0,1)$, 
then the child processes 
$\ChildProcNorm_1,\dots,\ChildProcNorm_{\GroupNum}$ are i.i.d.~normalized generalized gamma processes \citep[NGGP,][]{Jam(02), Lijoi_JRSSB_2007}. Further, if 
$\MotherMeanJump(\MotherJump)=\frac{\MotherProcTotMass}{\Gamma(1-\MotherProcSigma)} \MotherJump^{-1-\MotherProcSigma}\e^{-\MotherProcTau\MotherJump}$  where  $\MotherProcTotMass>0$, $\MotherProcTau\ge0$, and $\MotherProcSigma\in[0,1)$,
then $\MotherEPPFInt = (1-\MotherProcSigma)\left[1- \beta^{\frac1{\MotherProcSigma}}\e^{\beta}\Gamma\left(1-\frac1{\MotherProcSigma},\beta\right)\right]$, where $\beta_0 = \MotherProcTotMass / \MotherProcSigma \, \MotherProcTau^{\MotherProcSigma}$ and $\Gamma(\cdot,\cdot)$ is the upper incomplete gamma function, while $\pEPPFFirstInt$ and $\pEPPFSecondInt$ must be computed numerically.  Figure~\ref{fig:corr_examples} (center and right) displays the contour plot of the correlation between $\ChildProcNorm_\GroupIdx(A)$ and $\ChildProcNorm_{\GroupIdxAlt}(A)$ when the child processes are gamma or generalized gamma processes. 
\end{example}

\begin{example}[$\sigma$-stable processes]
Assume the child processes are generalized gamma processes with $\ChildProcTau=0$ (i.e., they are $\sigma$-stable processes) and $\MotherMeanJump(\MotherJump)=\frac{\MotherProcTotMass}{\Gamma(1-\MotherProcSigma)} \MotherJump^{-1-\MotherProcSigma}\e^{-\MotherProcTau\MotherJump}$ with $\MotherProcTau= 0$. Then, we have $\MotherEPPFInt=1-\MotherProcSigma$, $\pEPPFFirstInt=\ChildProcSigma(1-\MotherProcSigma)$, $\pEPPFFirstInt=1-\ChildProcSigma$, and the correlation admits a closed-form expression: 
$$
\Cor(\ChildProcNorm_{\GroupIdx}(A),\ChildProcNorm_{\GroupIdxAlt}(A))=\left\{\ChildProcSigma+\frac{m(A)-m(A)^2}{m(A, A)-m(A)^2}\frac{1-\ChildProcSigma}{1-\MotherProcSigma} \right\},
$$
which does not depend on $\MotherProcTotMass$ and $\ChildProcTotMass$, increases with $\ChildProcSigma$ and decreases in $\MotherProcSigma$. When $\Kernel(\cdot,\MotherLoc_\MotherIdx)=\delta_{\MotherLoc_\MotherIdx}(\cdot)$, the correlation simplifies to $\frac{1-\MotherProcSigma}{1-\ChildProcSigma\MotherProcSigma}$, which is the same expression as for the hierarchical NRMIs derived from $\sigma$-stable processes \citep{Camerlenghi_AoS_2019}.
In all other cases, the correlations of hierarchical NRMIs and HSNCP differ, because the construction of HSNCP implies no normalization of the mother process.
\end{example}

\subsection{Marginal and posterior distributions}

We now move to the Bayesian analysis of model \eqref{eq:latent_model}, that is the model for the latent parameters $\LatentMixt_{\GroupIdx \ObsIdx}$'s of the mixtures \eqref{eq:likelihood}. To this end, as customary in the analysis of NRMIs \citep{James_ScanJouStat_2009}, we introduce latent variables $\ChildAux_\GroupIdx \mid \ChildProc_\GroupIdx \ind \mathrm{Gamma}(\ObsNum_\GroupIdx, \ChildProc_\GroupIdx(\ChildSpace))$.  See also \cite{Favaro_StatSci_2013}.
Recall the notation introduced at the beginning of this section.
With a slight abuse of notation, we introduce another set of latent variables $\LatentChildMotherObsVec = (\LatentChildMotherObs_{\GroupIdx \ChildIdx}, \, \GroupIdx=1, \ldots, \GroupNum, \ChildIdx = 1, \ldots, \LatentMixtNumUnique_\ChildIdx)$ describing a partition of the unique values $\LatentMixtUnique_{\GroupIdx \ChildIdx}$'s. Informally, these play the same role of the variables $\LatentChildMother_{\GroupIdx \ChildIdx}$ introduced in \Cref{sec:hsncp_mom}. However, the $\LatentChildMother_{\GroupIdx \ChildIdx}$'s refer to the atoms of the $\ChildProc_\GroupIdx$'s, while the $\LatentChildMotherObs_{\GroupIdx \ChildIdx}$'s refer only to those atoms displayed in the sample. Define $\LatentChildMotherCounter_\MotherIdx = \sum_{\GroupIdx=1}^\GroupNum \sum_{\ChildIdx=1}^{\LatentMixtNumUnique_\GroupIdx} \mathbb I[\LatentChildMotherObs_{\GroupIdx \ChildIdx} = \MotherIdx]$.

\begin{theorem}\label{teo:marg}
    Let $\LatentMixtVec_1, \ldots, \LatentMixtVec_\GroupNum$ be a sample of size $\ObsNum = \ObsNum_1 + \cdots + \ObsNum_\GroupNum$ from \eqref{eq:latent_model}. Then, it is possible to introduce latent variables $\LatentChildMotherObsVec$, describing a partition of the unique values $\LatentMixtUnique_{\GroupIdx \ChildIdx}$, and $\ChildAuxVec \in \Rp^{\GroupNum}$, such that the joint distribution of the sample, $\LatentChildMotherObsVec$ and $\ChildAuxVec$ is
    \begin{multline}
    \label{eq:latent_joint}
\Law(\LatentMixtVec_{1},\dots,\LatentMixtVec_{\GroupNum},\LatentChildMotherObsVec, \ChildAuxVec) =  
\exp\left\{-\MotherLaplaceExp\left(\sum_{\GroupIdx=1}^\GroupNum\ChildLaplaceExp(\ChildAux_\GroupIdx)\right)\right\} \prod_{\GroupIdx=1}^{\GroupNum} \left\{\frac{\ChildAux_\GroupIdx^{\ObsNum_\GroupIdx - 1}}{\Gamma(\ObsNum_\GroupIdx)}  \prod_{ \ChildIdx=1}^{\LatentMixtNumUnique_\GroupIdx} \ChildMoment(\ChildAux_\GroupIdx, \LatentMixtCounter_{\GroupIdx \ChildIdx}) \right\} \\
        \times  \prod_{\MotherIdx=1}^{\LatentChildMotherNumUnique} \MotherMoment\left(\sum_{\GroupIdx=1}^\GroupNum\ChildLaplaceExp(\ChildAux_\GroupIdx), \LatentChildMotherCounter_\MotherIdx\right) m(\dd\LatentMixtUnique_{\GroupIdx \ChildIdx} \colon \LatentChildMotherObs_{{\GroupIdx \ChildIdx}} = \MotherIdx)
    \end{multline}
where $\MotherLaplaceExp$, $\MotherMoment$, $\ChildLaplaceExp$, $\ChildMoment$ are defined in Section~\ref{sec:theory}, and $m(\cdot)$ is defined in \eqref{eq:m_def}. Expression $m(\dd\LatentMixtUnique_{\GroupIdx \ChildIdx} \colon \LatentChildMotherObs_{{\GroupIdx \ChildIdx}} = \MotherIdx)$ is the joint marginal law of all $\LatentMixtUnique_{\GroupIdx \ChildIdx}$'s for which $\LatentChildMotherObs_{{\GroupIdx \ChildIdx}} = \MotherIdx$.
\end{theorem}

 Marginalizing with respect to $\LatentChildMotherObsVec$ and $\ChildAuxVec$ from the joint distribution of \Cref{teo:marg}, one obtains the joint marginal distribution of a sample $\LatentMixtVec_{1},\dots,\LatentMixtVec_{\GroupNum}$ from \eqref{eq:latent_model}.
Moreover, an application of Bayes' theorem yields the posterior distribution of $(\LatentChildMotherObsVec, \ChildAuxVec) \mid \LatentMixtVec$, which is proportional to \eqref{eq:latent_joint}. For our purposes, it is also useful to derive the ``full conditional'' distributions $\Law(\LatentChildMotherObsVec \mid \ChildAuxVec, \LatentMixtVec)$ and $\Law(\ChildAuxVec \mid \LatentChildMotherObsVec, \LatentMixtVec)$ that can be obtained from \eqref{eq:latent_joint}. In particular,
\begin{equation}\label{eq:post_t_u}
   \Law(\LatentChildMotherObsVec \mid \ChildAuxVec, \LatentMixtVec) \propto  \exp\left\{-\MotherLaplaceExp\left(\sum_{\GroupIdx=1}^\GroupNum\ChildLaplaceExp(\ChildAux_\GroupIdx)\right)\right\}  \prod_{\MotherIdx=1}^{\LatentChildMotherNumUnique} \MotherMoment\left(\sum_{\GroupIdx=1}^\GroupNum\ChildLaplaceExp(\ChildAux_\GroupIdx), \LatentChildMotherCounter_\MotherIdx\right) m(\dd\LatentMixtUnique_{\GroupIdx \ChildIdx} \colon \LatentChildMotherObs_{{\GroupIdx \ChildIdx}} = \MotherIdx)
\end{equation}
matches the posterior distribution for the cluster allocations in a standard nonparametric mixture model with data $(\LatentMixtUnique_{\GroupIdx \ChildIdx}, \, \GroupIdx=1, \ldots, \GroupNum, \,  \ChildIdx=1, \ldots, \LatentMixtNumUnique_\GroupIdx)$, mixture kernel $\Kernel(\cdot,  \cdot)$, base measure $\MotherMeanLoc$, and prior for the partition given by 
\[
\mathcal L(\LatentChildMotherObsVec \mid \ChildAuxVec) \propto \exp\left\{-\MotherLaplaceExp\left(\sum_{\GroupIdx=1}^\GroupNum\ChildLaplaceExp(\ChildAux_\GroupIdx)\right)\right\} \prod_{\MotherIdx=1}^{\LatentChildMotherNumUnique} \MotherMoment\left(\sum_{\GroupIdx=1}^\GroupNum\ChildLaplaceExp(\ChildAux_\GroupIdx), \LatentChildMotherCounter_\MotherIdx\right),
\]
which is a symmetric function of the counts $\zeta_j$'s. In particular, it is an exchangeable partition probability function \citep{Pitman_Prob_1995}.
See \cite{James_ScanJouStat_2009} for further details.
Hence, \eqref{eq:post_t_u} can be sampled via MCMC algorithms for traditional mixture models, such as the ones in \cite{Neal_JCGS_2000} or \cite{Favaro_StatSci_2013}.

Next, we turn to the study of the posterior distribution. By a disintegration argument, the posterior law of the $\ChildProc_\GroupIdx$'s is uniquely determined by the conditional law $\Law(\ChildProc_1, \ldots, \ChildProc_\GroupNum \mid \LatentMixtVec, \LatentChildMotherObsVec, \ChildAuxVec)$, having already established the posterior of $(\LatentChildMotherObsVec, \ChildAuxVec)$ before.
\begin{theorem}
\label{teo:post}
    Let $\LatentMixtVec_1, \ldots, \LatentMixtVec_\GroupNum$ be a sample of size $\ObsNum = \ObsNum_1 + \cdots + \ObsNum_\GroupNum$ from \eqref{eq:latent_model}. Then, the conditional distribution of $(\ChildProc_1, \ldots, \ChildProc_\GroupNum)$ given $(\LatentMixtVec, \LatentChildMotherObsVec, \ChildAuxVec)$ is equal to the law of the vector of random measures $(\ChildProc_1^*, \ldots, \ChildProc_\GroupNum^*)$ characterized as follows:
    \[
        \ChildProc_\GroupIdx^* = \sum_{\ChildIdx=1}^{\LatentMixtNumUnique_\GroupIdx}\ChildJump^*_{\GroupIdx\ChildIdx}\delta_{\LatentMixtUnique_{\GroupIdx\ChildIdx}} +\sum_{\MotherIdx=1}^{\LatentChildMotherNumUnique}\ChildProc^{(p)}_{\GroupIdx\MotherIdx} + \ChildProc_\GroupIdx^{(p)}
    \]
 whose elements are defined below. 
    \begin{itemize}
        \item[$(i)$] The $\ChildJump^*_{\GroupIdx\ChildIdx}$'s are independent positive random variables with density $f_{\ChildJump^*_{\GroupIdx\ChildIdx}}(s) \propto \e^{-\ChildAux_{\GroupIdx}s}s^{\LatentMixtCounter_{\GroupIdx\ChildIdx}}\ChildMeanJump(s)$.

        \item[$(ii)$] The $\ChildProc^{(p)}_{\GroupIdx \MotherIdx}$'s are independent random measures, such that 
        \[
         \ChildProc^{(p)}_{\GroupIdx\MotherIdx} \mid (\MotherJump^{(p)}_\MotherIdx, \MotherLoc^{(p)}_\MotherIdx)  \ind \CRM(\e^{- \ChildAux_\GroupIdx s}\ChildMeanJump(s)\dd s \, \MotherJump^{(p)}_\MotherIdx \Kernel(x, \MotherLoc^{(p)}_\MotherIdx) \dd x)
        \]
       where, as $\MotherIdx=1, \ldots , \LatentChildMotherNumUnique$, the $\MotherJump_\MotherIdx^{(p)}$'s and  $\MotherLoc^{(p)}_\MotherIdx$'s are independent random variables with densities  $f_{\MotherJump_{\MotherIdx}^{(p)}}(\MotherJump)$'s and $f_{\MotherLoc^{(p)}_\MotherIdx}(\MotherLoc)$, respectively, proportional to
         \[
            f_{\MotherJump_{\MotherIdx}^{(p)}}(\MotherJump) \propto \exp\{ \MotherJump\sum_{\GroupIdx=1}^{\GroupNum}\ChildLaplaceExp(\ChildAux_{\GroupIdx})\}\MotherJump^{\LatentChildMotherCounter_{\MotherIdx}}\MotherMeanJump(\MotherJump), \quad 
            f_{\MotherLoc^{(p)}_\MotherIdx}(\MotherLoc) \propto \prod_{(\GroupIdx, \ChildIdx): \LatentChildMotherObs_{\GroupIdx \ChildIdx} = \MotherIdx} \Kernel(\LatentMixtUnique_{\GroupIdx \ChildIdx}, \MotherLoc) \MotherMeanLoc(\dd \MotherLoc) .
        \]
        \item[$(iii)$] $(\ChildProc_1^{(p)}, \ldots, \ChildProc_\GroupNum^{(p)}) \sim \HSNCP( (\ChildMeanJump_1^{(p)},\ldots,\ChildMeanJump_{\GroupNum}^{(p)}), \MotherMeanJump^{(p)}, \MotherMeanLoc, \Kernel(\cdot,\cdot))$        
        with $\MotherMeanJump^{(p)}(\MotherJump) = \e^{-\MotherJump\sum_{\GroupIdx=1}^{\GroupNum}\ChildLaplaceExp(\ChildAux_\GroupIdx)}\MotherMeanJump(\MotherJump)$, and $\ChildMeanJump_{\GroupIdx}^{(p)}(s) = \e^{-\ChildAux_\GroupIdx s} \ChildMeanJump(s)$ for $\GroupIdx = 1, \ldots, \GroupNum$.
    \end{itemize}
\end{theorem}
Theorem \ref{teo:post} sheds light on the peculiar structure of the HSNCP model. In particular, the same two-level clustering property of the HSNCP is recovered also a posteriori, as we can interpret the $\tau_j^{(p)}$'s as latent cluster centers and observe that each $\ChildProc^{(p)}_{\GroupIdx\MotherIdx}$ has support concentrated around $\tau_j^{(p)}$: these correspond to those atoms $\phi_{\ell j}$ that belong to the $p$-th cluster but have not been discovered yet.
Moreover, Theorem \ref{teo:post} is the backbone of algorithm for posterior simulation described in Section~\ref{sec:mcmc}.

\subsection{The restaurant process and predictive distributions}\label{sec:pred}

Similarly to the HDP and hierarchical NRMI processes, the marginal distribution of $\LatentMixtVec_1, \ldots, \LatentMixtVec_\GroupNum$ can be expressed via a sequential allocation process known as the Chinese restaurant franchise \citep{Teh_JASA_2006}. 
\begin{theorem}\label{teo:pred}
Let $\LatentMixtVec_1, \ldots, \LatentMixtVec_\GroupNum$ be a sample of size $\ObsNum = \ObsNum_1 + \cdots + \ObsNum_\GroupNum$ from \eqref{eq:latent_model}. Then, for any $A\in{\mathcal B}(\ChildSpace)$, we have
\begin{align*}
\Prob(\LatentMixt_{\GroupIdx, \ObsNum_{\GroupIdx} + 1} \in A \mid \LatentMixtVec, \LatentChildMotherObsVec, \ChildAuxVec ) & = \frac{\ChildAux_\GroupIdx}{\Gamma(\ObsNum_\GroupIdx)} \sum_{\ChildIdx=1}^{\LatentMixtNumUnique_\GroupIdx}\frac{\ChildMoment(\ChildAux_\GroupIdx,\LatentMixtCounter_{\GroupIdx\ChildIdx}+1)}{\ChildMoment(\ChildAux_\GroupIdx,\LatentMixtCounter_{\GroupIdx\ChildIdx})}\delta_{\LatentMixtUnique_{\GroupIdx\ChildIdx}}(A)\\
&\ +\frac{\ChildAux_\GroupIdx}{\Gamma(\ObsNum_\GroupIdx)}\sum_{\MotherIdx=1}^{\LatentChildMotherNumUnique} \ChildMoment(\ChildAux_\GroupIdx,1)\frac{\MotherMoment\left(\sum_{\GroupIdx=1}^{\GroupNum}\ChildLaplaceExp(\ChildAux_{\GroupIdx}),\LatentChildMotherCounter_{\MotherIdx}+1\right)}
{\MotherMoment\left(\sum_{\GroupIdx=1}^{\GroupNum}\ChildLaplaceExp(\ChildAux_{\GroupIdx}),\LatentChildMotherCounter_{\MotherIdx}\right)}\int_A m(\dd x\mid\LatentMixtUnique_{\GroupIdx \ChildIdx} \colon \LatentChildMotherObs_{{\GroupIdx \ChildIdx}} = \MotherIdx)\\
&\ +\frac{\ChildAux_\GroupIdx}{\Gamma(\ObsNum_\GroupIdx)} \ChildMoment(\ChildAux_\GroupIdx,1)\MotherMoment\left(\sum_{\GroupIdx=1}^{\GroupNum}\ChildLaplaceExp(\ChildAux_{\GroupIdx}),1\right)\int_A m(\dd x)
\end{align*}
where $\MotherLaplaceExp$, $\MotherMoment$, $\ChildLaplaceExp$, $\ChildMoment$ are defined in Section~\ref{sec:theory}, $m(\dd x)$ is defined in \eqref{eq:m_def}, and 
$$
m(\dd x\mid \LatentMixtUnique_{\GroupIdx\ChildIdx}:\LatentChildMotherObs_{\GroupIdx\ChildIdx}=\MotherIdx)=\int_{\MotherSpace}\Kernel(x,\MotherLoc_{\MotherIdx})\frac{\prod_{(\GroupIdx,\ChildIdx):\LatentChildMotherObs_{\GroupIdx\ChildIdx}=\MotherIdx} \Kernel(\LatentMixtUnique_{\GroupIdx\ChildIdx}, \MotherLoc)\MotherMeanLoc(\dd \MotherLoc) \dd x}{\int_{\MotherSpace}\prod_{(\GroupIdx,\ChildIdx):\LatentChildMotherObs_{\GroupIdx\ChildIdx}=\MotherIdx} \Kernel(\LatentMixtUnique_{\GroupIdx\ChildIdx}, \MotherLoc')\MotherMeanLoc(\dd \MotherLoc') \dd x}
$$
is the marginal probability measure of $x$ (such that $x\mid\MotherLoc\sim \Kernel(\cdot,\MotherLoc), \MotherLoc\sim\MotherMeanLoc$) conditioning on all $\LatentMixtUnique_{\GroupIdx \ChildIdx}$'s for which $\LatentChildMotherObs_{{\GroupIdx \ChildIdx}} = \MotherIdx$.
\end{theorem}
The result in \Cref{teo:pred} suggests the following metaphor. Consider a restaurant franchise where tables are associated with dishes corresponding to the unique values $\LatentMixtUnique_{\GroupIdx \ChildIdx}$. 
Tables are arranged into several \emph{thematic rooms}, with themes being shared across restaurants of the franchise, such that table serving dish $\LatentMixtUnique_{\GroupIdx \ChildIdx}$ is in room $\LatentChildMotherObs_{\GroupIdx \ChildIdx}$ of the $\GroupIdx$-th restaurant. Contrary to the HDP franchise, here tables serve different dishes almost surely, but tables in the same room have similar dishes.
For instance, if the theme of a room is pizza, all the tables in that room serve dishes related to pizza, such as Margherita, Pepperoni, or Hawaiian. Each theme represents an atom of the mother process, each table in restaurant $\GroupIdx$ represents an atom of $\ChildProcNorm_{\GroupIdx}$, and each customer of restaurant $\GroupIdx$ represents a sample from $\ChildProcNorm_{\GroupIdx}$.
Then, the predictive law in \Cref{teo:pred} can be interpreted as follows. $\LatentMixt_{1,1}$ enters in the first thematic room of the first restaurant and samples dish $\LatentMixtUnique_{1, 1}$ from $m(\cdot)$ defined in \eqref{eq:m_def} with $p=1$, and we set $\LatentChildMotherObs_{1, 1} = 1$.
After customers $\LatentMixtVec_1, \ldots, \LatentMixtVec_{\GroupNum}$ have entered in the restaurant, customer $\LatentMixt_{\GroupIdx, \ObsNum_\GroupIdx + 1}$, conditionally to $(\LatentMixtVec, \LatentChildMotherObsVec)$, either: 
\begin{itemize}
\item[$(i)$] Sits at a previously occupied table, serving dish $\LatentMixtUnique_{\GroupIdx \ChildIdx}$ to $\LatentMixtCounter_{\GroupIdx \ChildIdx}$ customers, with probability proportional to $\frac{\ChildMoment(\ChildAux_\GroupIdx, \LatentMixtCounter_{\GroupIdx \ChildIdx}+1)}{\ChildMoment(\ChildAux_\GroupIdx, \LatentMixtCounter_{\GroupIdx \ChildIdx})}$. In this case, we set $\LatentMixt_{\GroupIdx, \ObsNum_\GroupIdx + 1} = \LatentMixtUnique_{\GroupIdx \ChildIdx}$ and leave $\LatentChildMotherObsVec$ unchanged.

\item[$(ii)$] Sits in a new table in room $\MotherIdx$, with probability proportional to $\ChildMoment(\ChildAux_\GroupIdx, 1) \frac{\MotherMoment\left(\sum_{\GroupIdx=1}^\GroupNum\ChildLaplaceExp(\ChildAux_\GroupIdx), \LatentChildMotherCounter_\MotherIdx + 1\right)}{\MotherMoment\left(\sum_{\GroupIdx=1}^\GroupNum\ChildLaplaceExp(\ChildAux_\GroupIdx), \LatentChildMotherCounter_\MotherIdx\right)}$. 
In this case, we sample $\LatentMixtUnique_{\GroupIdx, \LatentMixtNumUnique_\GroupIdx + 1}$ from $m(\cdot\mid\dd\LatentMixtUnique_{\GroupIdx \ChildIdx} \colon \LatentChildMotherObs_{{\GroupIdx \ChildIdx}} = \MotherIdx)$, increment the size of $\LatentChildMotherObsVec_\GroupIdx$ by one and set $\LatentChildMotherObs_{\GroupIdx, \LatentMixtNumUnique_\GroupIdx + 1} = \MotherIdx$,
$\LatentMixt_{\GroupIdx, \ObsNum_\GroupIdx + 1} = \LatentMixtUnique_{\GroupIdx, \LatentMixtNumUnique_\GroupIdx+ 1}$. Observe that $\LatentMixtUnique_{\GroupIdx, \LatentMixtNumUnique_\ell + 1}$ will be ``similar'' to the other dishes served in the $\MotherIdx$-th room, as its law can be seen as the predictive distribution in the Bayesian model $\LatentMixtUnique_{\GroupIdx \ChildIdx} \mid \MotherLoc_{\MotherIdx} \iid \Kernel(\cdot , \MotherLoc_{\MotherIdx})$ with prior $\MotherLoc_{\MotherIdx} \sim \MotherMeanLoc$.

\item[$(iii)$] Sits in a new room (and, a fortiori a new table), with probability proportional to $\ChildMoment(\ChildAux_\GroupIdx, 1) \MotherMoment\left(\sum_{\GroupIdx=1}^\GroupNum\ChildLaplaceExp(\ChildAux_\GroupIdx), 1\right)$. In this case, we sample $\LatentMixtUnique_{\GroupIdx, \LatentMixtNumUnique_\GroupIdx+ 1}$ from $m(\cdot)$, increment the size of $\LatentChildMotherObsVec_\GroupIdx$ by one and set $\LatentChildMotherObs_{\GroupIdx, \LatentMixtNumUnique_\GroupIdx + 1} = \LatentChildMotherNumUnique + 1$, $\LatentMixt_{\GroupIdx, \ObsNum_\GroupIdx + 1} = \LatentMixtUnique_{\GroupIdx, \LatentMixtNumUnique_\GroupIdx+ 1}$.
\end{itemize}

The clustering of the HSNCP mixture discussed in \Cref{sec:hsncp_mom} is described by the assignment of the customers to the rooms. 
It is also clear that the prior induced on the clustering is not dependent on the choice of $\Kernel(\cdot,\cdot)$ (although the posterior obviously is).

\section{Posterior inference on HSNCP mixtures}
\label{sec:post_inf}

In this section, we describe posterior inference for the mixture model \eqref{eq:likelihood} with $(\ChildProcNorm_1,\dots,\ChildProcNorm_{\GroupNum})\sim \NormHSNCP( \ChildMeanJump, \MotherMeanJump, \MotherMeanLoc, \Kernel(\cdot,\cdot))$.

\subsection{Prior elicitation for Gaussian mixtures}
\label{sec:prior_elicitation}

For simplicity, we focus here on univariate data and Gaussian mixtures, and describe an empirical Bayesian approach to set the hyperparameters in a data-dependent fashion. Specifically, we assume that our data follow a nonparametric location-mixture with mixing measure $\ChildProcNorm_{\GroupIdx}$, where  
$\ChildSpace=\R$, $\ObsMixt(\cdot \mid \ChildLoc) = \Gaussian(\cdot \mid \ChildLoc, \MixtVar)$ is the Gaussian density with mean $\phi$ and variance $\sigma^2$, the kernel encoding the proximity of components across and within groups as 
$\Kernel(\cdot, \MotherLoc_{\MotherIdx}) = \Gaussian(\cdot \mid \MotherLocMean_{\MotherIdx}, \MotherLocVar_{\MotherIdx})$ for $\MotherLoc_{\MotherIdx} = (\MotherLocMean_{\MotherIdx}, \MotherLocVar_{\MotherIdx}) \in \MotherSpace = \R \times \Rp$, and $\MotherMeanLoc(\MotherLocMean_{\MotherIdx}, \MotherLocVar_{\MotherIdx}) = \Gaussian (\MotherLocMean_{\MotherIdx} \mid \MotherLocMeanPriorMean, \MotherLocMeanPriorVar) \times \InverseGamma(\MotherLocVar_{\MotherIdx} \mid \MotherLocVarPriorShape, \MotherLocVarPriorScale)$,
where $\InverseGamma(\cdot\mid\alpha,\beta)$ is the inverse gamma density with mean $\beta/(\alpha-1)$.
We specialize the double-summation formulation of our model in \eqref{eq:mix_of_mix} as follows
\begin{equation}\label{eq:unidim}
\begin{aligned}
    \Obs_{\GroupIdx \ObsIdx} \mid \ChildProcNorm_\GroupIdx, \MixtVar  &\iid \tilde\ObsMixt_\GroupIdx(\cdot) = \frac{1}{\ChildJump_{\GroupIdx\middot\middot}} \sum_{\MotherIdx \geq 1} S_{\GroupIdx\MotherIdx\middot} \tilde \ObsMixt_{\GroupIdx \MotherIdx}(\cdot), \ \ \text{for } \tilde \ObsMixt_{\GroupIdx\MotherIdx}(\cdot) := \frac{1}{\ChildJump_{\GroupIdx\MotherIdx\middot}} \sum_{\ChildIdx \geq 1} \ChildJump_{\GroupIdx\MotherIdx\ChildIdx} \mathcal N(\cdot \mid \ChildLoc_{\GroupIdx \MotherIdx \ChildIdx}, \MixtVar) \\
 (\ChildProcNorm_1,\dots,\ChildProcNorm_{\GroupNum}) &\sim \NormHSNCP( \ChildMeanJump, \MotherMeanJump, \MotherMeanLoc, \Kernel(\cdot,\cdot))   \\
 \MixtVar &\sim \InverseGamma(\MixtVarPriorShape, \MixtVarPriorScale),
\end{aligned}
\end{equation}
with $(\ChildProcNorm_1,\dots,\ChildProcNorm_{\GroupNum})$ and $ \MixtVar$ a priori independent.
We first operate a centering of $G_0$ with respect to the observed data, namely by setting $\MotherLocMeanPriorMean$ as the sample mean of all observations, and set $\MotherLocMeanPriorVar$ so that $\max_{\GroupIdx, \ObsIdx}|\Obs_{\GroupIdx\ObsIdx}-\MotherLocMeanPriorMean|=2\MotherLocMeanPriorSd$, which ensures that the cluster centers cover the span of the data.
Next, we focus on $\MotherLocVarPriorShape$ and $\MotherLocVarPriorScale$. We take inspiration from the Bayesian distance clustering literature \citep{Duan21, natarajan2024cohesion}, and we consider the pairwise distances between datapoints. If there are well-separated clusters in the data, the kernel density estimate of the pairwise distances will be multimodal. The first (i.e., closest to zero) local minimum $\PriorElicitationDistance$ of the density estimate gives a rough estimate of the maximal distance between two datapoints belonging to the same cluster.
This suggests that two atoms $\ChildLoc_{\GroupIdx \MotherIdx \ChildIdx}, \ChildLoc_{\GroupIdx \MotherIdx \ChildIdxAlt}$ in \eqref{eq:unidim} should be closer than $\PriorElicitationDistance$ with high probability, as they are referring to the same clusters. Standard computations show that $|\ChildLoc_{\GroupIdx j \ChildIdx} -\ChildLoc_{\GroupIdx j \ChildIdxAlt}| / \sqrt{(2\MotherLocVarPriorScale / \MotherLocVarPriorShape)}$ is marginally distributed as a half-$t$ distribution with $2\MotherLocVarPriorShape$ degrees of freedom. We impose two conditions. First, $\Prob(|\ChildLoc_{\GroupIdx \MotherIdx \ChildIdx} -\ChildLoc_{\GroupIdx \MotherIdx \ChildIdxAlt}| \leq \PriorElicitationDistance) = 0.99$. Second, the expected distance between atoms $\ChildLoc_{\GroupIdx \MotherIdx \ChildIdx}$ and $\ChildLoc_{\GroupIdx \MotherIdx \ChildIdxAlt}$ is set equal to a fixed fraction of the mean distance between observations in the same cluster (i.e., those within $\PriorElicitationDistance$). If we denote this mean value by $m$, the condition can be written as $\E[|\ChildLoc_{\GroupIdx \MotherIdx \ChildIdx} -\ChildLoc_{\GroupIdx \MotherIdx \ChildIdxAlt}|]=\PriorElicitationCoef m$, where $\PriorElicitationCoef\in(0,1)$. In the remainder of this article, we fix $\PriorElicitationCoef=0.7$. 

To elicit the prior distribution of $\MixtVar$, we follow \cite{shi2019low} and set $\MixtVarPriorShape=3/2$ and let $\MixtVarPriorScale$ such that $\E\left[\Var\left(\Obs_{\GroupIdx\ObsIdx}\mid \MixtVar, \MotherLocVar_{\LatentChildMotherObs_{\GroupIdx\LatentMixtClusLabels_{\GroupIdx\ObsIdx}}}\right)\right] = s^2$, where $s^2$ is the empirical variance of the data.

Finally, having fixed the parameters in $\MotherMeanLoc(\MotherLocMean_{\MotherIdx},\MotherLocVar_{\MotherIdx})$ and $\Kernel(\cdot;\MotherLoc_{\MotherIdx})$, it is possible to exploit Theorem \ref{teo:corr} to fix the hyperparameters in $\ChildMeanJump$ and $\MotherMeanJump$ (as in the examples in Section~\ref{sec:a_priori_moments}) by considering the prior dependence between the random mixing measures.

\subsection{The MCMC algorithm}\label{sec:mcmc}

We propose a Gibbs sampler to approximate the posterior in the HSNCP mixture model. The algorithm state consists of: the labels $\LatentMixtClusLabelsVec =(\LatentMixtClusLabels_{\GroupIdx\ObsIdx},\GroupIdx=1,\dots,g,\ \ObsIdx=1,\dots, \ObsNum_{\GroupIdx})$ and $\LatentChildMotherObsVec$; the auxiliary variables $\ChildAuxVec$; the mother process' locations $(\MotherLoc_{\MotherIdx})_{\MotherIdx=1}^{\LatentChildMotherNumUnique}$; and the child processes $\ChildProc_{\GroupIdx}$, $l=1,\ldots,g$. The key result to develop our algorithm is Theorem \ref{teo:post}, which describes the full conditional of 
$(\ChildProc_1,\ldots, \ChildProc_g)$. 

According to $(ii)$ and $(iii)$ of Theorem \ref{teo:post}, the full conditional distribution of the child processes contains the sum of  $\LatentChildMotherNumUnique+1$ CRMs, which in turn have an infinite number of jumps and locations. We approximate these infinite support random measures 
with finite numbers of jumps and locations. Then we
adopt the Ferguson–Klass (FK) algorithm \citep{Ferguson_AnnMathStat_1972} to sample only the jumps whose magnitudes exceed a user-specified threshold $\FKThreshold$ and, then, of course, we also sample their corresponding locations. Implementation details are given in Section~\ref{sec:mcmc_detailed} of the Supplementary material. When $\FKThreshold$ is chosen sufficiently small, the truncation error is virtually negligible. See, e.g., \citet{arbel2017moment}, \citet{Campbell19}, and \citet{nguyen2024independent}. For a different truncation procedure see \cite{argiento2016posterior}.

Therefore, we classify each child process' jumps in different categories: the allocated jumps $\boldsymbol{\ChildJump}_{\GroupIdx}^{(a)}$, i.e., the unique values in $(\ChildJump_{\GroupIdx\LatentMixtClusLabels_{\GroupIdx\ObsIdx}})_{\ObsIdx=1}^{\ObsNum_{\GroupIdx}}$; 
the non-allocated jumps associated with allocated mother atoms in the FK approximation, i.e., the sets $\boldsymbol{\ChildJump}_{\GroupIdx\MotherIdx}^{(na)}=\{\ChildJump_{\GroupIdx\ChildIdx}: \LatentMixtClusLabels_{\GroupIdx\ObsIdx}\neq\ChildIdx,\ \ObsIdx=1,\dots, \ObsNum_{\GroupIdx};\LatentChildMother_{\GroupIdx\ChildIdx}=\MotherIdx; \ChildJump_{\GroupIdx\ChildIdx}\ge\FKThreshold\}$ for each $\MotherIdx=1,\dots,\LatentChildMotherNumUnique$; 
the non-allocated jumps associated with non-allocated mother atoms in the FK approximation, i.e., the set $\boldsymbol{\ChildJump}_{\GroupIdx}^{(na)}=\{\ChildJump_{\GroupIdx\ChildIdx}:\LatentChildMother_{\GroupIdx\ChildIdx}=\MotherIdx,\ \MotherIdx\ge\LatentChildMotherNumUnique+1;\ChildJump_{\GroupIdx\ChildIdx}\ge\FKThreshold\}$. We use similar notation for the corresponding locations as  $\boldsymbol{\ChildLoc}_{\GroupIdx}^{(a)}$, $\boldsymbol{\ChildLoc}_{\GroupIdx\MotherIdx}^{(na)}$, and $\boldsymbol{\ChildLoc}_{\GroupIdx}^{(na)}$.

The full MCMC algorithm is as follows.
\begin{enumerate}

\item Sample each label $\LatentMixtClusLabels_{\GroupIdx\ObsIdx}$ according to the probability that  $\LatentMixtClusLabels_{\GroupIdx\ObsIdx}$ is equal to $\ChildIdx$, which is proportional to  $\ChildJump_{\GroupIdx\ChildIdx}\ObsMixt(\Obs_{\GroupIdx\ObsIdx}\mid\ChildLoc_{\GroupIdx\ChildIdx})$ for each group $\GroupIdx$.

\item Sample each element of $\boldsymbol{\ChildLoc}_{\GroupIdx}^{(a)}$, from the density proportional to $\prod_{\ObsIdx:\LatentMixtClusLabels_{\GroupIdx\ObsIdx}=\ChildIdx}\ObsMixt(\Obs_{\GroupIdx\ObsIdx}\mid \ChildLoc_{\GroupIdx\ChildIdx})\Kernel(\ChildLoc_{\GroupIdx\ChildIdx},\MotherLoc_{\LatentChildMotherObs_{\GroupIdx\ChildIdx}})$ and sample each element of $\boldsymbol{\ChildJump}_{\GroupIdx}^{(a)}$ from the density  in Theorem \ref{teo:post} ($i$), for each group~$\GroupIdx$.

\item Sample $\LatentChildMotherObsVec$ from the law in \eqref{eq:post_t_u}. In particular, we use Algorithm 3 in \cite{Neal_JCGS_2000}. See Section~\ref{sec:mcmc_detailed} of the Supplementary material for further details.
\item Sample each $\MotherLoc_{\MotherIdx}$ and the elements of $\boldsymbol{\ChildLoc}_{\GroupIdx\MotherIdx}^{(na)}$ and  $\boldsymbol{\ChildJump}_{\GroupIdx\MotherIdx}^{(na)}$, $\GroupIdx=1,\ldots,\GroupNum$, according to Theorem \ref{teo:post} ($ii$). In particular, we use the FK algorithm to sample the elements of $\boldsymbol{\ChildLoc}_{\GroupIdx\MotherIdx}^{(na)}$ and  $\boldsymbol{\ChildJump}_{\GroupIdx\MotherIdx}^{(na)}$. See Section~\ref{sec:mcmc_detailed}  of the Supplementary material for further details.

\item Sample the elements of $\boldsymbol{\ChildLoc}_{\GroupIdx}^{(na)}$ and $\boldsymbol{\ChildJump}_{\GroupIdx}^{(na)}$, $\GroupIdx=1,\ldots,\GroupNum$, according to  Theorem \ref{teo:post} ($iii$) via the FK algorithm.

\item Sample each $\ChildAux_{\GroupIdx}$, $\GroupIdx=1,\ldots,\GroupNum$, independently from the Gamma density with shape $\ObsNum_{\GroupIdx}$ and rate $\ChildProc_{\GroupIdx}(\ChildSpace)$.
\end{enumerate}

\section{Numerical illustrations}
\label{sec:simulations}
In this section, we showcase the main aspects of the HSNCP mixture model via simulations.
We consider the setup of Section~\ref{sec:prior_elicitation} and select prior hyperparameters accordingly.

As a natural competitor of the HSNCP mixtures we consider the HDP mixture of Gaussian distributions, which corresponds to \eqref{eq:likelihood} with $\ChildProcNorm_\ell \mid \HDPMotherProc \iid \DP(\HDPChildTotMass, \HDPMotherProc)$ and $\HDPMotherProc \sim \DP(\HDPMotherTotMass,\HDPMeanLoc)$,
where $\DP(\cdot,\cdot)$ denotes a Dirichlet Process with total mass given by the first parameter and base measure given by the second, with $\HDPChildTotMass,\HDPMotherTotMass>0$. The base measure $\HDPMeanLoc$ is a normal-inverse gamma distribution: $\HDPMeanLoc(\HDPLocMean_{\MotherIdx},\HDPLocVar_{\MotherIdx})=\Gaussian(\HDPLocMean_{\MotherIdx}\mid\HDPLocPriorMean,\HDPLocVar_{\MotherIdx}/\HDPLocPriorDenom)\times\InverseGamma(\HDPLocVar_{\MotherIdx}\mid\HDPLocPriorShape,\HDPLocPriorScale)$. To allow the matching of the two mixture models, see Section~\ref{sec:simulations_appendix} of the Supplementary material, where we explain the prior hyperparameters elicitation procedure for the HDP.

In every simulation we run each MCMC algorithm for 5,000 iterations, after a burn-in period of 1,000 iterations. For the HSNCP mixture, we use the algorithm described in Section~\ref{sec:post_inf}, written in Julia, while for the HDP model we run the sampler implemented in \url{https://github.com/alessandrocolombi/hdp}.
We apply the \texttt{salso} package \citep[see][]{salso} to estimate clusters, minimizing the posterior expectation of the Variation of Information loss function.

\subsection{The illustrative example}
\label{sec:illustrative_example}

We begin by showing that the HSNCP mixture does not suffer from the trade-off between density and cluster estimation typically faced in traditional mixture models \citep{Beraha_arXiv_2024}, by comparing it with the HDP mixture. We resume the illustrative example in the introduction, where we
generate data from
\[
    \Obs_{\GroupIdx\ObsIdx} \iid \frac12\Gaussian(\Obs_{\GroupIdx\ObsIdx}\mid \mu_{\GroupIdx 1},1)+\frac12\Gaussian(\Obs_{\GroupIdx\ObsIdx}\mid \mu_{\GroupIdx 2},1), \quad \GroupIdx =1, 2, \ i=1, \ldots, n
\]
with $n\in\{50, 500\}$ and 
$\mu_{11}=-4.4$, $\mu_{12}=3.55$, $\mu_{21}=-3.6$, $\mu_{22}=4.45$.

Following the prior elicitation for the HSNCP prior in Section~\ref{sec:prior_elicitation}, we obtain the following hyperparameters: $\MotherLocVarPriorShape=1.658$, $\MotherLocVarPriorScale=0.529$, $\MotherLocMeanPriorMean=0$, $\MotherLocMeanPriorVar=10$, $\MixtVarPriorShape=3/2$, and $\MixtVarPriorScale=8.25$. 
We assume that mother and child processes are gamma processes by fixing $\rho(s)$ and $\rho_0(s)$ as in \Cref{ex:gamma} with $\ChildProcTotMass=1.076$ and $\MotherProcTotMass=1$, which gives $\Cor(\ChildProcNorm_{\GroupIdx}(A),\ChildProcNorm_{\GroupIdxAlt}(A))=0.5$
for $A=(-5,5)$.
Prior elicitation for the HDP mixture is described in Section~\ref{sec:simulations_appendix} of the Supplementary material. Following that procedure, we set the hyperparameters as follows: $\HDPLocPriorMean=0$, $\HDPLocPriorDenom=2$, $\HDPLocPriorShape=1.71$, $\HDPLocPriorScale=15.81$, $\HDPMotherTotMass=2$, and $\HDPChildTotMass=1$.

\begin{figure}[t]
\centering
\includegraphics[width=0.8\textwidth]{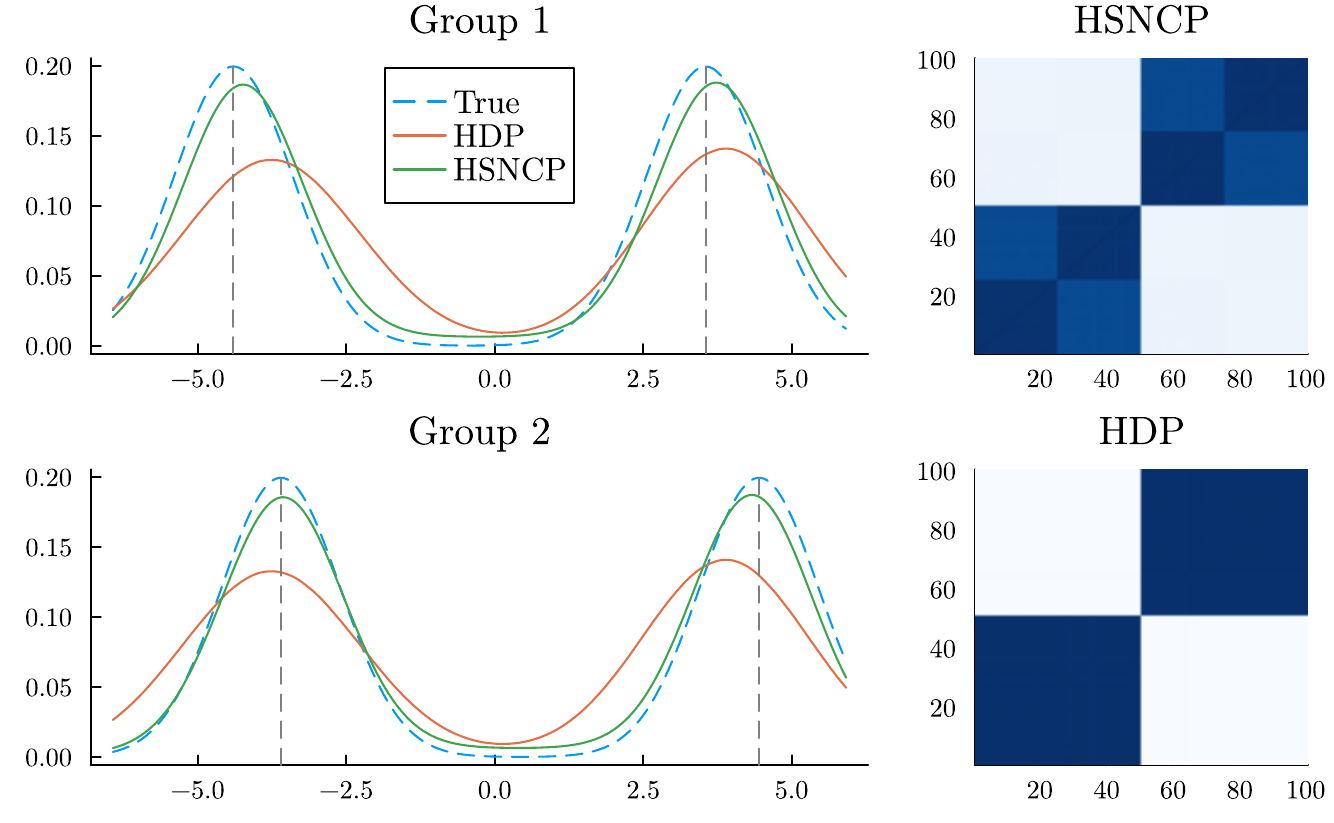}
\caption{Illustrative simulated example with $n_1 = n_2 = 50$.
The plots on the left display the true generating densities and the estimated densities from HSNCP and HDP mixtures. 
The right plots show the posterior similarity matrix of all data in the two models.} 
\label{fig:simulation_1_50}
\end{figure}

Figure~\ref{fig:illustrative} (from the introduction) and Figure~\ref{fig:simulation_1_50} display the posterior co-clustering matrices and predictive densities obtained using the two models when $n=500$ and $n=50$, respectively. 
Focusing on the HDP mixture, notice that when $n=50$ the clustering is correctly estimated but the density estimate is poor, while if $n=500$ the densities are recovered perfectly but the model fails to detect the clusters across different groups of data.
On the other hand, the HSNCP mixture shows exact density and cluster estimates in both scenarios.

This examples highlights the trade-off between cluster and density estimate inherited from the exact sharing of atoms across different groups of data, that is not peculiar to the HDP but shared by all models with common atoms across groups.
Such a trade-off is not faced by the HSNCP mixture thanks to its more flexible two-level clustering structure.

\subsection{Summary of further simulation studies}
Section~\ref{sec:simulations_appendix} of the Supplementary material contains several simulation studies illustrating key properties of our model. 
In Section~\ref{sec:sim_HDP_comp}, we extend the illustrative example from Section~\ref{sec:illustrative_example} to compare the posterior clustering ability of the HDP and the HSNCP priors. We consider scenarios where the means of the data-generating Gaussian mixtures differ between groups. The results show that the HSNCP mixture model allows for flexible across-group information sharing even when group-specific means differ, whereas the HDP sharply reduces such sharing as soon as the group differences are nonzero.
Section~\ref{sec:sim_sens_analysis} proves the strong effect of the kernel on the posterior inference of our model. In particular, we show that the posterior cluster estimates of the HSNCP mixture model are highly sensitive to the specification of the prior for the kernel variance $\MotherLocVar_{\MotherIdx}$, when $\Kernel(\cdot, \MotherLoc_{\MotherIdx}) = \Gaussian(\cdot \mid \MotherLocMean_{\MotherIdx}, \MotherLocVar_{\MotherIdx})$ for $\MotherLoc_{\MotherIdx} = (\MotherLocMean_{\MotherIdx}, \MotherLocVar_{\MotherIdx}) \in \MotherSpace = \R \times \Rp$. 
The example justifies the need for a prior elicitation procedure as the one described in Section~\ref{sec:prior_elicitation}.
Section~\ref{sec:sim_non_gaussian} illustrates the flexibility of our model in identifying clusters in \textit{misspecified}
settings, that is, when the generating process does not coincide with $\ObsMixt(\cdot \mid x)$; in particular, we consider examples where the 
data-generating mechanisms are skewed, multimodal, or with heavy tails.
This property is highly advantageous in practice. First, the practitioner does not need to decide the shape of the clusters in advance. Second, choosing computationally tractable $\ObsMixt$ and $\Kernel$, such as Gaussian densities, allows to obtain closed-form full conditionals for the MCMC algorithm described in Section~\ref{sec:mcmc}. This, in turn, can reduce computational time and improve mixing.

\section{Analysis of the Sloan Digital Sky Survey data}
\label{sec:sloan}
In this section we apply our model to a dataset from the Sloan Digital Sky Survey first data release \citep{Abazajian_AstroJou_2003}. The dataset contains, for $\sum_{\GroupIdx=1}^\GroupNum\ObsNum_{\GroupIdx}=24{,}312$ galaxies, measurements of the $u-r$ color, that is, the difference between ultraviolet and red color distributions. This quantity provides a robust indicator of galaxy type and star formation activity, as galaxies with smaller (bluer) $u-r$ values are typically younger and actively forming stars, while those with larger (redder) values host older stellar populations with little or no current star formation \citep{Balogh_AstroJou_2004}. Therefore, clustering galaxies according to their $u-r$ color allows us to identify and distinguish populations at different evolutionary stages. The galaxies are divided into $\GroupNum=25$ groups according to their luminosity type and environment. Further details on the observations can be found in \citet{Balogh_AstroJou_2004}, while alternative Bayesian analyses of the same dataset are provided in \citet{Stefanucci_StatComp_2021,Camerlenghi_arXiv_2024}.

We consider the HSNCP Gaussian mixture model \eqref{eq:unidim} with kernel
$\Kernel(\cdot, \MotherLoc_{\MotherIdx}) = \Gaussian(\cdot \mid \MotherLocMean_{\MotherIdx}, \MotherLocVar_{\MotherIdx})$ and the rest of the prior details as in Section~\ref{sec:prior_elicitation}. 
In particular, we assume generalized gamma process L\'evy densities for mother and child processes, i.e., we fix $\ChildMeanJump(s)$ and $\MotherMeanJump(s)$ as in \Cref{ex:gen_gamma}. 
Following the prior elicitation procedure described in Section~\ref{sec:prior_elicitation}, we set $\MotherLocMeanPriorMean=1.989$, $\MotherLocMeanPriorVar=1$, $\MixtVarPriorShape=3/2$, and $\MixtVarPriorScale=0.0892$, $\MotherLocVarPriorShape=12.0867$, and $\MotherLocVarPriorScale=0.6265$.
We compare our HSNCP mixture model against the HDP mixtures described in Section~\ref{sec:simulations}, which allows across-group information sharing (albeit in a more rigid way than the HSNCP). Following the prior elicitation procedure described in Appendix~\ref{sec:simulations_appendix}, we set $\HDPLocPriorMean=1.989$, $\HDPLocPriorDenom=1/4$, $\HDPLocPriorShape=2.331$, and $\HDPLocPriorScale=0.524$.
The last key parameters to be set are the total mass parameters of the HDP ($\HDPChildTotMass$, $\HDPMotherTotMass$) and the parameters of $\ChildMeanJump(s)$ and $\MotherMeanJump(s)$ in the HSNCP. We fix them so that $\Cor(\ChildProcNorm_\GroupIdx(A),\ChildProcNorm_{\GroupIdxAlt}(A))=0.75$ for all different $\GroupIdx,\GroupIdxAlt=1,\dots,\GroupNum$, where $A=(0.006584, 3.993)$ is the range of observations. For the HDP, we set $\HDPMotherTotMass = 1$ and $\HDPChildTotMass = 2$. For the HSNCP, we specify $\MotherMeanJump(s)$ with parameters $\MotherProcTotMass = 5$, $\MotherProcSigma = 0.1$, and $\MotherProcTau = 0.2$, and $\ChildMeanJump(s)$ with parameters $\ChildProcTotMass = 1.306$, $\ChildProcSigma = 0.05$, and $\ChildProcTau = 1$.

We run the MCMC algorithm for each model for a total of
50,000 iterations, discarding the first 10,000 and keeping one every ten iterations for a final sample
size of 5,000. Running our algorithm on a standard laptop took 3 hours and 19 minutes. Pointwise estimates of the clusters are obtained following the same procedure described in \Cref{sec:simulations}.

\begin{figure}
\centering
\includegraphics[width=\textwidth]{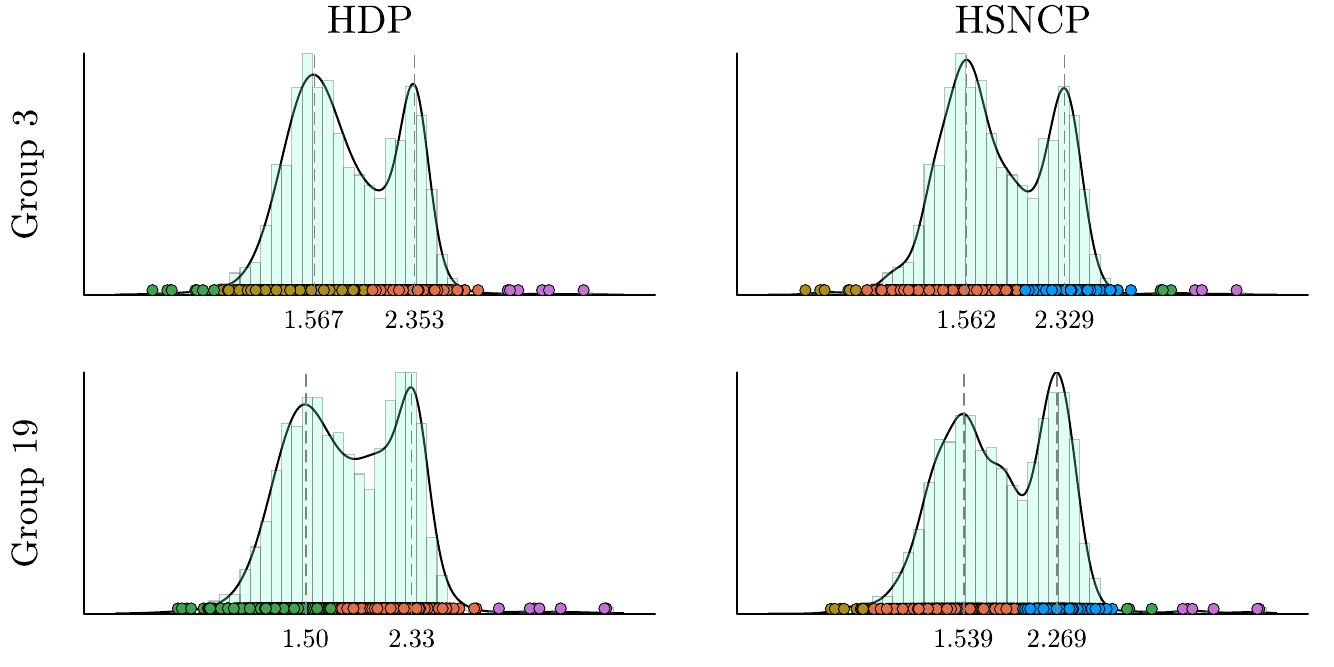}
\caption{Density estimation and pointwise estimate of the clusters for two selected groups. The vertical lines represent the peaks in the density estimations.}
\label{fig:sloan_selected_groups}
\end{figure}

\Cref{fig:sloan_selected_groups} shows the posterior density and cluster estimates for the HSNCP and HDP models in two selected groups. Looking at the leftmost peaks across the four panels, we observe a pattern similar to that in \Cref{fig:illustrative}: the HDP identifies peaks that are close but not identical, which prevents the observations from being clustered together across groups; in contrast, the HSNCP clusters them together by treating the peaks as sufficiently similar. Considering instead the rightmost peaks in all four panels, the behavior resembles that in \Cref{fig:simulation_1_50}: the HSNCP allows observations to be clustered together even when the peaks differ, whereas the HDP requires the peaks to be identical to cluster the observations across groups. Also, the HSNCP density estimates more closely follow the observed histograms. This behavior is consistent with the model construction described in \Cref{sec:hsncp_mom}, where the cluster-specific densities in the HSNCP are not constrained to be Gaussian and can capture multimodality and skewness.

A similar pattern is observed across all groups, as shown in \Cref{fig:sloan_HDP,fig:sloan_HSNCP} (see Appendix~\ref{sec:sloan_supp}). While the HSNCP estimates densities with distinct peaks for each group, the HDP yields estimates with peaks that are often aligned between each other. Consequently, the HSNCP clusters partition the observation space into largely non-overlapping intervals across groups, whereas the HDP often assigns different clusters in different groups to nearly the same intervals.  

\begin{figure}[t]
\centering
\includegraphics[width=\textwidth]{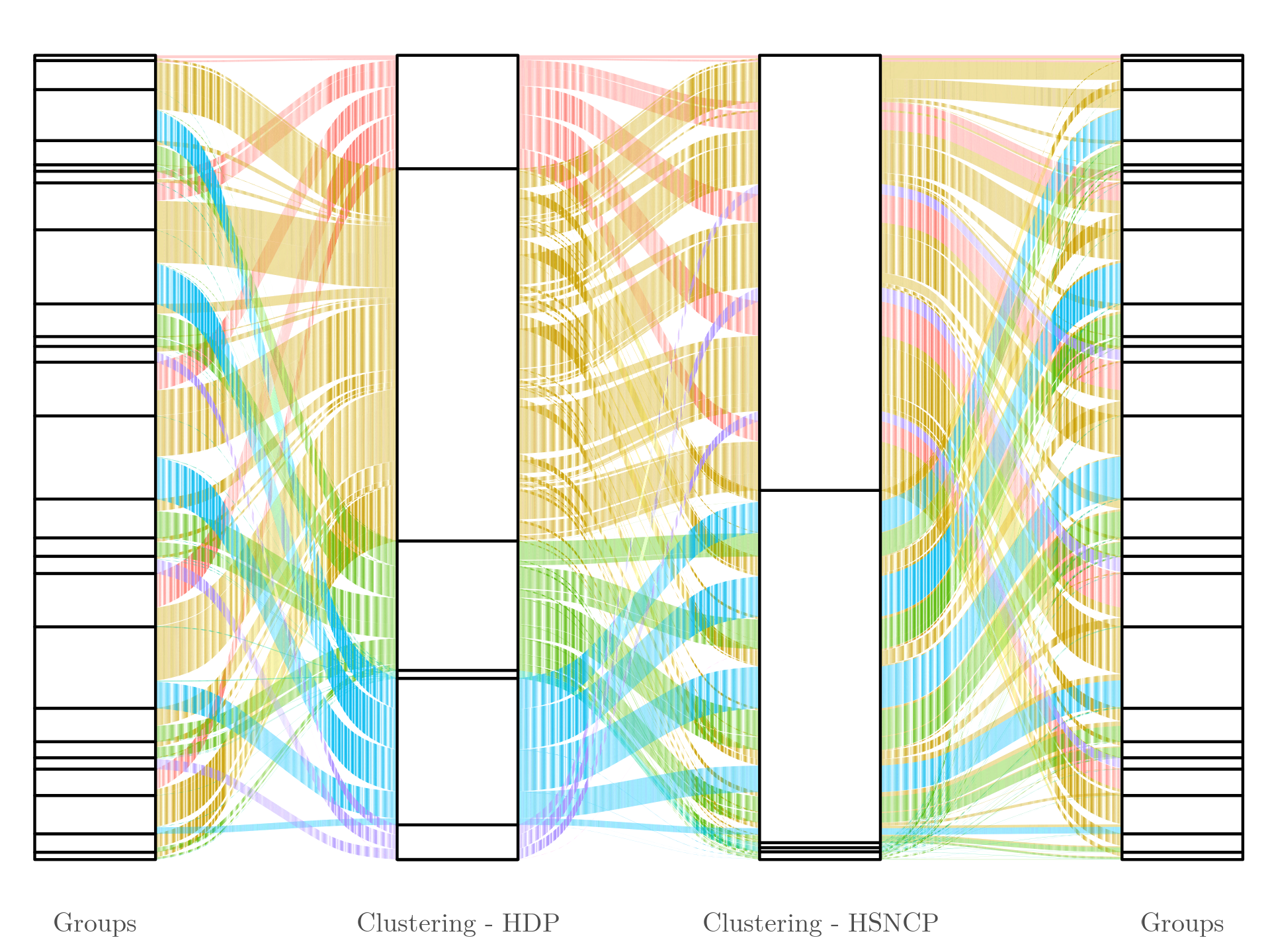}
\caption{Alluvial plot illustrating the relationship between the grouping of observations and the cluster estimates obtained under the HDP and HSNCP models. The color of each flow corresponds to the clustering induced by the HDP.}
\label{fig:sloan_alluvial}
\end{figure}

\begin{figure}[t]
\centering
\includegraphics[width=0.75\textwidth]{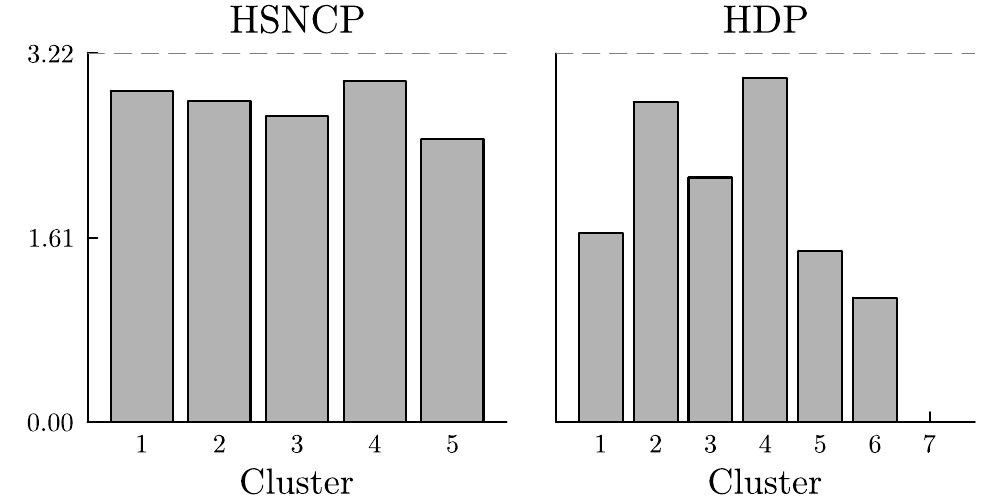}
\caption{Entropy of the group composition within each cluster identified by the HDP and HSNCP models. The gray horizontal line indicates the maximum possible entropy, which occurs when a cluster contains an equal number of observations from each group.}
\label{fig:sloan_entropy}
\end{figure}

The clustering induced by the HSNCP thus enables flexible information sharing across all groups, with each cluster typically containing observations from multiple --- often all --- groups. In contrast, the HDP promotes a more rigid form of across-group sharing.
\Cref{fig:sloan_alluvial} displays the alluvial plot comparing the cluster assignments under the HDP and HSNCP models and the grouping of the observations. An alluvial plot is a flow diagram used to visualize how elements are distributed or reallocated across different categorical groupings, highlighting correspondences between partitions.
\Cref{fig:sloan_clus_barplot} shows the stacked barplot of the estimated clusters, where each bar represents a cluster and the colored segments indicate the proportion of observations from each group, illustrating their internal group composition.
As shown in \Cref{fig:sloan_alluvial,fig:sloan_clus_barplot}, several clusters identified by the HDP contain observations from only a subset of groups, leading to lower entropy values, as reported in \Cref{fig:sloan_entropy}. Conversely, the clusters identified by the HSNCP generally include observations from all $\GroupNum$ groups; in some cases, they correspond to the union of multiple HDP clusters, particularly those containing observations only from specific groups. This behavior is reflected in \Cref{fig:sloan_entropy}, where the entropy of HSNCP clusters is consistently high, confirming their balanced group composition.

\section{Discussion}
We have introduced the Hierarchical Shot-Noise Cox Process (HSNCP) mixture model, a novel Bayesian nonparametric approach for partially exchangeable data. Unlike traditional hierarchical models such as the HDP of \citet{Teh_JASA_2006}, which allow borrowing of information only through common atoms, the HSNCP allows group-specific components to concentrate around shared centers through a kernel-based construction. This results in more flexible across-group information sharing, especially in the presence of subtle structural heterogeneity.

The HSNCP retains desirable theoretical and computational properties: it admits closed-form expressions for key prior quantities, supports tractable posterior inference via a conditional MCMC algorithm, and induces a two-level clustering structure that improves both interpretability and robustness. Empirical results from simulations and real data demonstrate improved performance in density estimation and clustering, particularly in settings where traditional models are either too rigid or fail to identify shared structure.
Moreover, our model and the associated MCMC algorithm scale well as the total size is large, as the application shows.

The posterior characterization provided in this work lays the foundation for marginal algorithms that avoid sampling latent variables --- a direction we leave for future research. Additionally, we are working on a nested extension of the HSNCP that enables clustering entire distributions based on their similarity, relaxing the requirement of exact equality imposed by standard nested processes.
Finally, extension to more complex dependence structures are of interest. For instance, for temporally evolving data, one option is to adopt the spatiotemporal shot-noise Cox process recently developed by \cite{bassetti2023spatiotemporal} as prior distribution for the atoms in the mixing measures.

\textbf{Software and Data Availability:} The code implementing the methods described in this paper, as well as scripts to reproduce the numerical experiments and data analysis together with the original datasets, is publicly available at: \url{https://github.com/AleCarminati/hsncp_cond_sampler}.


\spacingset{1.65}

\bibliography{bibliography}

\clearpage

\LARGE{
\begin{center}
\bf Supplementary material for:\\
``Hierarchical shot-noise Cox process mixtures for clustering across groups''
\end{center}
}

\appendix

\spacingset{1.8}

The supplementary material is organized as follows. Section \ref{app:proofs} collects all the proofs of the theoretical results in the main paper, Section \ref{sec:mcmc_detailed} describes the MCMC algorithm for posterior inference, Section \ref{sec:simulations_appendix} illustrates the HSNCP mixture model and the HDP on simulated datasets, and Section \ref{sec:sloan_supp} contains additional plots for the analysis of the Sloan Digital Sky Survey dataset. 

\section{Proofs}\label{app:proofs}
\subsection{Proof of Equation \eqref{eq:prior_expectation}}\label{sec:prior_expectation_proof}
We first exploit the law of total expectation and Equation (2) of \cite{James_ScanJouStat_2006}:
\begin{align*}
\E[\ChildProcNorm_{\GroupIdx}(A)]&=\E[\E[\ChildProcNorm_{\GroupIdx}(A)\mid \MotherProc]]=\E\left[\frac{\int_A\int_{\MotherSpace\times\Rp}\MotherJump\Kernel(x,\MotherLoc)\MotherProc(\dd\MotherJump,\dd\MotherLoc)\dd x}{\int_{\MotherSpace\times\Rp}\MotherJump\MotherProc(\dd\MotherJump,\dd\MotherLoc)}\right].\\
\intertext{Straightforward computations show that the reciprocal of $\int_{\MotherSpace\times\Rp}\MotherJump\MotherProc(\dd\MotherJump,\dd\MotherLoc)$ can be expressed as $\int_{\Rp}\exp\left\{-u\int_{\MotherSpace\times\Rp}\MotherJump'\MotherProc(\dd\MotherJump',\dd\MotherLoc')\right\}\dd u$, thus}
\E[\ChildProcNorm_{\GroupIdx}(A)]&=\int_{\Rp}\int_A\E\left[\int_{\MotherSpace\times\Rp}\MotherJump\Kernel(x,\MotherLoc)\exp\left\{-u\int_{\MotherSpace\times\Rp}\MotherJump' \MotherProc(\dd \MotherJump', \dd \MotherLoc') \right\}\MotherProc(\dd \MotherJump, \dd \MotherLoc)\right]\dd x\dd u.
\intertext{We apply Campbell-Little-Mecke (CLM) formula \citep{Baccelli_2024} and the definition of Laplace exponent for a Poisson process:}
\E[\ChildProcNorm_{\GroupIdx}(A)]&=\int_{A\times\MotherSpace}\Kernel(x,\MotherLoc)\MotherMeanLoc(\dd \MotherLoc)\dd x\int_{\Rp}\exp\{-\MotherLaplaceExp(u)\}\int_{\Rp}\MotherJump \e^{u\MotherJump}\MotherMeanJump(\MotherJump)\dd\MotherJump\dd u\\
&=m(A)\int_{\Rp}\exp\{-\MotherLaplaceExp(u)\}\MotherMoment(u,1)\dd u.
\intertext{Recalling that $\frac{\dd}{\dd u}\MotherLaplaceExp(u)=-\MotherMoment(u,1)$, we get}
&\E[\ChildProcNorm_{\GroupIdx}(A)]=m(A)\left[-\exp\{-\MotherLaplaceExp(u)\right]_0^{+\infty}=m(A).
\end{align*}

\subsection{Proof of Theorem \ref{teo:corr}}\label{sec:corr_proof}
In order to prove \Cref{teo:corr}, we first introduce two technical lemmas.

\begin{lemma}\label{lemma:corr_supp_1}
Let $(\ChildProcNorm_1,\ldots,\ChildProcNorm_{\GroupNum})\sim\NormHSNCP(\ChildMeanJump, \MotherMeanJump,\MotherMeanLoc,\Kernel(\cdot,\cdot))$. Then, for $A\in\mathcal B(\ChildSpace)$ and $\GroupIdx=1,\dots, \GroupNum$, 
$$\E[\E[\ChildProcNorm_{\GroupIdx}(A)\mid\MotherProc]^2]=m(A)^2\int_{\Rp}u\e^{-\MotherLaplaceExp(u)}(\MotherMoment(u,1))^2\dd u +m(A,A)\MotherEPPFInt,$$
where $\MotherEPPFInt=\int_{\Rp}u\e^{-\MotherLaplaceExp(u)}\MotherMoment(u,2)du$. Also,
$$\MotherEPPFInt=1-\int_{\Rp}u\e^{-\MotherLaplaceExp(u)}(\MotherMoment(u,1))^2\dd u.$$
\end{lemma}
\begin{proof}
Let $\MotherProc^2$ be the second power measure of $\MotherProc$, then
\begin{align*}
\E[\E[\ChildProcNorm_{\GroupIdx}(A)\mid\MotherProc]^2]&=\E\left[\int_{A^2}\int_{(\Rp\times\MotherSpace)^2}\frac{\MotherJump_1\MotherJump_2\Kernel(x_1,\MotherLoc_1)\Kernel(x_2,\MotherLoc_2)}{\left(\int_{\Rp\times\MotherSpace}\MotherJump'\MotherProc(\dd \MotherJump',\dd\MotherLoc')\right)^2}\MotherProc^2(\dd \MotherJump_1,\dd\MotherLoc_1,\dd \MotherJump_2,\dd\MotherLoc_2)\dd x_1\dd x_2\right].\\
\intertext{Straightforward computations show that the reciprocal of $\left(\int_{\MotherSpace\times\Rp}\MotherJump\MotherProc(\dd\MotherJump,\dd\MotherLoc)\right)^2$ can be expressed as $\int_{\Rp}u\exp\left\{-u\int_{\MotherSpace\times\Rp}\MotherJump'\MotherProc(\dd\MotherJump',\dd\MotherLoc')\right\}\dd u$, thus}
\E[\E[\ChildProcNorm_{\GroupIdx}(A)\mid\MotherProc]^2]&=\int_{\Rp}u\int_{A^2}\E\left[\int_{(\Rp\times\MotherSpace)^2}\MotherJump_1\MotherJump_2\Kernel(x_1,\MotherLoc_1)\Kernel(x_2,\MotherLoc_2)\e^{-u\int_{\Rp\times\MotherSpace}\MotherJump'\MotherProc(\dd \MotherJump',\dd\MotherLoc')}\right.\\
&\qquad\times\left.\MotherProc^2(\dd \MotherJump_1,\dd\MotherLoc_1,\dd \MotherJump_2,\dd\MotherLoc_2)\right]\dd x_1\dd x_2\dd u. 
\intertext{We exploit higher order CLM formula \citep{Baccelli_2024} and the definition of Laplace exponent for a Poisson process to obtain:}
\E[\E[\ChildProcNorm_{\GroupIdx}(A)\mid\MotherProc]^2]&=\int_{\Rp}u\int_{(A\times\Rp\times\MotherSpace)^2}\MotherJump_1\MotherJump_2\Kernel(x_1,\MotherLoc_1)\Kernel(x_2,\MotherLoc_2)\e^{-u(\MotherJump_1+\MotherJump_2)-\MotherLaplaceExp(u)}\\
&\qquad\times\MotherMeanJump(\dd \MotherJump_1)\MotherMeanJump(\dd \MotherJump_2)\MotherMeanLoc(\dd \MotherLoc_1)\MotherMeanLoc(\dd \MotherLoc_2)\dd x_1 \dd x_2\\
&\quad+\int_{\Rp}u\int_{A^2}\int_{\Rp\times\MotherSpace}\MotherJump^2\e^{-u\MotherJump}\Kernel(x_1,\MotherJump)\Kernel(x_2,\MotherJump)\e^{-\MotherLaplaceExp(u)}\MotherMeanJump(\dd \MotherJump)\MotherMeanLoc(\dd \MotherLoc)\dd x_1\dd x_2\dd u.
\end{align*}
The first result of the lemma follows by reordering the variables and the integrals, and the second result of the lemma follows by setting $A=\ChildSpace$.
\end{proof}

\begin{lemma}\label{lemma:corr_supp_2}
Let $(\ChildProcNorm_1,\ldots,\ChildProcNorm_{\GroupNum})\sim\NormHSNCP(\ChildMeanJump, \MotherMeanJump,\MotherMeanLoc,\Kernel(\cdot,\cdot))$. Then, for $A\in\mathcal B(\ChildSpace)$ and $\GroupIdx=1,\dots, \GroupNum$, 
\begin{align*}
\E[\ChildProcNorm_{\GroupIdx}(A)^2]&=m(A)^2\int_{\Rp}u\e^{-\MotherLaplaceExp(\ChildLaplaceExp(u))}(\MotherMoment(\ChildLaplaceExp(u),1))^2(\ChildMoment(u,1))^2\dd u+m(A,A)\pEPPFFirstInt+m(A)\pEPPFSecondInt,
\end{align*}
where $\pEPPFFirstInt=\int_{\Rp}u(\ChildMoment(u,1))^2\MotherMoment(\ChildLaplaceExp(u),2)\e^{-\MotherLaplaceExp(\ChildLaplaceExp(u))}du$ and $\pEPPFSecondInt=\int_{\Rp}u\ChildMoment(u,2)\MotherMoment(\ChildLaplaceExp(u),1)\e^{-\MotherLaplaceExp(\ChildLaplaceExp(u))}du$. Also,
$$
\pEPPFFirstInt+\pEPPFSecondInt+\int_{\Rp}u(\ChildMoment(u,1))^2(\MotherMoment(\ChildLaplaceExp(u),1))^2\e^{-\MotherLaplaceExp(\ChildLaplaceExp(u))}\dd u=1.
$$
\end{lemma}
\begin{proof}
Let $\ChildProcPP_{\GroupIdx}^2$ be the second power measure of $\ChildProcPP_{\GroupIdx}$, then
\begin{align*}
\E[\ChildProcNorm_{\GroupIdx}(A)^2]&=\E\left[\E\left[\int_{(A\times\Rp)^2}\frac1{\ChildProc_{\GroupIdx}(\ChildSpace)^2}s_1s_2\ChildProcPP_{\GroupIdx}^2(\dd s_1,\dd \ChildLoc_1,\dd s_2,\dd \ChildLoc_2)\mid\MotherProc\right]\right].\\
\intertext{Straightforward computations show that the reciprocal of $\ChildProc_{\GroupIdx}(\ChildSpace)^2$ can be expressed as $\int_{\Rp}u\exp\left\{-u\ChildProc_{\GroupIdx}(\ChildSpace)\right\}\dd u$, thus}
\E[\ChildProcNorm_{\GroupIdx}(A)^2]&=\int_{\Rp}u\E\left[\E\left[\int_{(A\times\Rp)^2}\exp\{-u\ChildProc_{\GroupIdx}(\ChildSpace)\}s_1s_2\ChildProcPP_{\GroupIdx}^2(\dd s_1,\dd \ChildLoc_1,\dd s_2,\dd \ChildLoc_2)\mid\MotherProc\right]\right]\dd u.
\intertext{We exploit higher order CLM formula \citep{Baccelli_2024} and the definition of Laplace exponent for a Poisson process to obtain:}
\E[\ChildProcNorm_{\GroupIdx}(A)^2]&=\int_{\Rp\times A^2}\E\left[\int_{(\Rp\times\MotherSpace)^2}\MotherJump_1\MotherJump_2\Kernel(\ChildLoc_1,\MotherLoc_1)\Kernel(\ChildLoc_2,\MotherLoc_2)\e^{-\ChildLaplaceExp(u)\int_{\Rp\times\MotherSpace}\MotherJump'\MotherProc(\dd\MotherJump',\dd\MotherLoc')}\MotherProc^2(\dd\MotherJump_1,\dd\MotherLoc_1,\dd\MotherJump_2,\dd\MotherLoc_2) \right]\\
&\qquad\times u(\ChildMoment(u,1))^2\dd\ChildLoc_1\dd\ChildLoc_2\dd u\\
&\quad+\int_{\Rp}u\ChildMoment(u,2)\int_A\E\left[\e^{-\ChildLaplaceExp(u)\int_{\Rp\times\MotherSpace}\MotherJump\MotherProc(\dd\MotherJump,\dd\MotherLoc)}\int_{\Rp\times\MotherSpace}\MotherJump\Kernel(\ChildLoc,\MotherLoc)\MotherProc(\dd\MotherJump,\dd \MotherLoc)\right]\dd\ChildLoc\dd u.
\end{align*}
The first result of the lemma follows by applying higher order CLM formula to the first term and CLM formula to the second term. The second result of the lemma follows by setting $A=\ChildSpace$.
\end{proof}

Now we proceed to prove \Cref{teo:corr}. The formula for the correlation is
$$
\Cor(\ChildProcNorm_{\GroupIdx}(A),\ChildProcNorm_{\GroupIdxAlt}(A))=\frac{\Cov(\ChildProcNorm_{\GroupIdx}(A),\ChildProcNorm_{\GroupIdxAlt}(A))}{\sqrt{\Var(\ChildProcNorm_{\GroupIdx}(A))}\sqrt{\Var(\ChildProcNorm_{\GroupIdxAlt}(A))}}.
$$
Exploiting Equation~\eqref{eq:prior_expectation} and \Cref{lemma:corr_supp_1}, we write the numerator as
$$
\Cov(\ChildProcNorm_{\GroupIdx}(A),\ChildProcNorm_{\GroupIdxAlt}(A))=\E[\E[\ChildProcNorm_{\GroupIdx}(A)\mid\MotherProc]\E[\ChildProcNorm_{\GroupIdxAlt}(A)\mid\MotherProc]]-\E[\ChildProcNorm_{\GroupIdx}(A)]\E[\ChildProcNorm_{\GroupIdxAlt}(A)]=[m(A,A)-m(A)^2]\MotherEPPFInt.
$$
Exploiting Equation~\eqref{eq:prior_expectation} and \Cref{lemma:corr_supp_2}, we write the denominator as
$$
\Var(\ChildProcNorm_{\GroupIdx}(A))=\E[\ChildProcNorm_{\GroupIdx}(A)^2]-\E[\ChildProcNorm_{\GroupIdx}(A)]^2 =[m(A)-m(A)^2]\pEPPFSecondInt+[m(A,A)-m(A)^2]\pEPPFFirstInt
$$
and the result of the theorem follows.

\subsection{Interpretation of the three integrals in Equation~\eqref{eq:prior_corr}}\label{sec:corr_interpretation}
In this section we clarify the notation and interpretation of the three integrals appearing in Equation~\eqref{eq:prior_corr}.

As shown in Equation (36) of \cite{Pitman_LecNotes_2003}, the integral $\MotherEPPFInt$ is the marginal probability that two cluster allocation labels $\LatentChildMother_{\GroupIdx\ChildIdx}$ and $\LatentChildMother_{\GroupIdxAlt\ChildIdxAlt}$ are equal. The subscript 1 encodes that the corresponding partition of $\{\LatentMixtUnique_{\GroupIdx\ChildIdx}, \LatentMixtUnique_{\GroupIdxAlt\ChildIdxAlt}\}$ consists of a single block $\{\LatentMixtUnique_{\GroupIdx\ChildIdx}, \LatentMixtUnique_{\GroupIdxAlt\ChildIdxAlt}\}$. In terms of the restaurant franchise metaphor in \Cref{sec:pred}, $\MotherEPPFInt$ corresponds to the probability that two tables are in rooms associated with the same theme.

Building on \Cref{teo:marg}, we can prove that $\pEPPFFirstInt$ corresponds to the marginal probability that $\LatentMixt_{\GroupIdx\ObsIdx}\neq\LatentMixt_{\GroupIdx\ObsIdxAlt}$ and $\LatentChildMotherObs_{\GroupIdx\LatentMixtClusLabels_{\GroupIdx\ObsIdx}}=\LatentChildMotherObs_{\GroupIdx\LatentMixtClusLabels_{\GroupIdx\ObsIdxAlt}}$. The first term in the subscript has the same interpretation as in $\MotherEPPFInt$, indeed the partition of $\{\LatentMixtUnique_{\GroupIdx\LatentMixtClusLabels_{\GroupIdx\ObsIdx}}, \LatentMixtUnique_{\GroupIdx\LatentMixtClusLabels_{\GroupIdx\ObsIdxAlt}}\}$ consists of a single block. The second term encodes that the partition of $\{\LatentMixt_{\GroupIdx\ObsIdx}, \LatentMixt_{\GroupIdx\ObsIdxAlt}\}$ consists of two singleton blocks $\{\LatentMixt_{\GroupIdx\ObsIdx}\}$ and $\{\LatentMixt_{\GroupIdx\ObsIdxAlt}\}$. In terms of the restaurant franchise metaphor, $\pEPPFFirstInt$ corresponds to the probability that two customers in the same restaurant choose the same thematic room but sit at different tables.

By a similar argument, $\pEPPFSecondInt$ corresponds to the marginal probability that $\LatentMixt_{\GroupIdx\ObsIdx}=\LatentMixt_{\GroupIdx\ObsIdxAlt}$. The notation follows the same convention as described for $\pEPPFFirstInt$. In terms of the restaurant franchise metaphor, $\pEPPFSecondInt$ corresponds to the probability that two customers in the same restaurant sit at the same tables.

\subsection{Proof of \Cref{teo:marg,teo:post}}\label{sec:marg_post_proof}
We first introduce a technical lemma at the core of the proofs of \Cref{teo:marg,teo:post}.

\begin{lemma}\label{lemma:marg_post_supp}
Let $f_1,\dots,f_{\GroupNum}:\ChildSpace\to\Rp$ be measurable functions, $(\LatentMixtVec_1, \ldots, \LatentMixtVec_\GroupNum)$ be a sample of size $\ObsNum = \ObsNum_1 + \cdots + \ObsNum_\GroupNum$ from \eqref{eq:latent_model}, and  $\Kernel(\cdot,\MotherLoc)$ be a probability density function for each $\MotherLoc\in\MotherSpace$. Then,
\begin{align*}
&\E\left[\e^{-\sum_{\GroupIdx=1}^{\GroupNum}\int_{\ChildSpace}f_{\GroupIdx}(x)\ChildProc_{\GroupIdx}(\dd x)}\Law(\LatentMixtVec,\ChildAuxVec\mid\ChildProcList)\right]\\
&\quad=\sum_{\LatentChildMotherObsVec\in(\cdot)}\prod_{\GroupIdx=1}^{\GroupNum}\frac{\ChildAux_{\GroupIdx}^{\ObsNum_{\GroupIdx}-1}}{\Gamma(\ObsNum_{\GroupIdx})}\prod_{\ChildIdx=1}^{\LatentMixtNumUnique_{\GroupIdx}}\ChildMoment(f_{\GroupIdx}(\LatentMixtUnique_{\GroupIdx\ChildIdx})+\ChildAux_{\GroupIdx},\LatentMixtCounter_{\GroupIdx\ChildIdx})\\
&\qquad\times\exp\left\{-\int_{\MotherSpace}\MotherLaplaceExp\left(\sum_{\GroupIdx=1}^{\GroupNum}\int_{\ChildSpace}\ChildLaplaceExp(f_{\GroupIdx}(x)+\ChildAux_{\GroupIdx})\Kernel(x,\MotherLoc)\dd x\right)\MotherMeanLoc(\dd\MotherLoc)\right\}\\
&\qquad\times\prod_{\MotherIdx=1}^{\LatentChildMotherNumUnique}\int_{\MotherSpace}\MotherMoment\left(\sum_{\GroupIdx=1}^{\GroupNum}\int_{\ChildSpace}\ChildLaplaceExp(f_{\GroupIdx}(x)+\ChildAux_{\GroupIdx})\Kernel(x,\MotherLoc_{\MotherIdx})\dd x,\LatentChildMotherCounter_{\MotherIdx}\right)\prod_{(\GroupIdx,\ChildIdx):\LatentChildMotherObs_{\GroupIdx\ChildIdx}=\MotherIdx}\Kernel(\LatentMixtUnique_{\GroupIdx\ChildIdx},\MotherLoc_{\MotherIdx})\MotherMeanLoc(\dd\MotherLoc_{\MotherIdx})
\end{align*}
where the sum is over all possible combinations of the latent variables $\LatentChildMotherObsVec$, and $\ChildAuxVec=(\ChildAux_{1},\ldots,\ChildAux_{\GroupNum})$ with $\ChildAux_{\GroupIdx}\mid\ChildProc_{\GroupIdx}\ind\mathrm{Gamma}(\ObsNum_{\GroupIdx},\ChildProc_{\GroupIdx}(\ChildSpace))$ for $\GroupIdx=1,\ldots,\GroupNum$.
\end{lemma}
\begin{proof}
Conditioning on $\ChildProc_1,\ldots,\ChildProc_{\GroupNum}$, $\LatentMixtVec$ and $\ChildAuxVec$ are independent. Thus, we have
\begin{align*}
&\E\left[\e^{-\sum_{\GroupIdx=1}^{\GroupNum}\int_{\ChildSpace}f_{\GroupIdx}(x)\ChildProc_{\GroupIdx}(\dd x)}\Law(\LatentMixtVec,\ChildAuxVec\mid\ChildProcList)\right]\\
&\quad=\E\left[\prod_{\GroupIdx=1}^{\GroupNum}\frac{\ChildAux_{\GroupIdx}^{\ObsNum_{\GroupIdx}-1}}{\Gamma(\ObsNum_{\GroupIdx})}\e^{-\int_{\Rp\times\ChildSpace}s(f_{\GroupIdx}(\ChildLoc)+\ChildAux_{\GroupIdx})\ChildProcPP_{\GroupIdx}(\dd s,\dd \ChildLoc)}\prod_{\ChildIdx=1}^{\LatentMixtNumUnique_{\GroupIdx}}\ChildProc(\LatentMixtUnique_{\GroupIdx\ChildIdx})^{\LatentMixtCounter_{\GroupIdx\ChildIdx}}\right]\\
&\quad=\E\left[\prod_{\GroupIdx=1}^{\GroupNum}\frac{\ChildAux_{\GroupIdx}^{\ObsNum_{\GroupIdx}-1}}{\Gamma(\ObsNum_{\GroupIdx})}\E\left[\e^{-\int_{\Rp\times\ChildSpace}s(f_{\GroupIdx}(\ChildLoc)+\ChildAux_{\GroupIdx})\ChildProcPP_{\GroupIdx}(\dd s,\dd \ChildLoc)}\int_{(\Rp)^{\LatentMixtNumUnique_{\GroupIdx}}}\prod_{\ChildIdx=1}^{\LatentMixtNumUnique_{\GroupIdx}}s_{\GroupIdx\ChildIdx}^{\LatentMixtCounter_{\GroupIdx\ChildIdx}}\ChildProcPP_{\GroupIdx}^{\LatentMixtNumUnique_{\GroupIdx}}(\dd s_{\GroupIdx1},\dd\LatentMixtUnique_{\GroupIdx1},\ldots,\dd s_{\GroupIdx\LatentMixtNumUnique_{\GroupIdx}},\dd\LatentMixtUnique_{\GroupIdx\LatentMixtNumUnique_{\GroupIdx}})\mid\MotherProc\right]\right].
\intertext{Since $k$ is a probability density kernel  $\LatentMixtUnique_{\GroupIdx\ChildIdx}\neq \LatentMixtUnique_{\GroupIdxAlt\ChildIdxAlt}$ for $(\GroupIdx,\ChildIdx)\neq(\GroupIdxAlt,\ChildIdxAlt)$, therefore we can apply higher order CLM formula \citep{Baccelli_2024} to obtain:}
&\E\left[\e^{-\sum_{\GroupIdx=1}^{\GroupNum}\int_{\ChildSpace}f_{\GroupIdx}(x)\ChildProc_{\GroupIdx}(\dd x)}\Law(\LatentMixtVec,\ChildAuxVec\mid\ChildProcList)\right]\\
&\quad=\E\left[\prod_{\GroupIdx=1}^{\GroupNum}\exp\left\{-\int_{\ChildSpace}\ChildLaplaceExp(f_{\GroupIdx}(\ChildLoc)+\ChildAux_{\GroupIdx})\int_{\Rp\times\MotherSpace}\Kernel(\ChildLoc,\MotherLoc)\MotherJump\MotherProc(\dd\MotherJump,\dd\MotherLoc)\dd\ChildLoc\right\}\prod_{\ChildIdx=1}^{\LatentMixtNumUnique_{\GroupIdx}}\int_{\Rp\times\MotherSpace}\Kernel(\LatentMixtUnique_{\GroupIdx\ChildIdx},\MotherLoc)\MotherJump\MotherProc(\dd\MotherJump,\dd\MotherLoc)\right]\\
&\qquad\times \prod_{\GroupIdx=1}^{\GroupNum}\frac{\ChildAux_{\GroupIdx}^{\ObsNum_{\GroupIdx}-1}}{\Gamma(\ObsNum_{\GroupIdx})}\prod_{\ChildIdx=1}^{\LatentMixtNumUnique_{\GroupIdx}}\ChildMoment(f_{\GroupIdx}(\LatentMixtUnique_{\GroupIdx\ChildIdx})+\ChildAux_{\GroupIdx},\LatentMixtCounter_{\GroupIdx\ChildIdx}).
\intertext{Following Equation~14.E.3 and Lemma~14.E.4 from \citet{Baccelli_2024}, we can introduce the vector of latent variables $\LatentChildMotherObsVec$:}
&\E\left[\e^{-\sum_{\GroupIdx=1}^{\GroupNum}\int_{\ChildSpace}f_{\GroupIdx}(x)\ChildProc_{\GroupIdx}(\dd x)}\Law(\LatentMixtVec,\ChildAuxVec\mid\ChildProcList)\right]\\
&\quad=\sum_{\LatentChildMotherObsVec\in(\cdot)}\E\left[\int_{(\Rp\times\MotherSpace)^{\LatentChildMotherNumUnique}}\exp\left\{-\sum_{\GroupIdx=1}^{\GroupNum}\int_{\ChildSpace}\ChildLaplaceExp(f_{\GroupIdx}(\ChildLoc)+\ChildAux_{\GroupIdx})\int_{\Rp\times\MotherSpace}\Kernel(\ChildLoc,\MotherLoc)\MotherJump\MotherProc(\dd\MotherJump,\dd\MotherLoc)\dd\ChildLoc\right\}\right.\\
&\qquad\times\left.\prod_{\MotherIdx=1}^{\LatentChildMotherNumUnique}\prod_{(\GroupIdx,\ChildIdx):\LatentChildMotherObs_{\GroupIdx\ChildIdx}=\MotherIdx}\MotherJump_{\MotherIdx}^{\LatentChildMotherCounter_{\MotherIdx}}\Kernel(\LatentMixtUnique_{\GroupIdx\ChildIdx},\MotherLoc_{\MotherIdx})\MotherProc^{(\LatentChildMotherNumUnique)}(\dd\MotherJump_1,\dd\MotherLoc_1,\ldots,\dd\MotherJump_{\LatentChildMotherNumUnique},\dd\MotherLoc_{\LatentChildMotherNumUnique})\right]\\
&\qquad\times \prod_{\GroupIdx=1}^{\GroupNum}\frac{\ChildAux_{\GroupIdx}^{\ObsNum_{\GroupIdx}-1}}{\Gamma(\ObsNum_{\GroupIdx})}\prod_{\ChildIdx=1}^{\LatentMixtNumUnique_{\GroupIdx}}\ChildMoment(f_{\GroupIdx}(\LatentMixtUnique_{\GroupIdx\ChildIdx})+\ChildAux_{\GroupIdx},\LatentMixtCounter_{\GroupIdx\ChildIdx})
\end{align*}
where the sum is extended over all possible configurations of $\LatentChildMotherObsVec$, and $\MotherProc^{(\LatentChildMotherNumUnique)}$ is the $\LatentChildMotherNumUnique$th factorial moment measure of $\MotherProc$. The statement of the lemma follows from applying higher order CLM formula.
\end{proof}

\Cref{teo:marg} follows from \Cref{lemma:marg_post_supp} with $f_1,\dots,f_{\GroupNum}$ set to null functions. To prove \Cref{teo:post}, we observe that elementary properties of conditional expectations yield
\begin{align*}
&\E\left[\e^{-\sum_{\GroupIdx=1}^{\GroupNum}\int_{\ChildSpace}f_{\GroupIdx}(x)\ChildProc_{\GroupIdx}(\dd x)}\mid \LatentMixtVec,\ChildAuxVec,\LatentChildMotherObsVec\right]=\frac{\E\left[\e^{-\sum_{\GroupIdx=1}^{\GroupNum}\int_{\ChildSpace}f_{\GroupIdx}(x)\ChildProc_{\GroupIdx}(\dd x)}\Law(\LatentMixtVec,\ChildAuxVec,\LatentChildMotherObsVec\mid\ChildProcList)\right]}{\Law(\LatentMixtVec,\ChildAuxVec,\LatentChildMotherObsVec)}.
\intertext{Thanks to \Cref{teo:marg} and \Cref{lemma:marg_post_supp}, we get:}
&\E\left[\e^{-\sum_{\GroupIdx=1}^{\GroupNum}\int_{\ChildSpace}f_{\GroupIdx}(x)\ChildProc_{\GroupIdx}(\dd x)}\mid \LatentMixtVec,\ChildAuxVec,\LatentChildMotherObsVec\right]\\
&\quad=\prod_{\MotherIdx=1}^{\LatentChildMotherNumUnique}\int_{\MotherSpace}\frac{\MotherMoment\left(\sum_{\GroupIdx=1}^{\GroupNum}\int_{\ChildSpace}\ChildLaplaceExp(f_{\GroupIdx}(x)+\ChildAux_{\GroupIdx})\Kernel(x,\MotherLoc_{\MotherIdx})\dd x,\LatentChildMotherCounter_{\MotherIdx}\right)}
{\MotherMoment\left(\sum_{\GroupIdx=1}^{\GroupNum}\ChildLaplaceExp(\ChildAux_{\GroupIdx}),\LatentChildMotherCounter_{\MotherIdx}\right)}
\frac{\prod_{(\GroupIdx,\ChildIdx):\LatentChildMotherObs_{\GroupIdx\ChildIdx}=\MotherIdx}\Kernel(\LatentMixtUnique_{\GroupIdx\ChildIdx},\MotherLoc_{\MotherIdx})}{\int_{\MotherSpace}\prod_{(\GroupIdx,\ChildIdx):\LatentChildMotherObs_{\GroupIdx\ChildIdx}=\MotherIdx}\Kernel(\LatentMixtUnique_{\GroupIdx\ChildIdx},\MotherLoc_{\MotherIdx}')\MotherMeanLoc(\dd\MotherLoc_{\MotherIdx}')}\MotherMeanLoc(\dd\MotherLoc_{\MotherIdx})\\
&\qquad\times\frac{\exp\left\{-\int_{\MotherSpace}\MotherLaplaceExp\left(\sum_{\GroupIdx=1}^{\GroupNum}\int_{\ChildSpace}\ChildLaplaceExp(f_{\GroupIdx}(x)+\ChildAux_{\GroupIdx})\Kernel(x,\MotherLoc)\dd x\right)\MotherMeanLoc(\dd\MotherLoc)\right\}}{\exp\left\{-\MotherLaplaceExp\left(\sum_{\GroupIdx=1}^{\GroupNum}\ChildLaplaceExp(\ChildAux_{\GroupIdx})\right)\right\}}\notag\\
&\qquad\times\prod_{\GroupIdx=1}^{\GroupNum}\prod_{\ChildIdx=1}^{\LatentMixtNumUnique_{\GroupIdx}}\frac{\MotherMoment(f_{\GroupIdx}(\LatentMixtUnique_{\GroupIdx\ChildIdx})+\ChildAux_{\GroupIdx},\LatentMixtCounter_{\GroupIdx\ChildIdx})}{\MotherMoment(\ChildAux_{\GroupIdx},\LatentMixtCounter_{\GroupIdx\ChildIdx})}.
\intertext{Straightforward  computations show that:}
&\E\left[\e^{-\sum_{\GroupIdx=1}^{\GroupNum}\int_{\ChildSpace}f_{\GroupIdx}(x)\ChildProc_{\GroupIdx}(\dd x)}\mid \LatentMixtVec,\ChildAuxVec,\LatentChildMotherObsVec\right]\notag\\
&\quad=\prod_{\MotherIdx=1}^{\LatentChildMotherNumUnique}\int_{\Rp\times\MotherSpace}\exp\left\{-\MotherJump\sum_{\GroupIdx=1}^{\GroupNum}\int_{\Rp\times\ChildSpace}(1-\e^{-sf_{\GroupIdx}(x)})\e^{-s\ChildAux_{\GroupIdx}}\ChildMeanJump(s)\dd s\Kernel(x,\MotherLoc_{\MotherIdx})\dd x\right\}\\
&\qquad\times \frac{\prod_{(\GroupIdx,\ChildIdx):\LatentChildMotherObs_{\GroupIdx\ChildIdx}=\MotherIdx}\Kernel(\LatentMixtUnique_{\GroupIdx\ChildIdx},\MotherLoc_{\MotherIdx})\MotherMeanLoc(\dd\MotherLoc_{\MotherIdx})}{\int_{\MotherSpace}\prod_{(\GroupIdx,\ChildIdx):\LatentChildMotherObs_{\GroupIdx\ChildIdx}=\MotherIdx}\Kernel(\LatentMixtUnique_{\GroupIdx\ChildIdx},\MotherLoc_{\MotherIdx}')\MotherMeanLoc(\dd\MotherLoc_{\MotherIdx}')}
\frac{\exp\left\{-\MotherJump_{\MotherIdx}\sum_{\GroupIdx=1}^{\GroupNum}\ChildLaplaceExp(\ChildAux_{\GroupIdx}) \right\}\MotherJump_{\MotherIdx}^{\LatentChildMotherCounter_{\MotherIdx}}\MotherMeanJump(\MotherJump_{\MotherIdx})}{\int_{\Rp}\exp\left\{-\MotherJump'_{\MotherIdx}\sum_{\GroupIdx=1}^{\GroupNum}\ChildLaplaceExp(\ChildAux_{\GroupIdx}) \right\}(\MotherJump'_{\MotherIdx})^{\LatentChildMotherCounter_{\MotherIdx}}\MotherMeanJump(\MotherJump'_{\MotherIdx})}\dd\MotherJump_{\MotherIdx}\\
&\qquad\times\prod_{\GroupIdx=1}^{\GroupNum}\prod_{\ChildIdx=1}^{\LatentMixtNumUnique_{\GroupIdx}}\int_{\Rp}\e^{-sf_{\GroupIdx}(\LatentMixtUnique_{\GroupIdx\ChildIdx})}\frac{\e^{-\ChildAux_{\GroupIdx}s}s^{\LatentMixtCounter_{\GroupIdx\ChildIdx}}\ChildMeanJump(s)}{\int_{\Rp}\e^{-\ChildAux_{\GroupIdx}s'}s'^{\LatentMixtCounter_{\GroupIdx\ChildIdx}}\ChildMeanJump(s')\dd s'}\dd s\\
&\qquad\times\exp\left\{-\int_{\Rp\times\MotherSpace}\left(1-\e^{-\MotherJump\sum_{\GroupIdx=1}^{\GroupNum}\int_{\Rp\times\ChildSpace}(1-\e^{-sf_{\GroupIdx}(x)})\e^{-s\ChildAux_{\GroupIdx}}\ChildMeanJump(s)\dd s\Kernel(x,\MotherLoc)\dd x}\right)\right.\\
&\qquad\quad\times\left.\e^{-\MotherJump\sum_{\GroupIdx=1}^{\GroupNum}\ChildLaplaceExp(\ChildAux_{\GroupIdx})}\MotherMeanJump(\MotherJump)\dd\MotherJump\MotherMeanLoc(\dd\MotherLoc)\right\}.
\end{align*}
The right hand side of the last equation consists of  three factors. The first one equals
\begin{align*}
&\prod_{\MotherIdx=1}^{\LatentChildMotherNumUnique}\int_{\Rp\times\MotherSpace}\exp\left\{-\MotherJump\sum_{\GroupIdx=1}^{\GroupNum}\int_{\Rp\times\ChildSpace}(1-\e^{-sf_{\GroupIdx}(x)})\e^{-s\ChildAux_{\GroupIdx}}\ChildMeanJump(s)\dd s\Kernel(x,\MotherLoc_{\MotherIdx})\dd x\right\}\\
&\quad\times \frac{\prod_{(\GroupIdx,\ChildIdx):\LatentChildMotherObs_{\GroupIdx\ChildIdx}=\MotherIdx}\Kernel(\LatentMixtUnique_{\GroupIdx\ChildIdx},\MotherLoc_{\MotherIdx})\MotherMeanLoc(\dd\MotherLoc_{\MotherIdx})}{\int_{\MotherSpace}\prod_{(\GroupIdx,\ChildIdx):\LatentChildMotherObs_{\GroupIdx\ChildIdx}=\MotherIdx}\Kernel(\LatentMixtUnique_{\GroupIdx\ChildIdx},\MotherLoc_{\MotherIdx}')\MotherMeanLoc(\dd\MotherLoc_{\MotherIdx}')}
\frac{\exp\left\{-\MotherJump_{\MotherIdx}\sum_{\GroupIdx=1}^{\GroupNum}\ChildLaplaceExp(\ChildAux_{\GroupIdx}) \right\}\MotherJump_{\MotherIdx}^{\LatentChildMotherCounter_{\MotherIdx}}\MotherMeanJump(\MotherJump_{\MotherIdx})}{\int_{\Rp}\exp\left\{-\MotherJump'_{\MotherIdx}\sum_{\GroupIdx=1}^{\GroupNum}\ChildLaplaceExp(\ChildAux_{\GroupIdx}) \right\}(\MotherJump'_{\MotherIdx})^{\LatentChildMotherCounter_{\MotherIdx}}\MotherMeanJump(\MotherJump'_{\MotherIdx})}\dd\MotherJump_{\MotherIdx},
\end{align*}
which can be interpreted as a  Laplace transform. Indeed,  for each $\MotherIdx=1,\ldots,\LatentChildMotherNumUnique$, the exponential term can be seen as the Laplace transform of the processes 
\[
\ChildProc^{(p)}_{1\MotherIdx},\ldots,\ChildProc^{(p)}_{\GroupNum\MotherIdx}\mid\MotherJump_\MotherIdx,\MotherLoc_\MotherIdx,\ChildAuxVec\sim\CRM(\e^{-\ChildAux_{\GroupIdx}s}\ChildMeanJump(s)\dd s \MotherJump_\MotherIdx\Kernel(x,\MotherJump_\MotherIdx)\dd x), \]
and the marginal densities of $\MotherLoc_\MotherIdx$ and $\MotherJump_\MotherIdx$ are proportional to $\MotherMeanLoc(\MotherLoc_{\MotherIdx})\prod_{(\GroupIdx,\ChildIdx):\LatentChildMotherObs_{\GroupIdx\ChildIdx}=\MotherIdx}\Kernel(\LatentMixtUnique_{\GroupIdx\ChildIdx},\MotherLoc_{\MotherIdx})$ and $\exp\left\{-\MotherJump_{\MotherIdx}\sum_{\GroupIdx=1}^{\GroupNum}\ChildLaplaceExp(\ChildAux_{\GroupIdx}) \right\}\MotherJump_{\MotherIdx}^{\LatentChildMotherCounter_{\MotherIdx}}\MotherMeanJump(\MotherJump_{\MotherIdx})$, respectively. The second factor coincides with
$$
\prod_{\GroupIdx=1}^{\GroupNum}\prod_{\ChildIdx=1}^{\LatentMixtNumUnique_{\GroupIdx}}\int_{\Rp}\e^{-sf_{\GroupIdx}(\LatentMixtUnique_{\GroupIdx\ChildIdx})}\frac{\e^{-\ChildAux_{\GroupIdx}s}s^{\LatentMixtCounter_{\GroupIdx\ChildIdx}}\ChildMeanJump(s)}{\int_{\Rp}\e^{-\ChildAux_{\GroupIdx}s'}s'^{\LatentMixtCounter_{\GroupIdx\ChildIdx}}\ChildMeanJump(s')\dd s'}\dd s , 
$$
where, for each $\GroupIdx=1,\ldots,\GroupNum$ and $\ChildIdx=1,\ldots,\LatentMixtNumUnique_{\GroupIdx}$, the exponential $\e^{-sf_{\GroupIdx}(\LatentMixtUnique_{\GroupIdx\ChildIdx})}$ can be seen as the Laplace transform of a point process that, conditioning on $\ChildJump_{\GroupIdx\ChildIdx}$, has a single fixed point in $(\ChildJump_{\GroupIdx\ChildIdx}, \LatentMixtUnique_{\GroupIdx\ChildIdx})$, and the marginal density of $\ChildJump_{\GroupIdx\ChildIdx}$ is proportional to $\e^{-\ChildAux_{\GroupIdx}s}s^{\LatentMixtCounter_{\GroupIdx\ChildIdx}}\ChildMeanJump(s)$. The third factor amounts to be
$$
\exp\left\{-\int_{\Rp\times\MotherSpace}\left(1-\e^{-\MotherJump\sum_{\GroupIdx=1}^{\GroupNum}\int_{\Rp\times\ChildSpace}(1-\e^{-sf_{\GroupIdx}(x)})\e^{-s\ChildAux_{\GroupIdx}}\ChildMeanJump(s)\dd s\Kernel(x,\MotherLoc)\dd x}\right)\e^{-\MotherJump\sum_{\GroupIdx=1}^{\GroupNum}\ChildLaplaceExp(\ChildAux_{\GroupIdx})}\MotherMeanJump(\MotherJump)\dd\MotherJump\MotherMeanLoc(\dd\MotherLoc)\right\},
$$
and it is the Laplace transform of the processes 
\[\ChildProc^{(p)}_{1},\ldots\ChildProc^{(p)}_{\GroupNum}\mid\MotherProc^{(p)},\ChildAuxVec\sim\CRM\left(\e^{-\ChildAux_{\GroupIdx}s}\ChildMeanJump(s)\dd s\sum_{\MotherIdx\ge1}\MotherJump_{\MotherIdx}^{(p)}\Kernel(x,\MotherLoc_{\MotherIdx}^{(p)})\dd x\right),
\]
where $\left(\MotherJump_{\MotherIdx}^{(p)}\right)_{\MotherIdx\ge1}$ and $\left(\MotherLoc_{\MotherIdx}^{(p)}\right)_{\MotherIdx\ge1}$ are the jumps and locations of $\MotherProc^{(p)}$ and $\MotherProc^{(p)}\mid\ChildAuxVec\sim\PP\left(\e^{-\MotherJump\sum_{\GroupIdx=1}^{\GroupNum}\ChildLaplaceExp(\ChildAux_{\GroupIdx})}\MotherMeanJump(\MotherJump)\dd\MotherJump\MotherMeanLoc(\dd\MotherLoc)\right)$. Thus, the statement of \Cref{teo:post} follows. 

\subsection{Proof of Theorem \ref{teo:pred}}\label{sec:pred_proof}
For simplicity of exposition, we focus on the case $\GroupIdx=1$ in the following proof. The argument for $\GroupIdx=2,\ldots,\GroupNum$ is analogous. 

The almost sure discreteness of $\ChildProcNorm_{1}$ implies that the event $\LatentMixt_{1,\ObsNum_{1}+1}\in A$ can arise in two distinct ways: (\textit{i}) $\LatentMixt_{1,\ObsNum_{1}+1}$ coincides with a unique value $\LatentMixtUnique_{1\ChildIdx}$ already observed in $\LatentMixtVec_{\GroupIdx}$, with $\LatentMixtUnique_{1\ChildIdx}\in A$; or (\textit{ii}) $\LatentMixt_{1,\ObsNum_{1}+1}$ corresponds to a new unique value $\LatentMixtUnique_{1,\LatentMixtNumUnique_{1}+1}$, not previously observed in $\LatentMixtVec_{\GroupIdx}$, with  $\LatentMixtUnique_{1,\LatentMixtNumUnique_{1}+1}\in A$. Since $\Kernel$ is a probability density kernel, case (\textit{ii}) implies that  $\LatentMixt_{1,\ObsNum_{1}+1}\neq\LatentMixtUnique_{1\ChildIdx}$ for all $\ChildIdx=1,\ldots,\LatentMixtNumUnique_{1}$. According to the restaurant metaphor of \Cref{sec:pred}, the event $\LatentMixt_{1,\ObsNum_{1}+1}\in A$ corresponds to a new customer entering restaurant 1 and sitting at a table where the served dish belongs to $A$. In case (\textit{i}), the table is already occupied by other customers, while in case (\textit{ii}) the table is empty.

We first consider case (\textit{i}) with a fixed $\ChildIdx\in\{1,\ldots,\LatentMixtNumUnique_{1}\}$. Without loss of generality, we consider $\ChildIdx=1$. The probability associated to such a case is the product of two terms: the Dirac's delta $\delta_{\LatentMixtUnique_{1,1}}(A)$, which guarantees that $\LatentMixtUnique_{1,1}\in A$, and the ratio between $\Law(\LatentMixt_{1,\ObsNum_{1}+1},\LatentMixtVec,\ChildAuxVec,\LatentChildMotherObsVec)$ and $\Law(\LatentMixtVec,\ChildAuxVec,\LatentChildMotherObsVec)$, which can be computed exploiting Equation~\eqref{eq:latent_joint}. In particular, the ratio becomes:
\begin{align*}
&\frac{\exp\left\{-\MotherLaplaceExp\left(\sum_{\GroupIdx=1}^\GroupNum\ChildLaplaceExp(\ChildAux_{\GroupIdx})\right)\right\}\prod_{\MotherIdx=1}^{\LatentChildMotherNumUnique} \MotherMoment\left(\sum_{\GroupIdx=1}^\GroupNum\ChildLaplaceExp(\ChildAux_{\GroupIdx}), \LatentChildMotherCounter_\MotherIdx\right) m(\dd\LatentMixtUnique_{\GroupIdx \ChildIdx} \colon \LatentChildMotherObs_{{\GroupIdx \ChildIdx}} = \MotherIdx)}
{\exp\left\{-\MotherLaplaceExp\left(\sum_{\GroupIdx=1}^\GroupNum\ChildLaplaceExp(\ChildAux_{\GroupIdx})\right)\right\}\prod_{\MotherIdx=1}^{\LatentChildMotherNumUnique} \MotherMoment\left(\sum_{\GroupIdx=1}^\GroupNum\ChildLaplaceExp(\ChildAux_{\GroupIdx}), \LatentChildMotherCounter_\MotherIdx\right) m(\dd\LatentMixtUnique_{\GroupIdx \ChildIdx} \colon \LatentChildMotherObs_{{\GroupIdx \ChildIdx}} = \MotherIdx)}\\
&\quad\times\frac{\frac{\ChildAux_1^{\ObsNum_{1}}}{\Gamma(\ObsNum_1+1)}\ChildMoment(\ChildAux_{1}, \LatentMixtCounter_{1,1}+1)\prod_{\ChildIdx=2}^{\LatentMixtNumUnique_{1}} \ChildMoment(\ChildAux_{1}, \LatentMixtCounter_{1 \ChildIdx})}{\frac{\ChildAux_1^{\ObsNum_{1} - 1}}{\Gamma(\ObsNum_1)}\prod_{\ChildIdx=1}^{\LatentMixtNumUnique_{1}} \ChildMoment(\ChildAux_{1}, \LatentMixtCounter_{1 \ChildIdx})}
\frac{\prod_{\GroupIdx=2}^{\GroupNum} \left\{\frac{\ChildAux_\GroupIdx^{\ObsNum_{\GroupIdx} - 1}}{\Gamma(\ObsNum_\GroupIdx)}  \prod_{\ChildIdx=1}^{\LatentMixtNumUnique_{\GroupIdx}} \ChildMoment(\ChildAux_{\GroupIdx}, \LatentMixtCounter_{\GroupIdx \ChildIdx}) \right\}}
{\prod_{\GroupIdx=2}^{\GroupNum} \left\{\frac{\ChildAux_\GroupIdx^{\ObsNum_{\GroupIdx} - 1}}{\Gamma(\ObsNum_\GroupIdx)}  \prod_{\ChildIdx=1}^{\LatentMixtNumUnique_{\GroupIdx}} \ChildMoment(\ChildAux_{\GroupIdx}, \LatentMixtCounter_{\GroupIdx \ChildIdx}) \right\}}
\end{align*}
and the first term of \Cref{teo:pred} follows. 

Then, we consider case (\textit{ii}). We augment the space of variables with the latent variable $\LatentChildMotherObs_{1,\LatentMixtNumUnique_1+1}$, which leads to two subcases: (\textit{ii}.1) $\LatentChildMotherObs_{1,\LatentMixtNumUnique_1+1}=\MotherIdx$ for some $\MotherIdx\in\{1,\ldots,\LatentChildMotherNumUnique\}$, meaning   $\LatentChildMotherObs_{1,\LatentMixtNumUnique_1+1}$ is associated with an already observed atom of the mother process; and (\textit{ii}.2) $\LatentChildMotherObs_{1,\LatentMixtNumUnique_1+1}=\LatentChildMotherNumUnique+1$, meaning  $\LatentChildMotherObs_{1,\LatentMixtNumUnique_1+1}$ is associated with a previously unobserved atom of the mother process. In the restaurant metaphor of \Cref{sec:pred}, case (\textit{ii}.1) corresponds to the new customer sitting at a table in a thematic room already occupied by other customers (seated at different tables in the same room), whereas case (\textit{ii}.2) corresponds to the new customer sitting at a table in a thematic room that is completely empty. For case (\textit{ii}.1) we consider, without loss of generality, $\MotherIdx=1$. The probability associated to this case is the ratio between $\Law(\LatentMixt_{1,\ObsNum_{1}+1},\LatentChildMotherObs_{1,\LatentMixtNumUnique_1+1}=1,\LatentMixtVec,\ChildAuxVec,\LatentChildMotherObsVec)$ and $\Law(\LatentMixtVec,\ChildAuxVec,\LatentChildMotherObsVec)$, integrated over $A$:
\begin{align*}
&\int_A \frac{\exp\left\{-\MotherLaplaceExp\left(\sum_{\GroupIdx=1}^\GroupNum\ChildLaplaceExp(\ChildAux_\GroupIdx)\right)\right\} \prod_{\GroupIdx=2}^{\GroupNum} \left\{\frac{\ChildAux_\GroupIdx^{\ObsNum_\GroupIdx - 1}}{\Gamma(\ObsNum_\GroupIdx)}  \prod_{ \ChildIdx=1}^{\LatentMixtNumUnique_\GroupIdx} \ChildMoment(\ChildAux_\GroupIdx, \LatentMixtCounter_{\GroupIdx \ChildIdx})\right\}}
{\exp\left\{-\MotherLaplaceExp\left(\sum_{\GroupIdx=1}^\GroupNum\ChildLaplaceExp(\ChildAux_\GroupIdx)\right)\right\} \prod_{\GroupIdx=2}^{\GroupNum} \left\{\frac{\ChildAux_\GroupIdx^{\ObsNum_\GroupIdx - 1}}{\Gamma(\ObsNum_\GroupIdx)}  \prod_{ \ChildIdx=1}^{\LatentMixtNumUnique_\GroupIdx} \ChildMoment(\ChildAux_\GroupIdx, \LatentMixtCounter_{\GroupIdx \ChildIdx})\right\}}\\
&\quad\times\frac{\prod_{\MotherIdx=2}^{\LatentChildMotherNumUnique} \MotherMoment\left(\sum_{\GroupIdx=1}^\GroupNum\ChildLaplaceExp(\ChildAux_\GroupIdx), \LatentChildMotherCounter_\MotherIdx\right) m(\dd\LatentMixtUnique_{\GroupIdx \ChildIdx} \colon \LatentChildMotherObs_{{\GroupIdx \ChildIdx}} = \MotherIdx)}{\prod_{\MotherIdx=2}^{\LatentChildMotherNumUnique} \MotherMoment\left(\sum_{\GroupIdx=1}^\GroupNum\ChildLaplaceExp(\ChildAux_\GroupIdx), \LatentChildMotherCounter_\MotherIdx\right) m(\dd\LatentMixtUnique_{\GroupIdx \ChildIdx} \colon \LatentChildMotherObs_{{\GroupIdx \ChildIdx}} = \MotherIdx)}\\
&\quad\times\frac{\frac{\ChildAux_1^{\ObsNum_{1}}}{\Gamma(\ObsNum_1+1)}\ChildMoment(\ChildAux_{1},1)\prod_{\ChildIdx=1}^{\LatentMixtNumUnique_{1}} \ChildMoment(\ChildAux_{1}, \LatentMixtCounter_{1 \ChildIdx})}{\frac{\ChildAux_1^{\ObsNum_{1} - 1}}{\Gamma(\ObsNum_1)}\prod_{\ChildIdx=1}^{\LatentMixtNumUnique_{1}} \ChildMoment(\ChildAux_{1}, \LatentMixtCounter_{1 \ChildIdx})}\\
&\quad\times\frac{ \MotherMoment\left(\sum_{\GroupIdx=1}^\GroupNum\ChildLaplaceExp(\ChildAux_\GroupIdx), \LatentChildMotherCounter_1+1\right)\int_{\MotherSpace}\Kernel(x,\MotherLoc_1)\prod_{(\GroupIdx,\ChildIdx):\LatentChildMotherObs_{\GroupIdx\ChildIdx}=1}\Kernel(\LatentMixtUnique_{\GroupIdx\ChildIdx},\MotherLoc_1)\MotherMeanLoc(\dd\MotherLoc_1)}{ \MotherMoment\left(\sum_{\GroupIdx=1}^\GroupNum\ChildLaplaceExp(\ChildAux_\GroupIdx), \LatentChildMotherCounter_1\right) m(\dd\LatentMixtUnique_{\GroupIdx \ChildIdx} \colon \LatentChildMotherObs_{{\GroupIdx \ChildIdx}} = 1)}\dd x
\end{align*}
and the second term of \Cref{teo:pred} follows. The probability associated to case (\textit{ii}.2) is the ratio between $\Law(\LatentMixt_{1,\ObsNum_{1}+1},\LatentChildMotherObs_{1,\LatentMixtNumUnique_1+1}=\LatentChildMotherNumUnique+1,\LatentMixtVec,\ChildAuxVec,\LatentChildMotherObsVec)$ and $\Law(\LatentMixtVec,\ChildAuxVec,\LatentChildMotherObsVec)$, integrated over $A$:
\begin{align*}
&\int_A \frac{\exp\left\{-\MotherLaplaceExp\left(\sum_{\GroupIdx=1}^\GroupNum\ChildLaplaceExp(\ChildAux_\GroupIdx)\right)\right\} \prod_{\GroupIdx=2}^{\GroupNum} \left\{\frac{\ChildAux_\GroupIdx^{\ObsNum_\GroupIdx - 1}}{\Gamma(\ObsNum_\GroupIdx)}  \prod_{ \ChildIdx=1}^{\LatentMixtNumUnique_\GroupIdx} \ChildMoment(\ChildAux_\GroupIdx, \LatentMixtCounter_{\GroupIdx \ChildIdx})\right\}}
{\exp\left\{-\MotherLaplaceExp\left(\sum_{\GroupIdx=1}^\GroupNum\ChildLaplaceExp(\ChildAux_\GroupIdx)\right)\right\} \prod_{\GroupIdx=2}^{\GroupNum} \left\{\frac{\ChildAux_\GroupIdx^{\ObsNum_\GroupIdx - 1}}{\Gamma(\ObsNum_\GroupIdx)}  \prod_{ \ChildIdx=1}^{\LatentMixtNumUnique_\GroupIdx} \ChildMoment(\ChildAux_\GroupIdx, \LatentMixtCounter_{\GroupIdx \ChildIdx})\right\}}\\
&\quad\times\frac{\prod_{\MotherIdx=1}^{\LatentChildMotherNumUnique} \MotherMoment\left(\sum_{\GroupIdx=1}^\GroupNum\ChildLaplaceExp(\ChildAux_\GroupIdx), \LatentChildMotherCounter_\MotherIdx\right) m(\dd\LatentMixtUnique_{\GroupIdx \ChildIdx} \colon \LatentChildMotherObs_{{\GroupIdx \ChildIdx}} = \MotherIdx)}{\prod_{\MotherIdx=1}^{\LatentChildMotherNumUnique} \MotherMoment\left(\sum_{\GroupIdx=1}^\GroupNum\ChildLaplaceExp(\ChildAux_\GroupIdx), \LatentChildMotherCounter_\MotherIdx\right) m(\dd\LatentMixtUnique_{\GroupIdx \ChildIdx} \colon \LatentChildMotherObs_{{\GroupIdx \ChildIdx}} = \MotherIdx)}\\
&\quad\times\frac{\frac{\ChildAux_1^{\ObsNum_{1}}}{\Gamma(\ObsNum_1+1)}\ChildMoment(\ChildAux_{1},1)\prod_{\ChildIdx=1}^{\LatentMixtNumUnique_{1}} \ChildMoment(\ChildAux_{1}, \LatentMixtCounter_{1 \ChildIdx})}{\frac{\ChildAux_1^{\ObsNum_{1} - 1}}{\Gamma(\ObsNum_1)}\prod_{\ChildIdx=1}^{\LatentMixtNumUnique_{1}} \ChildMoment(\ChildAux_{1}, \LatentMixtCounter_{1 \ChildIdx})}\\
&\quad\times \MotherMoment\left(\sum_{\GroupIdx=1}^\GroupNum\ChildLaplaceExp(\ChildAux_\GroupIdx),1\right)\int_{\MotherSpace}\Kernel(x,\MotherLoc)\MotherMeanLoc(\dd\MotherLoc)\dd x
\end{align*}
and the third term of \Cref{teo:pred} follows.

\section{Further details on the MCMC algorithm}\label{sec:mcmc_detailed}
Here we report further details on the MCMC algorithm described in Section~\ref{sec:mcmc}.

\paragraph*{Ferguson-Klass algorithm} Given a CRM $\GenCRM=\sum_{\GenIdx\ge1}\GenJump_{\GenIdx}\delta_{\GenLoc_{\GenIdx}}$, the FK algorithm samples an approximate version of $\GenCRM$ by retaining only the jumps greater than a fixed threshold $\FKThreshold$, along with their corresponding locations. The algorithm introduces a sequence of latent variables $\FKExpVar_1,\FKExpVar_2, \FKExpVar_3,\ldots$ such that the increments $\FKExpVar_1,\FKExpVar_2-\FKExpVar_1, \FKExpVar_3-\FKExpVar_2,\ldots$ are i.i.d.~exponential random variables with unit mean. The $\GenIdx$-th jump $\GenJump_{\GenIdx}$ is obtained by solving the equation $\GenIntensity([\GenJump_{\GenIdx},+\infty),\GenSpace)=\FKExpVar_{\GenIdx}$, where $\GenIntensity$ is the mean measure associated to the CRM. This procedure ensures that the sampled jumps are in decreasing order, i.e., $\GenJump_1\ge\GenJump_2\ge\ldots$. The algorithm stops when $\GenJump_{\GenIdx}$ falls below a positive threshold $\FKThreshold$, thus keeping only the jumps bigger than $\FKThreshold$. The corresponding locations are drawn independently from the base measure of the (homogeneous) CRM.

\paragraph*{Supporting algorithm} Throughout the MCMC execution, we also run a supporting algorithm that maintains a valid and coherent state at all iterations. In particular, this step warrants no gaps in the labels $\LatentChildMotherObsVec$ by relabeling elements to obtain that $\LatentChildMotherCounter_{\MotherIdx}\ge0\iff \MotherIdx\le \LatentChildMotherNumUnique$. It also guarantees consistency between the labels $\LatentMixtClusLabelsVec$ and $\LatentChildMotherObsVec$ and the sets $\boldsymbol{\ChildLoc}_{\GroupIdx}^{(a)}$, $\boldsymbol{\ChildLoc}_{\GroupIdx\MotherIdx}^{(na)}$, $\boldsymbol{\ChildLoc}_{\GroupIdx}^{(na)}$ and $\boldsymbol{\ChildJump}_{\GroupIdx}^{(a)}$, $\boldsymbol{\ChildJump}_{\GroupIdx\MotherIdx}^{(na)}$, $\boldsymbol{\ChildJump}_{\GroupIdx}^{(na)}$, for all $\MotherIdx$, $\GroupIdx=1,\ldots,\GroupNum$. For instance, in step 1., if an $\ChildIdx$ is sampled such that $\ChildLoc_{\GroupIdx\ChildIdx}\in\boldsymbol{\ChildLoc}_{\GroupIdx}^{(na)}$, then the supporting algorithm moves $\ChildLoc_{\GroupIdx\ChildIdx}$ to $\boldsymbol{\ChildLoc}_{\GroupIdx}^{(a)}$ and samples $\LatentChildMotherObs_{\GroupIdx\ChildIdx}$ according to step 3.~of the MCMC algorithm.

\paragraph*{Details for step 3.} In step 3., we sample each element of $\LatentChildMotherObsVec$ using Algorithm 3 from \cite{Neal_JCGS_2000}. In particular, the full conditional probabilities can be derived from \eqref{eq:post_t_u} and are as follows:
\begin{align*}
&\Prob(\LatentChildMotherObs_{\GroupIdx\ChildIdx}=\MotherIdx\mid -)\\
&\qquad\propto\begin{cases}
\frac{\MotherMoment(\sum_{\GroupIdx=1}^{\GroupNum}\ChildLaplaceExp(\ChildAux_\GroupIdx),\LatentChildMotherCounter_{\MotherIdx}^{-\GroupIdx\ChildIdx}+1)}{\MotherMoment(\sum_{\GroupIdx=1}^{\GroupNum}\ChildLaplaceExp(\ChildAux_\GroupIdx),\LatentChildMotherCounter_{\MotherIdx}^{-\GroupIdx\ChildIdx})} m(\LatentMixtUnique_{\GroupIdx\ChildIdx}\mid \LatentMixtUnique_{\GroupIdxAlt\ChildIdxAlt}:\LatentChildMotherObs_{\GroupIdxAlt\ChildIdxAlt}=\MotherIdx\wedge (\GroupIdxAlt,\ChildIdxAlt)\neq (\GroupIdx,\ChildIdx)) &  \MotherIdx\le\LatentChildMotherNumUnique\\
\MotherMoment(\sum_{\GroupIdx=1}^{\GroupNum}\ChildLaplaceExp(\ChildAux_\GroupIdx),1)m(\LatentMixtUnique_{\GroupIdx\ChildIdx}) & \MotherIdx=\LatentChildMotherNumUnique+1
\end{cases}
\end{align*}
where $\LatentChildMotherCounter_{\MotherIdx}^{-\GroupIdx\ChildIdx}$ is the cardinality of the set $\{\LatentMixtUnique_{\GroupIdxAlt\ChildIdxAlt}:\LatentChildMotherObs_{\GroupIdxAlt\ChildIdxAlt}=\MotherIdx\wedge (\GroupIdxAlt,\ChildIdxAlt)\neq (\GroupIdx,\ChildIdx)\}$. Whenever $\MotherIdx=\LatentChildMotherNumUnique+1$ is chosen, we generate $\MotherLoc_{\LatentChildMotherNumUnique+1}$ from its marginal prior. After this Gibbs step, the supporting algorithm discards all $\MotherLoc_{\MotherIdx}$'s such that $\LatentChildMotherCounter_{\MotherIdx}=0$.

\paragraph*{Details and alternatives for step 5.} To sample each element of $\boldsymbol{\ChildLoc}_{\GroupIdx}^{(na)}$ and $\boldsymbol{\ChildJump}_{\GroupIdx}^{(na)}$ we exploit the FK algorithm twice. First, we use it to sample an approximate version of the mother process described in Theorem~\ref{teo:post} ($iii$). Let $\bm{\MotherJump}^{(na)}$ and $\bm{\MotherLoc}^{(na)}$ be the finite dimensional vector of jumps exceeding the threshold $\FKThreshold$ and their corresponding locations. We then apply the FK algorithm the second time to sample $\boldsymbol{\ChildJump}_{\GroupIdx}^{(na)}$ and $\boldsymbol{\ChildLoc}_{\GroupIdx}^{(na)}$. Specifically, the elements of $\boldsymbol{\ChildLoc}_{\GroupIdx}^{(na)}$ are drawn from a finite mixture model with mixing function $\Kernel$, weights obtained from the normalization of $\bm{\MotherJump}^{(na)}$, and locations $\bm{\MotherLoc}^{(na)}$.

As shown in Theorem~\ref{teo:post}, the elements of $\boldsymbol{\ChildLoc}_{\GroupIdx}^{(na)}$ are generated according to a SNCP.
Therefore, an alternative approach to step 5, though we did not use it in the current MCMC, is to apply the FK algorithm only once, to sample $\boldsymbol{\ChildJump}_{\GroupIdx}^{(na)}$ and $\boldsymbol{\ChildLoc}_{\GroupIdx}^{(na)}$. In this case, the locations $\boldsymbol{\ChildLoc}_{\GroupIdx}^{(na)}$ can be sampled using simulation algorithms for SNCPs. See \cite{Moller_StatInfSpat_2003} for further details.

\section{Simulation studies}
\label{sec:simulations_appendix}
In this section, we present additional simulation studies to further characterize the HSNCP mixture model. We consider the model in \eqref{eq:unidim} with prior hyperparameters as in Section~\ref{sec:prior_elicitation}, compared with the HPD Gaussian mixtures, that is \eqref{eq:likelihood} with $\ChildProcNorm_\ell \mid \HDPMotherProc \iid \DP(\HDPChildTotMass, \HDPMotherProc)$ and $\HDPMotherProc \sim \DP(\HDPMotherTotMass,\HDPMeanLoc)$. The base measure $\HDPMeanLoc$ is a normal-inverse gamma distribution: $\HDPMeanLoc(\HDPLocMean_{\MotherIdx},\HDPLocVar_{\MotherIdx})=\Gaussian(\HDPLocMean_{\MotherIdx}\mid\HDPLocPriorMean,\HDPLocVar_{\MotherIdx}/\HDPLocPriorDenom)\times\InverseGamma(\HDPLocVar_{\MotherIdx}\mid\HDPLocPriorShape,\HDPLocPriorScale)$. To ensure a fair comparison with the HSNCP mixture model, we choose the HDP prior hyperparameters to match the two models. First, we set the total mass parameters $\HDPChildTotMass$ and $\HDPMotherTotMass$ so that the correlation between $\ChildProcNorm_{\GroupIdx}(A)$ and $\ChildProcNorm_{\GroupIdxAlt}(A)$ under the HDP matches that of the HSNCP when $A$ is the interval containing all the observations. The expression of the correlation in the HDP case is given in Example 1 of \cite{Camerlenghi_AoS_2019}. We match the prior distribution of the standard deviation for each mixture component in the HDP with the prior distribution of $\MixtSd+\MotherLocSd_{\MotherIdx}$ in the HSNCP. In this way, the prior distribution for the standard deviation of each cluster is the same for both models. Then, to align the prior variance of the atoms' mean, we set $\HDPLocPriorDenom$ so that the expected value of the prior variance of $\HDPLocMean_{\MotherIdx}$ is equal to $\MotherLocMeanPriorVar$, i.e., $\frac{\HDPLocPriorScale}{\HDPLocPriorShape-1}\frac1{\HDPLocPriorDenom}=\MotherLocMeanPriorVar$. Finally, we match the prior means by setting $\MotherLocMeanPriorMean=\HDPLocPriorMean$.

In every simulation we run each MCMC algorithm for 5,000 iterations, after a burn-in period of 1,000 iterations. For the HSNCP mixture, we use the algorithm described in Section~\ref{sec:post_inf}, while for the HDP model we run the sampler implemented in  \url{https://github.com/alessandrocolombi/hdp}.
We apply the \texttt{salso} package \citep[see][]{salso} to estimate clusters, minimizing the posterior expectation of the Variation of Information loss function.

\subsection{Comparison with the HDP}\label{sec:sim_HDP_comp}

In this simulation study we highlight the flexibility of the clustering structure of the HSNCP prior w.r.t. the hierarchical Dirichlet process prior \citep{Teh_JASA_2006}. We consider $\GroupNum=2$ groups and generate $\ObsNum_1=\ObsNum_2=500$ observations from the following mixture of Gaussian densities:
$$
\frac13\Gaussian(\Obs_{\GroupIdx\ObsIdx}\mid \mu_{\GroupIdx 1},1)+\frac13\Gaussian(\Obs_{\GroupIdx\ObsIdx}\mid \mu_{\GroupIdx 2},1)+\frac13\Gaussian(\Obs_{\GroupIdx\ObsIdx}\mid \mu_{\GroupIdx 3},1), \quad \GroupIdx=1,2
$$
where $\mu_{\GroupIdx1}\iid\Gaussian(-8,\sigma_T^2)$, $\mu_{\GroupIdx2}\iid\Gaussian(0,\sigma_T^2)$, $\mu_{\GroupIdx3}\iid\Gaussian(8,\sigma_T^2)$ for $\GroupIdx=1,2$. For each value of $\sigma_T^2$, we generate 100 datasets from the above mixture and compute the posterior cluster estimate obtained using the HSNCP and HDP mixture models. 

Following the prior elicitation procedure described in Section~\ref{sec:prior_elicitation}, we fix $\MotherLocMeanPriorMean=0$ and $\MotherLocMeanPriorVar=10$. The prior hyperparameters for the $\MotherLocVar_{\MotherIdx}$'s and for $\MixtVar$ are chosen following the same procedure, based on a dataset generated with $\sqrt{\sigma^2_T}=0.286$, resulting in $\MotherLocVarPriorShape=4.52$, $\MotherLocVarPriorScale=1.341$, $\MixtVarPriorShape=3/2$ and $\MixtVarPriorScale=20.97$.
We consider three HSNCP priors, each with different definitions of $\ChildMeanJump$ and $\MotherMeanJump$:
\begin{itemize}
\item HSNCP-1: $\ChildMeanJump$ and $\MotherMeanJump$ are given by \Cref{ex:gamma} with $\ChildProcTotMass=1.91$ and $\MotherProcTotMass=3$.
\item HSNCP-2: $\ChildMeanJump$ is given by \Cref{ex:gamma} and $\MotherMeanJump$ is given by \Cref{ex:gen_gamma} with $\ChildProcTotMass=2.8$, $\MotherProcTotMass=1$, $\MotherProcSigma=0.1$, and $\MotherProcTau=1$.
\item HSNCP-3: $\ChildMeanJump$ is given by \Cref{ex:gen_gamma} and $\MotherMeanJump$ is given by \Cref{ex:gamma} with $\ChildProcTotMass=1.7$, $\ChildProcSigma=0.05$, $\ChildProcTau=1$, and $\MotherProcTotMass=5$.
\end{itemize}

In all three cases, the hyperparameters of $\ChildMeanJump$ and $\MotherMeanJump$ are selected so that the correlation $\Cor(\ChildProcNorm_{\GroupIdx}(A),\ChildProcNorm_{\GroupIdxAlt}(A))$ equals $0.5$ for $A=(-10,10)$.

The HDP prior hyperparameters are chosen according to the procedure described at the beginning of this section, yielding $\HDPLocPriorMean=0$, $\HDPLocPriorDenom=4.85$, $\HDPLocPriorShape=1.613$, $\HDPLocPriorScale=29.71$, $\HDPChildTotMass=1$, and $\HDPMotherTotMass=2$.

\begin{figure}[ht]
\centering
\includegraphics[width=0.65\textwidth]{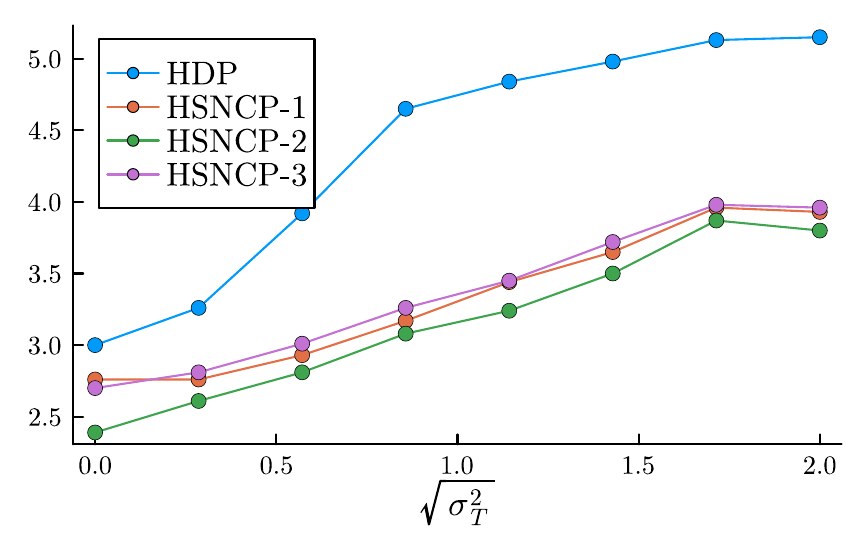}
\caption{Average (over 100 datasets) number of the clusters identified by the HDP and the three HSNCP mixtures for the simulated example in Appendix~\ref{sec:sim_HDP_comp} for each value of the parameter $\sqrt{\sigma^2_T}$.}
\label{fig:simulation_2}
\end{figure}
In Figure~\ref{fig:simulation_2} we report the average (over the 100 simulated datasets) number of clusters estimated by the HDP and the three HSNCP mixtures for each value of $\sigma_T^2$. The graph shows that, in general, the HSCNP prior identifies fewer clusters than the HDP for nonzero values of $\sigma_T^2$. Moreover, the figure illustrates that, while different choices for $\ChildMeanJump$ and $\MotherMeanJump$ can lead to slight variations in the estimated number of clusters, these differences remain minor compared to the more substantial gap between the HSNCP and HDP models.

Table~\ref{tab:simulation_1_ari} show the average 
adjusted Rand index \citep[ARI,][]{Hubert_JouClass_1985} over the 100 simulated datasets, comparing the clusters estimated by the three HSNCP mixtures to the ones obtained by a K-means algorithm applied on all observations with $K=3$, for different values of $\sqrt{\sigma^2_T}$. The results in Table~\ref{tab:simulation_1_ari} show that the HSNCP consistently identifies meaningful clustering structures, as indicated by the high ARI values across all settings.

\begin{table*}[ht]
\caption{Adjusted Rand index between the cluster estimates under the HSCNP mixtures and the K-means estimate with $K=3$ for the simulated example in Appendix~\ref{sec:sim_HDP_comp}.}
\label{tab:simulation_1_ari}
\begin{center}
\begin{tabular}{|l||c|c|c|c|c|c|c|c|c|}
\hline
$\sqrt{\sigma^2_T}$ & 0 & 0.286 & 0.571 & 0.857 & 1.14 & 1.43 & 1.71 & 2\\
\hline
HSNCP-1 & 0.880 & 0.880 & 0.908 & 0.918 & 0.927 & 0.863 & 0.806 & 0.797\\
\hline
HSNCP-2 & 0.695 & 0.800 & 0.861 & 0.901 & 0.902 & 0.845 & 0.831 & 0.782\\
\hline
HSNCP-3 & 0.850 & 0.900 & 0.937 & 0.933 & 0.925 & 0.868 & 0.816 & 0.802\\
\hline
\end{tabular}
\end{center}
\end{table*}

\subsection{Sensitivity analysis}
\label{sec:sim_sens_analysis} 
In this simulation study we consider $\Kernel(\cdot,\MotherLoc_{\MotherIdx})=\Gaussian(\cdot\mid\MotherLocMean_{\MotherIdx}, \MotherLocVar_{\MotherIdx})$ where $\MotherLoc_{\MotherIdx}=(\MotherLocMean_{\MotherIdx}, \MotherLocVar_{\MotherIdx})\in\MotherSpace=\R\times\Rp$ and we analyze the sensitivity of the across-group information sharing with respect to the prior hyperparameters of the $\MotherLocVar_{\MotherIdx}$'s. We consider $\GroupNum=2$ and we generate 30 datasets with sizes $\ObsNum_1=\ObsNum_2=700$. The generating densities are 
$\mathcal{N}(\cdot\mid2,1)$ for group~1 and $\mathcal{N}(\cdot\mid-2,1)$ for group~2. 

Following the prior elicitation procedure described in Section~\ref{sec:prior_elicitation}, we fix $\MotherLocMeanPriorMean=0$, $\MotherLocMeanPriorVar=4$, $\MixtVarPriorShape=3/2$, and $\MixtVarPriorScale=1/2(s^2-\MotherLocVarPriorScale/(\MotherLocVarPriorShape-1))$, where $s^2$ is the empirical variance of the observations. Here we investigate how the hyperparameters $\MotherLocVarPriorShape$ and $\MotherLocVarPriorScale$ of the marginal prior IG$(\MotherLocVarPriorShape,\MotherLocVarPriorScale)$ of the mother process atoms $\MotherLocVar_{\MotherIdx}$'s affect the model's ability to cluster together observations from the two Gaussian distributions associated with mean values 2 and -2.
In particular, the model assigns these observations to the same cluster if it identifies an atom of the mother process $\MotherProc$ with $\MotherLocMean_{\MotherIdx}$ around 0 and $\MotherLocVar_{\MotherIdx}$ large enough to allow for two associated child locations $\ChildLoc_{1\ChildIdx}=-2$ and $\ChildLoc_{2\ChildIdx'}=2$.
To this end, we consider a grid of values for $\PriorElicitationDistance$, and for each of them we compute the corresponding prior hyperparameters $(\MotherLocVarPriorShape,\MotherLocVarPriorScale)$ for the $\MotherLocVar_{\MotherIdx}$'s according to the procedure described in Section~\ref{sec:prior_elicitation}. We assume the L\'evy densities $\ChildMeanJump(s)$ and $\MotherMeanJump(\MotherJump)$ as in \Cref{ex:gamma}. The total mass parameters $\ChildProcTotMass$ and $\MotherProcTotMass$ are chosen so that the correlation $\Cor(\ChildProcNorm_{\GroupIdx}(A),\ChildProcNorm_{\GroupIdxAlt}(A))$ equals $0.5$ for $A=(-4,4)$. 

\begin{figure}[t]
\centering
\includegraphics[width=0.65\linewidth]{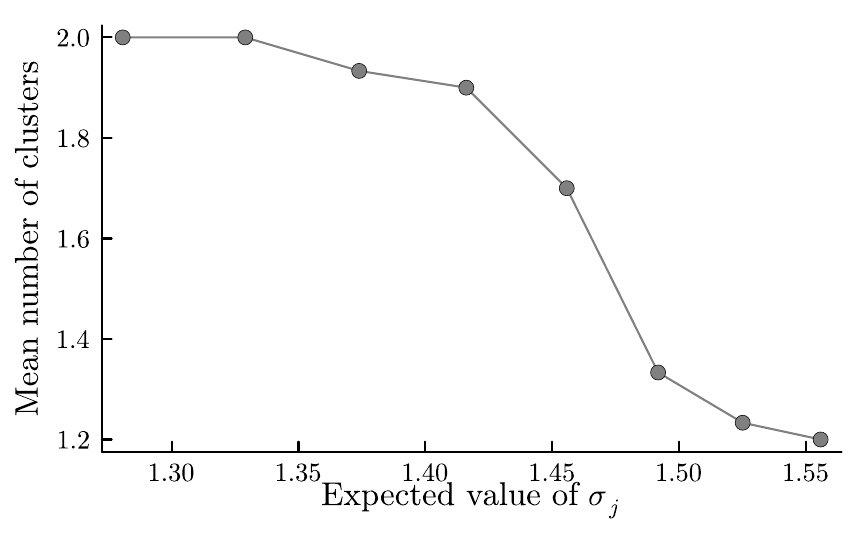}
\caption{Average (over 30 simulated datasets) number of estimated clusters of the HSNCP mixture for different prior hyperparameters of the $\MotherLocVar_\MotherIdx$'s, as a function of the prior mean of $\MotherLocSd_{\MotherIdx}$.}
\label{fig:simulation_3}
\end{figure}

Figure~\ref{fig:simulation_3} depicts the average number of detected clusters (over the 30 generated datasets) obtained for different prior hyperparameters $\MotherLocVarPriorShape,\MotherLocVarPriorScale$ of $\MotherLocVar_{\MotherIdx}$. 
The graph shows that this number abruptly decreases from 2 to 1 as the prior expected value of $\MotherLocSd_{\MotherIdx}=\sqrt{\MotherLocSd_{\MotherIdx}^2}$ increases, 
suggesting that clustering is highly sensitive to the prior specification of the $\MotherLocVar_{\MotherIdx}$'s. This sensitivity arises because the posterior distribution of these parameters does not directly depend on the observations, making the prior specification more influential. 
Fixing this marginal prior, we define the degree of flexibility in information sharing across groups, which is a choice made to reflect prior belief.

We highlight that, following the procedure described in Section~\ref{sec:prior_elicitation}, computing the closest-to-zero local minimum $d^*$ of
the kernel density estimate of the pairwise distances between pairs of observations, instead of considering a grid of values for $d^*$, we would obtain $\PriorElicitationDistance=2.7$. The HSNCP mixture corresponding to this value identifies two clusters in all the 30 repeated simulated data.

\subsection{Misspecified mixtures: an example with non-Gaussian components}
\label{sec:sim_non_gaussian}
\begin{figure}[t]
\centering
\includegraphics[width=\textwidth]{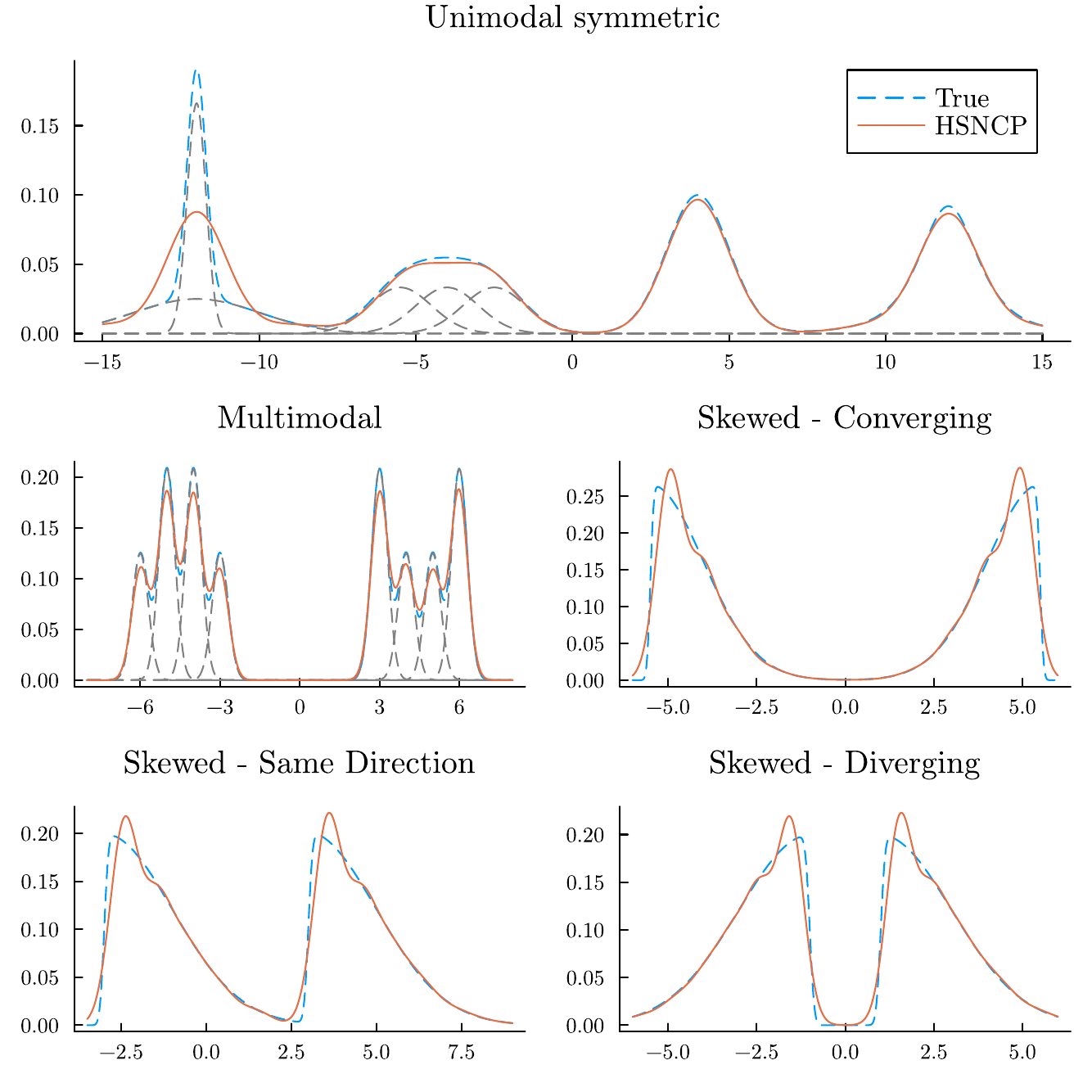}
\caption{True generating densities and the average (over 30 replicated datasets) posterior predictive densities obtained by the HSNCP mixture. If a mixture component is composed of multiple Gaussian densities, those components are represented with grey dotted lines.}
\label{fig:simulation_4_densities}
\end{figure}

In this simulation study, we illustrates the flexibility of our model in identifying clusters in \emph{misspecified} settings, that is, when the generating process is a finite mixture model $\sum_{h=1}^H w_h f_h(\cdot)$ with $f_h(\cdot)$'s non-Gaussian densities ($w_h>0$ for all $h$ and $\sum_{h=1}^H w_h=1$). We restrict to the case of one group only, i.e., $g=1$. In particular, we consider examples where the $f_h$'s are skewed, multimodal, or with heavy tails. In particular, we consider the following scenarios for the true generating densities,  shown as dotted lines in Figure~\ref{fig:simulation_4_densities}: 
\begin{enumerate}
\item Unimodal symmetric: we set $H=4$, with mixing proportions $w_1=w_2=w_3=w_4=0.25$. The component densities are:
\begin{align*}
f_1(\cdot)&=\frac12\Gaussian(\cdot\mid -12,0.09)+\frac12\Gaussian(\cdot\mid -12,4)\\
f_2(\cdot)&=\frac13\Gaussian(\cdot\mid -5.5,1)+\frac13\Gaussian(\cdot\mid -4,1)+\frac13\Gaussian(\cdot\mid -2.5,1)\\
f_3(\cdot)&=\Gaussian(\cdot\mid 4,1)\\ 
f_4(\cdot)&=\StudentT_3(\cdot\mid 12,1) 
\end{align*}
where $\StudentT_{\nu}(\cdot\mid, \mu, \sigma)$ is the pdf of a  $t$-distribution with $\nu$ degrees of freedom, mean $\mu$ and scale $\sigma$.
\item Multimodal: we set $H=2$, with $w_1=w_2=1/2$ and
\begin{align*}
f_1(\cdot)&=\frac3{16}\Gaussian(\cdot\mid -6,0.09)+\frac5{16}\Gaussian(\cdot\mid -5,0.09)+\frac5{16}\Gaussian(\cdot\mid -4,0.09)+\frac3{16}\Gaussian(\cdot\mid -3,0.09)\\
f_2(\cdot)&=\frac5{16}\Gaussian(\cdot\mid 6,0.09)+\frac3{16}\Gaussian(\cdot\mid 5,0.09)+\frac3{16}\Gaussian(\cdot\mid -4,0.09)+\frac5{16}\Gaussian(\cdot\mid -3,0.09)
\end{align*}
\item Skewed: we consider $H=2$, $w_1=w_2=1/2$, with two skew-normal components $\SkewedGaussian(\cdot\mid\xi,\omega,\alpha)$, where $\xi$ is the location, $\omega$ is the scale, and $\alpha$ is the shape. We analyze three configurations, classified on the direction of the skewedness (see the plots of the true densities in Figure~\ref{fig:simulation_4_densities}):
\begin{enumerate}
\item Converging: $f_1(\cdot)=\SkewedGaussian(\cdot\mid-5.5,1.5,20)$, and $f_2(\cdot)=\SkewedGaussian(\cdot\mid5.5,1.5,-20)$.
\item Diverging: $f_1(\cdot)=\SkewedGaussian(\cdot\mid-3,2,20)$, and $f_2(\cdot)=\SkewedGaussian(\cdot\mid3,2,20)$.
\item Same direction:$f_1(\cdot)=\SkewedGaussian(\cdot\mid-1,2,-20)$, and $f_2(\cdot)=\SkewedGaussian(\cdot\mid1,2,20)$.
\end{enumerate}  
\end{enumerate}
For each of the above densities, we simulate 30 different datasets with $\ObsNum_{1}=3{,}000$ datapoints.

We follow the prior elicitation procedure for the HSNCP prior described in \Cref{sec:prior_elicitation}. We set $\MotherLocMeanPriorMean=0$, $\MotherLocMeanPriorVar=64$, allowing the locations of the mother process $\MotherProc$ to span the full range of observations. For each of the five scenarios, one generated dataset is used to determine the value of $\PriorElicitationDistance$ and the corresponding values of $\MotherLocVarPriorShape$, $\MotherLocVarPriorScale$, and $\MixtVarPriorScale$. The hyperparameters do not change across the 30 replicated datasets for each scenario.
We assume $\ChildMeanJump(s)$ and $\MotherMeanJump(s)$ to be as defined in \Cref{ex:gamma} with total mass parameters $\MotherProcTotMass=5$ and $\ChildProcTotMass=1$. 

In this simulation study, we define an \textit{oracle} cluster estimate by assigning each observation to the true mixture component $f_h$ that attains the highest density at that observation. That is, observation $\Obs_{1\ObsIdx}$ is assigned to component $f_h$ if and only if $f_h(\Obs_{1\ObsIdx})>f_{h'}(\Obs_{1\ObsIdx})$ for all $h'=1,\ldots,H$ such that $h'\neq h$. In Table~\ref{tab:simulation_4_ari} we report the adjusted Rand index between the oracle cluster estimate and the one under the HSNCP model. The results show values close to one across all scenarios, indicating that the HSNCP recovers clustering configurations nearly identical to the oracle. Figure~\ref{fig:simulation_4_densities} displays the average posterior predictive densities. We observe that the HSNCP achieves excellent density estimation in most scenarios.

These results highlight that the HSCNP mixture can accurately recover the mixture components and their shapes, even when the true components densities $f_h(\cdot)$'s differ from $\ObsMixt(\cdot\mid x)$. This flexibility is highly advantageous in practice: while standard mixture models often require careful exploratory analysis to select an appropriate mixture component distribution, the HSNCP can adapt to a wide range of mixture component shapes using a default choice for $\ObsMixt(\cdot\mid x)$, such as Gaussian densities. Moreover, such choices are beneficial from a computational perspective, allowing for closed form full conditional distributions in the MCMC algorithm. This improves both computational efficiency and mixing, avoiding the need for numerical approximations and Metropolis-Hastings steps.

\begin{table*}[t]
\caption{Adjusted Rand index between the HSNCP mixture estimated clusters and the \textit{oracle} cluster estimate.}
\label{tab:simulation_4_ari}
\begin{center}
\begin{tabular}{|l|c|}
\hline
Experiment & Mean ARI\\
\hline
1. Unimodal symmetric & 0.751\\
2. Multimodal & 1.000\\
3.(a) Skewed - Converging & 0.999\\
3.(b) Skewed - Diverging & 0.983\\
3.(c) Skewed - Same direction & 0.839\\
\hline
\end{tabular}
\end{center}
\end{table*}

\FloatBarrier

\section{Additional plots for the application in \Cref{sec:sloan}}\label{sec:sloan_supp}
\begin{figure}
\centering
\includegraphics[width=0.8\linewidth]{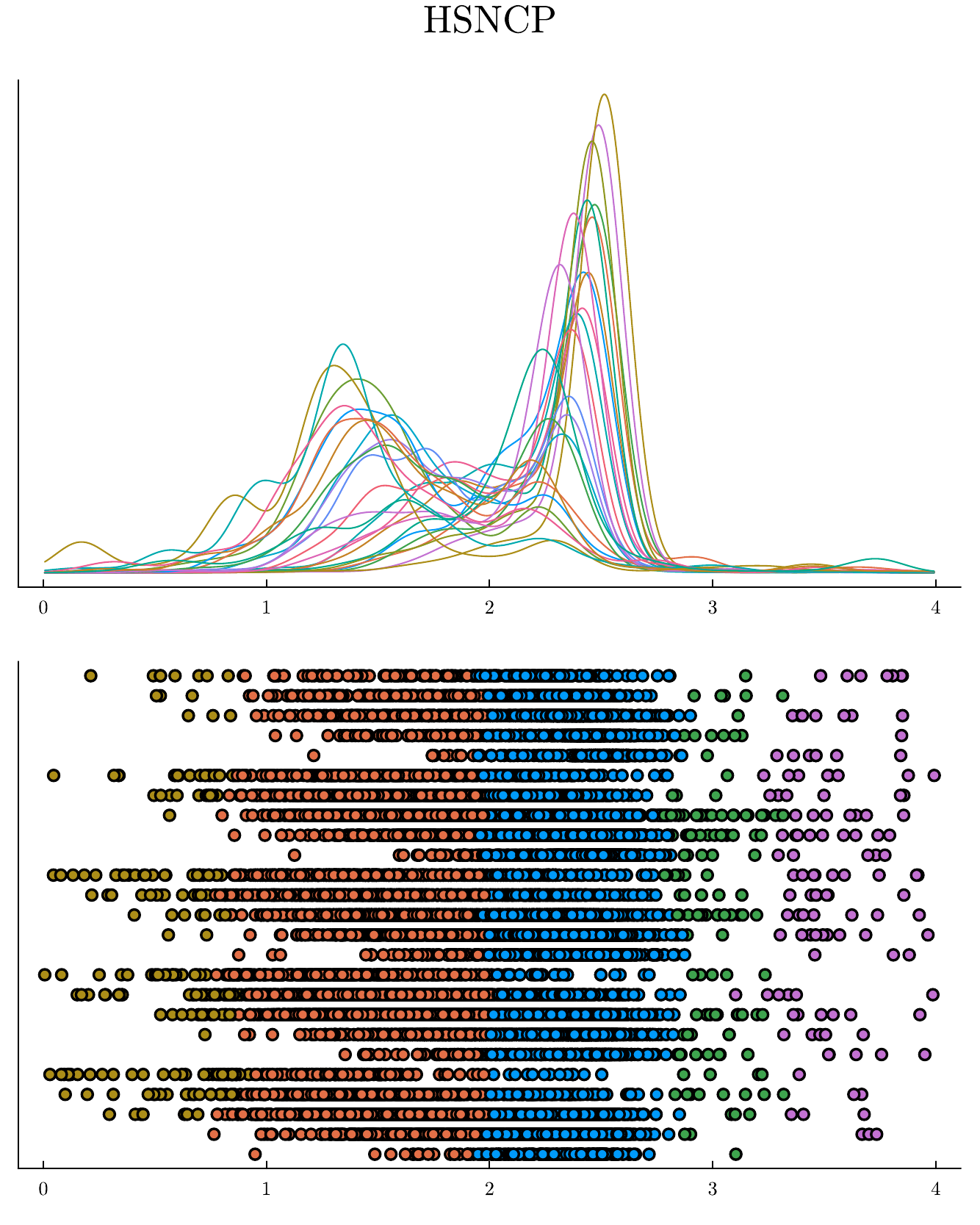}
\caption{Estimated density (top) and clustering structure (bottom) under the HSNCP model.}
\label{fig:sloan_HSNCP}
\end{figure}

\begin{figure}
\centering
\includegraphics[width=0.8\linewidth]{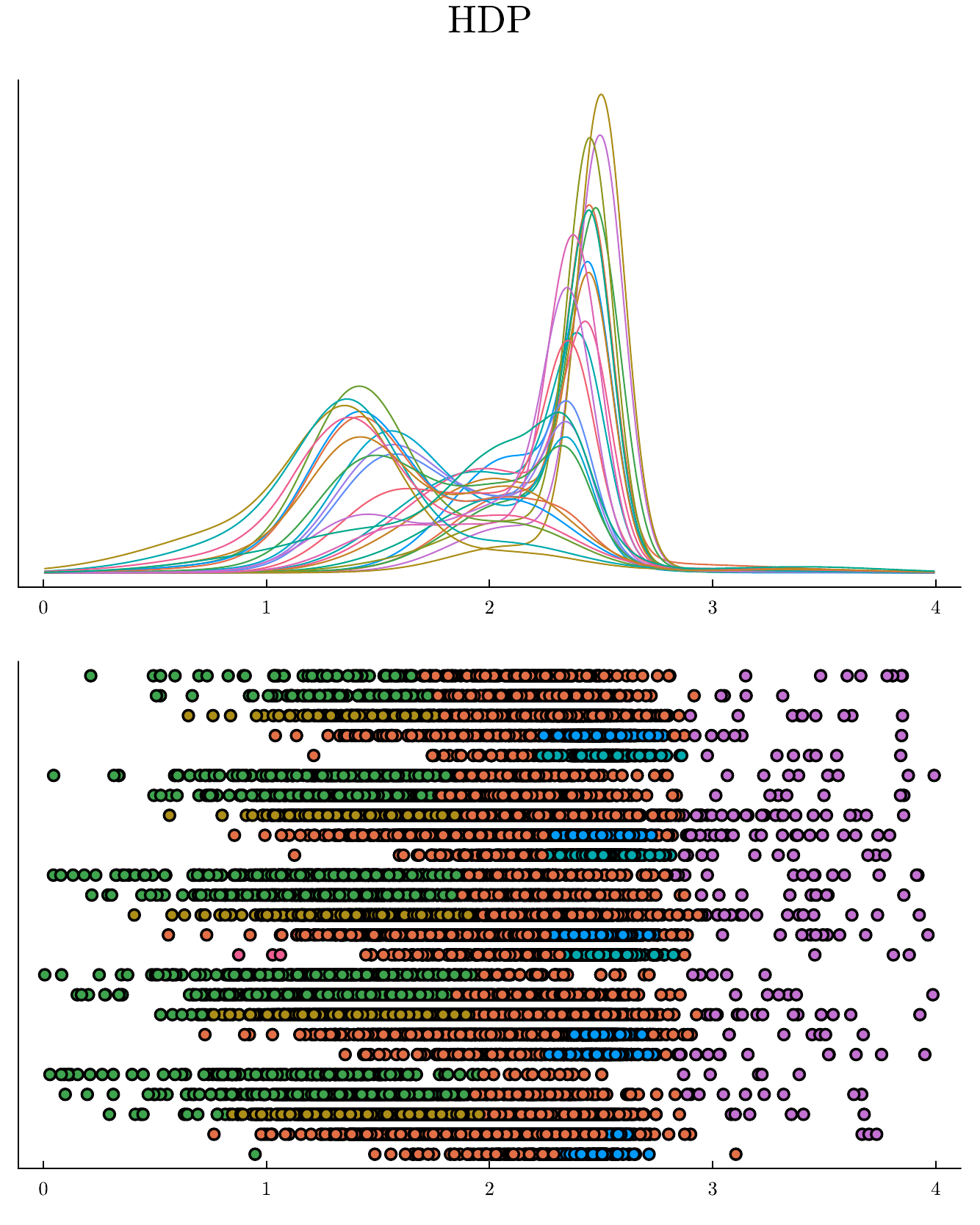}
\caption{Estimated density (top) and clustering structure (bottom) under the HDP model.}
\label{fig:sloan_HDP}
\end{figure}

\begin{figure}
\centering
\includegraphics[width=\textwidth]{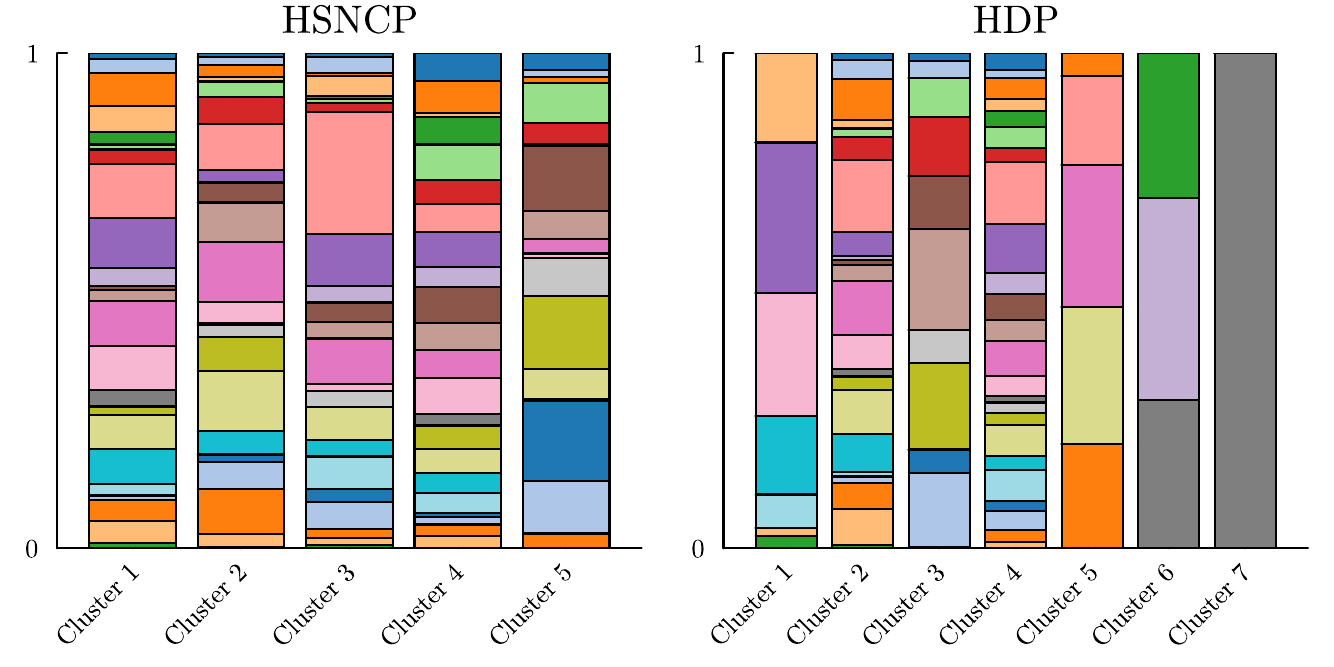}
\caption{Stacked barplot of the estimated clusters, with colors indicating the contribution of each group.}
\label{fig:sloan_clus_barplot}
\end{figure}

This section contains additional plots for the application in \Cref{sec:sloan}. \Cref{fig:sloan_HSNCP,fig:sloan_HDP} display the estimated densities and cluster estimates
obtained under the HSNCP and HDP models, respectively. In each figure, the top panel shows the posterior density estimates, where each line corresponds to a group, while the bottom panel illustrates the estimated cluster configuration, where each point represents a galaxy and each row corresponds to a group, with colors indicating the cluster assignments.
\Cref{fig:sloan_clus_barplot} summarizes the overall composition of the estimated clusters through a stacked barplot, where each bar represents a cluster and the colored segments indicate the proportion of galaxies from each group within that cluster.

\FloatBarrier
\end{document}

%% file: preamble.tex
\PassOptionsToPackage{hyphens}{url}
\PassOptionsToPackage{dvipsnames,svgnames,x11names}{xcolor}
\documentclass[12pt]{article}

\usepackage{amsthm,amsmath,amsfonts,amssymb}
\usepackage{graphicx}
\usepackage{enumerate}
\usepackage{natbib}
\usepackage{url} 
\usepackage{bm}
\usepackage{hyperref}
\hypersetup{hidelinks,unicode}
\usepackage{cleveref}
\usepackage{placeins}

\usepackage{iftex}
\ifPDFTeX
  \usepackage[T1]{fontenc}
  \usepackage[utf8]{inputenc}
  \usepackage{textcomp} 
\else 
  \usepackage{unicode-math}
  \defaultfontfeatures{Scale=MatchLowercase}
  \defaultfontfeatures[\rmfamily]{Ligatures=TeX,Scale=1}
\fi
\usepackage{lmodern}
\ifPDFTeX\else  
\fi
\IfFileExists{upquote.sty}{\usepackage{upquote}}{}
\IfFileExists{microtype.sty}{
  \usepackage[]{microtype}
  \UseMicrotypeSet[protrusion]{basicmath} 
}{}
\makeatletter
\@ifundefined{KOMAClassName}{
  \IfFileExists{parskip.sty}{%
    \usepackage{parskip}
  }{
    \setlength{\parindent}{0pt}
    \setlength{\parskip}{6pt plus 2pt minus 1pt}}
}{
  \KOMAoptions{parskip=half}}
\makeatother
\usepackage{xcolor}
\makeatletter
\ifx\paragraph\undefined\else
  \let\oldparagraph\paragraph
  \renewcommand{\paragraph}{
    \@ifstar
      \xxxParagraphStar
      \xxxParagraphNoStar
  }
  \newcommand{\xxxParagraphStar}[1]{\oldparagraph*{#1}\mbox{}}
  \newcommand{\xxxParagraphNoStar}[1]{\oldparagraph{#1}\mbox{}}
\fi
\ifx\subparagraph\undefined\else
  \let\oldsubparagraph\subparagraph
  \renewcommand{\subparagraph}{
    \@ifstar
      \xxxSubParagraphStar
      \xxxSubParagraphNoStar
  }
  \newcommand{\xxxSubParagraphStar}[1]{\oldsubparagraph*{#1}\mbox{}}
  \newcommand{\xxxSubParagraphNoStar}[1]{\oldsubparagraph{#1}\mbox{}}
\fi
\makeatother

\usepackage{longtable,booktabs,array}
\usepackage{calc} 
\usepackage{etoolbox}
\makeatletter
\patchcmd\longtable{\par}{\if@noskipsec\mbox{}\fi\par}{}{}
\makeatother
\IfFileExists{footnotehyper.sty}{\usepackage{footnotehyper}}{\usepackage{footnote}}
\makesavenoteenv{longtable}
\usepackage{graphicx}
\makeatletter
\def\maxwidth{\ifdim\Gin@nat@width>\linewidth\linewidth\else\Gin@nat@width\fi}
\def\maxheight{\ifdim\Gin@nat@height>\textheight\textheight\else\Gin@nat@height\fi}
\makeatother
\setkeys{Gin}{width=\maxwidth,height=\maxheight,keepaspectratio}
\makeatletter
\def\fps@figure{htbp}
\makeatother

\makeatletter
\@ifpackageloaded{caption}{}{\usepackage{caption}}
\AtBeginDocument{%
\ifdefined\contentsname
  \renewcommand*\contentsname{Table of contents}
\else
  \newcommand\contentsname{Table of contents}
\fi
\ifdefined\listfigurename
  \renewcommand*\listfigurename{List of Figures}
\else
  \newcommand\listfigurename{List of Figures}
\fi
\ifdefined\listtablename
  \renewcommand*\listtablename{List of Tables}
\else
  \newcommand\listtablename{List of Tables}
\fi
\ifdefined\figurename
  \renewcommand*\figurename{Figure}
\else
  \newcommand\figurename{Figure}
\fi
\ifdefined\tablename
  \renewcommand*\tablename{Table}
\else
  \newcommand\tablename{Table}
\fi
}
\@ifpackageloaded{float}{}{\usepackage{float}}
\floatstyle{ruled}
\@ifundefined{c@chapter}{\newfloat{codelisting}{h}{lop}}{\newfloat{codelisting}{h}{lop}[chapter]}
\floatname{codelisting}{Listing}

\makeatother
\makeatletter
\@ifpackageloaded{caption}{}{\usepackage{caption}}
\@ifpackageloaded{subcaption}{}{\usepackage{subcaption}}
\makeatother

\ifLuaTeX
  \usepackage{selnolig}  
\fi
\usepackage[]{natbib}
\usepackage{bookmark}

\IfFileExists{xurl.sty}{\usepackage{xurl}}{} 
\hypersetup{
  pdftitle={Hierarchical shot-noise Cox process mixtures for clustering across groups},
  pdfauthor={Alessandro Carminati; Mario Beraha; Federico Camerlenghi; Alessandra Guglielmi},
  pdfkeywords={partial exchangeability; normalized random measures; dependent random measures; density estimation},
  colorlinks=true,
  linkcolor={blue},
  filecolor={Maroon},
  citecolor={Blue},
  urlcolor={Blue},
  pdfcreator={LaTeX via pandoc}}

\usepackage{comment}

\usepackage{enumitem}

\usepackage[normalem]{ulem}

\usepackage{authblk}
\date{}
\title{\bf Hierarchical shot-noise Cox process mixtures for clustering across groups}

\author[1]{Alessandro Carminati\thanks{The authors gratefully acknowledge support from the Italian Ministry of Education, University and Research (MUR), ‘Dipartimenti di Eccellenza’ grant 2023-2027 and from MUR - Prin 2022 - Grant no. 2022CLTYP4, funded by the European Union—Next Generation EU.}}
\author[2]{Mario Beraha}
\author[2]{Federico Camerlenghi}
\author[1]{Alessandra Guglielmi}
\affil[1]{Department of Mathematics, Politecnico di Milano}
\affil[2]{Department of Economics, Management and Statistics, University of Milano–Bicocca}
\date{\today}

\providecommand{\keywords}[1]{
  \small	
  \textbf{\textit{Keywords:}} #1
  \normalsize
}

\theoremstyle{plain}

\newtheorem{theorem}{Theorem}[section]
\newtheorem{lemma}[theorem]{Lemma}
\theoremstyle{definition}

\newtheorem{example}{Example}

\theoremstyle{remark}

\newcommand{\iid}{\overset{iid}{\sim}}
\newcommand{\ind}{\overset{ind}{\sim}}
\newcommand{\E}{\mathbb E}
\newcommand{\Var}{\mathrm{Var}}
\newcommand{\Cov}{\mathrm{Cov}}
\newcommand{\Cor}{\mathrm{Cor}}
\newcommand{\Prob}{\mathbb P}

\newcommand{\Law}{\mathcal L}
\newcommand{\Gaussian}{\mathcal{N}}
\newcommand{\StudentT}{t}
\newcommand{\SkewedGaussian}{\mathcal{SN}}
\newcommand{\InverseGamma}{\mathrm{IG}}
\newcommand{\e}{\operatorname{e}}
\newcommand{\dd}{\mathrm d}
\newcommand{\middot}{\mathbin{\scalebox{0.4}{$\bullet$}}}

\newcommand{\GenSpace}{\mathbb X}
\newcommand{\GenPP}{\tilde N}
\newcommand{\GenCRM}{\tilde\mu}
\newcommand{\GenNRMI}{\tilde p}
\newcommand{\GenIntensity}{\nu}
\newcommand{\GenMeanJump}{\rho}
\newcommand{\GenMeanMass}{C}
\newcommand{\GenMeanLoc}{G_0}
\newcommand{\GenJump}{\gamma}
\newcommand{\GenLoc}{\tau}
\newcommand{\GenIdx}{j}

\newcommand{\MotherSpace}{\Theta}
\newcommand{\ChildSpace}{\mathbb X}
\newcommand{\R}{\mathbb R}
\newcommand{\Rp}{\R_+}
\newcommand{\N}{\mathbb N}

\renewcommand{\mid}{\,|\,}

\newcommand{\MotherProc}{\Lambda}
\newcommand{\MotherMeanJump}{\rho_0}
\newcommand{\MotherMeanLoc}{G_0}
\newcommand{\MotherIntensity}{\nu_0}
\newcommand{\MotherJump}{\gamma}
\newcommand{\MotherLoc}{\tau}
\newcommand{\MotherIdx}{j}
\newcommand{\MotherLaplaceExp}{\psi_0}
\newcommand{\MotherMoment}{\kappa_0}

\newcommand{\ChildProc}{\tilde\mu}
\newcommand{\ChildProcList}{\ChildProc_1,\dots,\ChildProc_\GroupNum}
\newcommand{\ChildMeanJump}{\rho}

\newcommand{\ChildMeanLoc}{\eta_\MotherProc}

\newcommand{\ChildJump}{S}

\newcommand{\ChildLoc}{\phi}
\newcommand{\ChildIdx}{h}
\newcommand{\ChildIdxAlt}{h'}
\newcommand{\ChildProcNorm}{\tilde p}
\newcommand{\ChildLaplaceExp}{\psi}
\newcommand{\ChildMoment}{\kappa}
\newcommand{\ChildAux}{U}
\newcommand{\ChildAuxVec}{\bm U}
\newcommand{\ChildProcPP}{\tilde N}
\newcommand{\Kernel}{k}

\newcommand{\pEPPFFirstInt}{\mathcal I_{1,2}}
\newcommand{\pEPPFSecondInt}{\mathcal I_{1,1}}
\newcommand{\MotherEPPFInt}{\mathcal I_{1}}

\newcommand{\GroupIdx}{\ell}
\newcommand{\GroupIdxAlt}{\ell'}
\newcommand{\GroupNum}{g}
\newcommand{\Obs}{Y}
\newcommand{\ObsIdx}{i}
\newcommand{\ObsIdxAlt}{i'}
\newcommand{\ObsNum}{n}
\newcommand{\ObsMixt}{f}

\newcommand{\LatentMixt}{X}
\newcommand{\LatentMixtVec}{\bm X}
\newcommand{\LatentMixtNumUnique}{K}
\newcommand{\LatentMixtUnique}{X^*}
\newcommand{\LatentMixtCounter}{\xi}
\newcommand{\LatentMixtClusLabels}{c}
\newcommand{\LatentMixtClusLabelsVec}{\bm{\LatentMixtClusLabels}}
\newcommand{\LatentChildMother}{t}

\newcommand{\LatentChildMotherObs}{T}
\newcommand{\LatentChildMotherObsVec}{\bm \LatentChildMotherObs}
\newcommand{\LatentChildMotherCounter}{\zeta}
\newcommand{\LatentChildMotherNumUnique}{|\LatentChildMotherObsVec|}

\newcommand{\FKThreshold}{\varepsilon}

\newcommand{\FKExpVar}{\eta}

\newcommand{\PP}{\mathrm{PP}}
\newcommand{\CRM}{\mathrm{CRM}}

\newcommand{\DP}{\mathrm{DP}}
\newcommand{\HDP}{\mathrm{HDP}}
\newcommand{\HSNCP}{\mathrm{HSNCP}}
\newcommand{\NormHSNCP}{\mathrm{normHSNCP}}

\newcommand{\MotherProcTotMass}{a_0}
\newcommand{\MotherProcSigma}{\sigma_0}
\newcommand{\MotherProcTau}{\tau_0}
\newcommand{\MotherLocMean}{\mu}
\newcommand{\MotherLocSd}{\sigma}
\newcommand{\MotherLocVar}{\MotherLocSd^2}
\newcommand{\MotherLocMeanPriorMean}{\MotherLocMean_0}
\newcommand{\MotherLocMeanPriorVar}{\MotherLocVar_0}
\newcommand{\MotherLocMeanPriorSd}{\MotherLocSd_0}
\newcommand{\MotherLocVarPriorShape}{\alpha_0}
\newcommand{\MotherLocVarPriorScale}{\beta_0}
\newcommand{\ChildProcTotMass}{a}
\newcommand{\ChildProcSigma}{\sigma}
\newcommand{\ChildProcTau}{\tau}
\newcommand{\MixtSd}{\sigma_f}
\newcommand{\MixtVar}{\MixtSd^2}
\newcommand{\MixtVarPriorShape}{\alpha_f}
\newcommand{\MixtVarPriorScale}{\beta_f} 
\newcommand{\PriorElicitationDistance}{d^*}
\newcommand{\PriorElicitationCoef}{\lambda}

\newcommand{\HDPLocMean}{\mu}
\newcommand{\HDPLocSd}{\sigma}
\newcommand{\HDPLocVar}{\HDPLocSd^2}
\newcommand{\HDPChildTotMass}{b}
\newcommand{\HDPMotherTotMass}{b_0}
\newcommand{\HDPMeanLoc}{F_0}
\newcommand{\HDPMotherProc}{\tilde p_0}
\newcommand{\HDPLocPriorMean}{\HDPLocMean_{\HDP}}
\newcommand{\HDPLocPriorDenom}{\lambda_{\HDP}}
\newcommand{\HDPLocPriorShape}{\alpha_{\HDP}}
\newcommand{\HDPLocPriorScale}{\beta_{\HDP}}